\documentclass[oldversion]{aa}

\usepackage{graphicx}
\usepackage{txfonts}
\usepackage[round]{natbib}
\usepackage{amssymb}
\usepackage[figuresright]{rotating}
\bibpunct{(}{)}{;}{a}{}{,}
\def\ion#1#2{{\rm #1}\,{\sc #2}}

\begin{document}

\title{Molecular gas in high-velocity clouds: revisited scenario
\thanks{Based on observations collected with the IRAM 30~m telescope at Pico
Veleta, Spain, in June 20--23, 2006.}}

%\subtitle{}

\author{M. Dessauges-Zavadsky\inst{1},
	F. Combes\inst{2},
        \and
        D. Pfenniger\inst{1}
        }

\offprints{miroslava.dessauges@obs.unige.ch}

\institute{Observatoire de Gen\`eve, Universit\'e de Gen\`eve, 51 Ch. des 
           Maillettes, 1290 Sauverny, Switzerland
           \and
           Observatoire de Paris, LERMA, 61 Av. de l'Observatoire, 75014 Paris, 
           France}

\date{Received; accepted}
 
\abstract{We report a new search for $^{12}$CO(1--0) emission in high-velocity 
clouds (HVCs) performed with the IRAM 30~m millimeter-wave telescope. This 
search was motivated by the recent detection of cold dust emission ($T\sim 
10.7$~K) in the HVCs of Complex~C, implying a total gas column density 5 times 
larger than the column density measured in \ion{H}{i} and suggesting undetected 
gas, presumably in molecular form. Despite a spatial resolution which is three 
times better and sensitivity twice as good compared to previous studies, no CO 
emission is detected in the HVCs of Complex~C down to a best 5~$\sigma$ limit 
of 0.16 K~km~s$^{-1}$ at a $22''$ resolution. The non-detection of both the 
$^{12}$CO(1--0) emission and of the diffuse H$_2$ absorption with the Far 
Ultraviolet Spectroscopic Explorer does not provide any evidence in favor of 
large amounts of molecular gas in these HVCs and hence in favor of the infrared 
findings. We discuss different configurations which, however, allow us to 
reconcile the negative CO result with the presence of molecular gas and cold 
dust emission. H$_2$ column densities higher than our detection limit, 
$N({\rm H}_2) = 3\times 10^{19}$ cm$^{-2}$, are expected to be confined in very 
small and dense clumps with 20 times smaller sizes than the 0.5 pc clumps 
resolved in our observations according to the results obtained in cirrus 
clouds, and might thus still be highly diluted. As a consequence, the 
inter-clump gas at the 1 pc scale, as resolved in our data, has a volume 
density lower than 20 cm$^{-3}$ and already appears as too diffuse to excite 
the CO molecules. The observed physical conditions in the HVCs of Complex~C 
also play an important role against CO emission detection. The sub-solar 
metallicity of $0.1-0.3$ dex affects the H$_2$ formation rate onto dust grains, 
and it has been shown that the CO-to-H$_2$ conversion factor in low metallicity 
media is 60 times higher than at the solar metallicity, leading for a given 
H$_2$ column density to a 60 times weaker integrated CO intensity. And the very 
low dust temperature estimated in these HVCs implies the possible presence of 
gas cold enough ($< 20$~K) to cause CO condensation onto dust grains under 
interstellar medium pressure conditions and thus CO depletion in gas-phase 
observations.

}

\maketitle

%
%________________________________________________________________

\section{Introduction}

A meaningful definition of intermediate-velocity clouds (IVCs) and 
high-velocity clouds (HVCs) is hard to establish. Generally speaking, these
neutral gas clouds of the Milky Way halo refer to dynamically significant gas 
moving several kiloparsecs above the Galactic plane with intermediate ($30 \leq 
|\upsilon_{\rm LSR}| \leq 90$ km~s$^{-1}$) and high ($|\upsilon_{\rm LSR}| \geq 
90$ km~s$^{-1}$) radial velocities, respectively. Such radial velocities cannot 
be ascribed to any reasonable model of differential Galactic rotation. IVCs and 
HVCs are found in large complexes with angular sizes of $10-90$ degrees and 
cover $10-37$\% of the sky \citep{murphy95,hartmann97}. Since their discovery 
in 21~cm observations by \citet{muller63}, HVCs have been the target of 
numerous studies \citep[see][ for a review]{wakker97a,woerden04}, and it became 
rapidly clear that HVCs may have significant implications on the Galaxy as well 
as on more general issues related to Galactic formation and evolution and to 
the intergalactic medium (IGM). Thus, the knowledge of both IVC and HVC 
nature or origin may lead to fundamental insights.

The origin of IVCs and HVCs, however, is still a matter of debate. The 
difficulty in determining their distances and metallicities is the major 
obstacle to pining down their origin. Indeed, the values of many of the 
physical parameters that can be derived from observations are proportional to a 
power of the distance (e.g. ${\rm mass}\propto D^2$, ${\rm size}\propto D$, 
${\rm volume~density}\propto D^{-1}$), and the metallicity provides indications 
on a Galactic or extragalactic origin. Recent studies with the Far Ultraviolet
Spectroscopic Explorer (FUSE) and the Space Telescope Imaging Spectrograph 
(STIS) tend to show that the chemical composition of IVCs and HVCs is not
uniform, with metallicities varying from 0.1 to 1.0 solar
\citep{wakker99,richter01a,richter01c,gibson00,gibson01,wakker01b,richter03a}.
This strongly suggests that the Galactic halo cloud phenomenon is diverse and 
can be separated into three different categories. 

First, some HVC complexes are proven to be associated with the Milky Way in 
lying only a few kpc from the disk due to absorptions detected in front of 
Galactic halo stars \citep[e.g.][]{woerden99} and having near-solar 
metallicities. This supports a Galactic origin and more specifically the 
``Galactic fountain'' model where gas is ejected in the halo by the star 
formation feedback mechanisms \citep{shapiro76,houck90}. Second, some HVCs
result from the gas left over from the formation of the Milky Way 
\citep{oort70}. They are dynamically associated with nearby galaxies and are 
now raining down to the Galactic disk. This is the case, for example, of the 
dominant HVC feature, the so-called Magellanic Stream, at $\sim 50$ kpc and 
with a metallicity between the Small and Large Magellanic Clouds 
\citep[$\sim 0.3$ solar,][]{lu98}, which was identified as tidal debris from 
the interaction between the Magellanic Clouds and the Milky Way 
\citep{mathewson74}. And third, other HVCs with low metallicities of $\sim 0.1$ 
solar have very likely an extragalactic --~dwarf or intergalactic~-- origin. 
\citet{blitz99} suggested that some HVCs may represent material that failed to 
form galaxies. This material would be in gravitationally-bound systems, where 
most of the gravity would be provided by dark matter. These ``mini-halos'' 
would collect into filaments, the nearby ones falling into the Local Group. 
HVCs might then play a major role in the galaxy formation process. Explanations 
for the origins of IVCs mirror those offered for the HVCs, since there are many 
reasons to consider the origins of HVCs and IVCs together. Most of the IVCs 
fall into the ``Galactic fountain'' category \citep{richter03b}, because they 
appear to be located between 0.3 and 2.1 kpc away from the Galactic plane 
\citep{wakker01b} and have roughly solar metallicities.

Access to the total mass content of IVCs and HVCs is another key piece of 
information that can help pin down their origin. The vast majority of our 
knowledge on these clouds comes from 21~cm \ion{H}{i} surveys, but what is 
their molecular content~? For HVCs with an extragalactic origin, the molecular 
gas may be a possible counterpart of their dark matter content. The higher and 
higher angular resolution \ion{H}{i} maps have shown considerable structure in 
HVCs down to the scale of the resolution \citep{schwarz81,wakker91,wakker02}. 
\citet{wakker01a} found that this cloud structure is hierarchical and has 
structural similarities: the brightest knots are embedded in the brightest 
parts of the smoother background structure. This is the basic feature of 
fractals which allows a large amount of mass in the knots/cores within the 
clouds to remain hidden. The concept of fractals was first applied to HVC maps 
by \citet{vogelaar94}. The resolved \ion{H}{i} cores within the HVCs have sizes 
down to at least 1 arcmin, \ion{H}{i} column densities of a few times $10^{20}$ 
cm$^{-2}$, brightness temperatures of up to 30 K, and velocity widths of 
typically $\sim 8-10$ km~s$^{-1}$. They hence appear to consist of cold gas, 
and most of the velocity width seen at lower resolution is due to random 
motions of these cores. The central densities in these cores can reach $>80 
D^{-1}_{\rm kpc}$ cm$^{-3}$, where $D_{\rm kpc}$ is the distance of the HVC in 
kpc. Depending on the distance, these central densities may be high enough to 
form molecular gas. 

The search for molecular gas in IVCs and HVCs was not very successful before UV 
measurements became possible. Indeed, except for two IVCs, the Draco Cloud 
\citep{mebold85} and IV\,21 \citep{weiss99,heithausen01}, which are observed in 
the disk-halo interface at less than 500 pc and may well be ``normal'' 
molecular clouds at exceptionally large $z$ heights, no CO emission or 
absorption has been detected in Galactic halo clouds so far. \citet{wakker97b} 
attempted to observe the CO emission toward six dense HVC cores, but no CO 
emission was detected down to a best 5~$\sigma$ limit of 0.077 K~km~s$^{-1}$ in 
their observations done with the NRAO 12~m telescope at Kitt Peak at an angular 
resolution of 1 arcmin. Similarly, \citet{combes00} used the absorption 
technique toward 27 quasars known to be strong millimetric continuum sources, 
and obtained only one tentative detection of HCO$^+$ in a HVC. Molecular gas 
was found for the first time in HVC gas in absorption through the H$_2$ 
electronic Lyman and Werner bands toward UV-bright background sources by 
\citet{richter99}. Further H$_2$ detections in absorption followed with the 
advent of FUSE in HVCs \citep{richter01a,richter01b,sembach01,wakker06} 
and in IVCs, half of which currently show the presence of H$_2$ 
\citep{richter03b,wakker06,gillmon06}. 

Additional strong evidence for molecular gas in HVCs was recently provided by 
\citet[][ hereafter MD05]{miville05a}. They demonstrated the first detection of 
dust emission in the HVCs located on the edge of Complex~C by comparing 21~cm 
data from the Green Bank Telescope (GBT) with very sensitive infrared 
observations from the Spitzer Space Telescope and from IRIS 
\citep[Improved Reprocessing of the IRAS Survey,][]{miville05b}. They found 
clear correlations between \ion{H}{i} and the infrared emission at 24, 60, 100, 
and 160 $\mu$m. Previously dust emission was unsuccessfully looked for in HVCs 
using the Infrared Astronomical Satellite (IRAS) data \citep{wakker86}. The 
detected dust is found to be significantly colder, $T_{\rm HVC} = 
10.7^{+0.9}_{-0.8}$~K, than in the local interstellar medium, $T_{\rm ISM} = 
17.5$~K, in accordance with its association with cold gas. This dust emission, 
in addition, suggests a total column density of gas more than 5 times higher 
than the observed \ion{H}{i}. This cold gas component, presumably in the form 
of molecular hydrogen, may thus be an important contribution to the total mass 
of HVCs.

We further investigate the molecular content of HVCs, presenting a new survey 
for CO emission performed with the IRAM 30~m telescope toward two main HVC 
\ion{H}{i} cores of Complex~C for which the dust emission was detected by MD05. 
Relative to the study by \citet{wakker97b}, we benefit from improved 
sensitivity, plus three times better angular resolution. This allowed us to 
explore the expected clumpy structure of the molecular gas, since the signal 
from the molecular clumps embedded in the \ion{H}{i} cores is less diluted and 
the probability of detecting the CO emission is increased. In 
Sect.~\ref{observations} we describe our observations and in 
Sect.~\ref{results} we present the results. The implications of this new CO 
emission survey in HVCs are discussed in Sect.~\ref{discussion}, before 
concluding in Sect.~\ref{conclusions}.

%To conclude on the origin of HVCs, measurements of their their distances (and 
%therefore masses) and metallicities are essential.

%
%________________________________________________________________

\begin{figure*}[t]
\centering
\includegraphics[width=9.5cm]{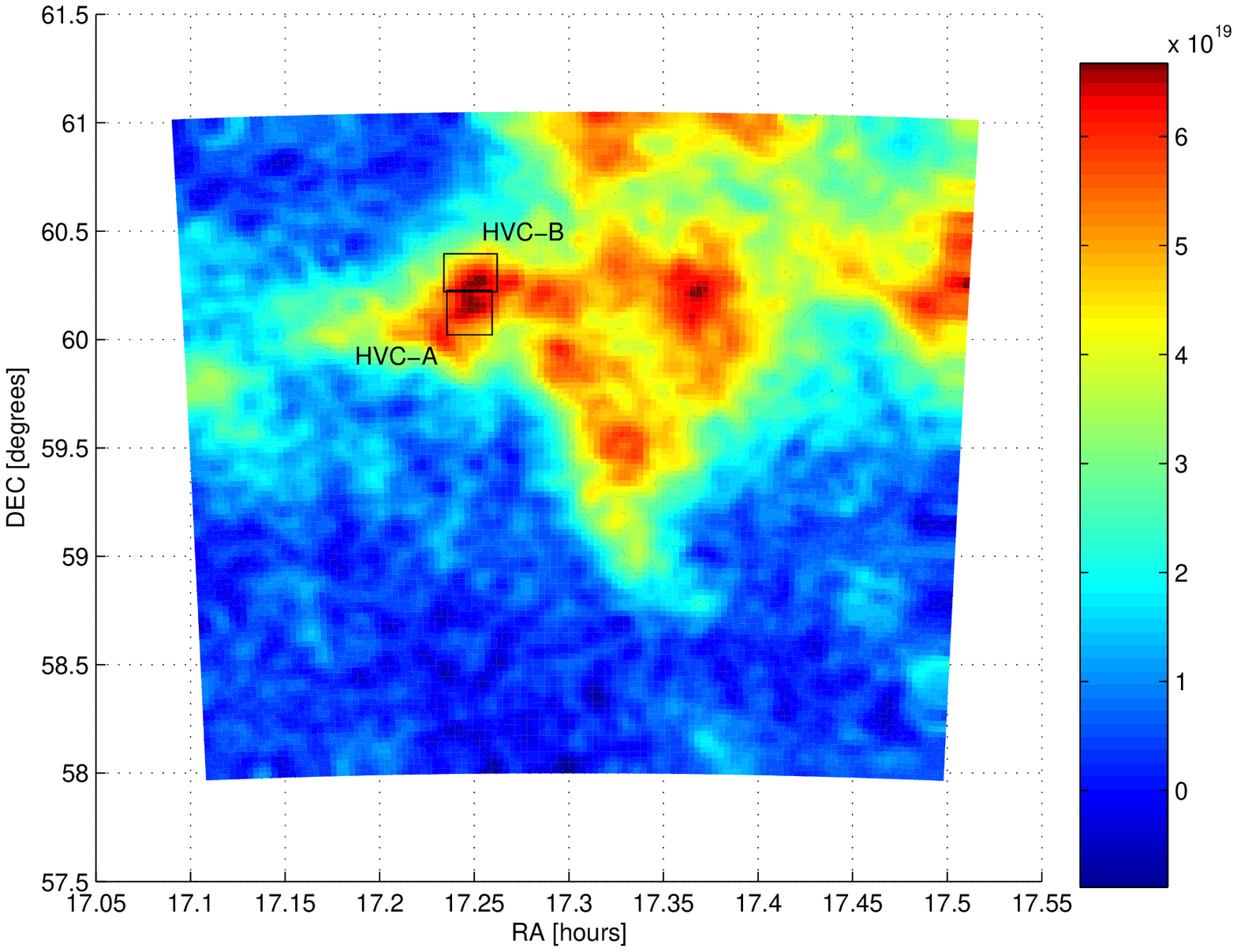}\hfill
\includegraphics[width=8.5cm]{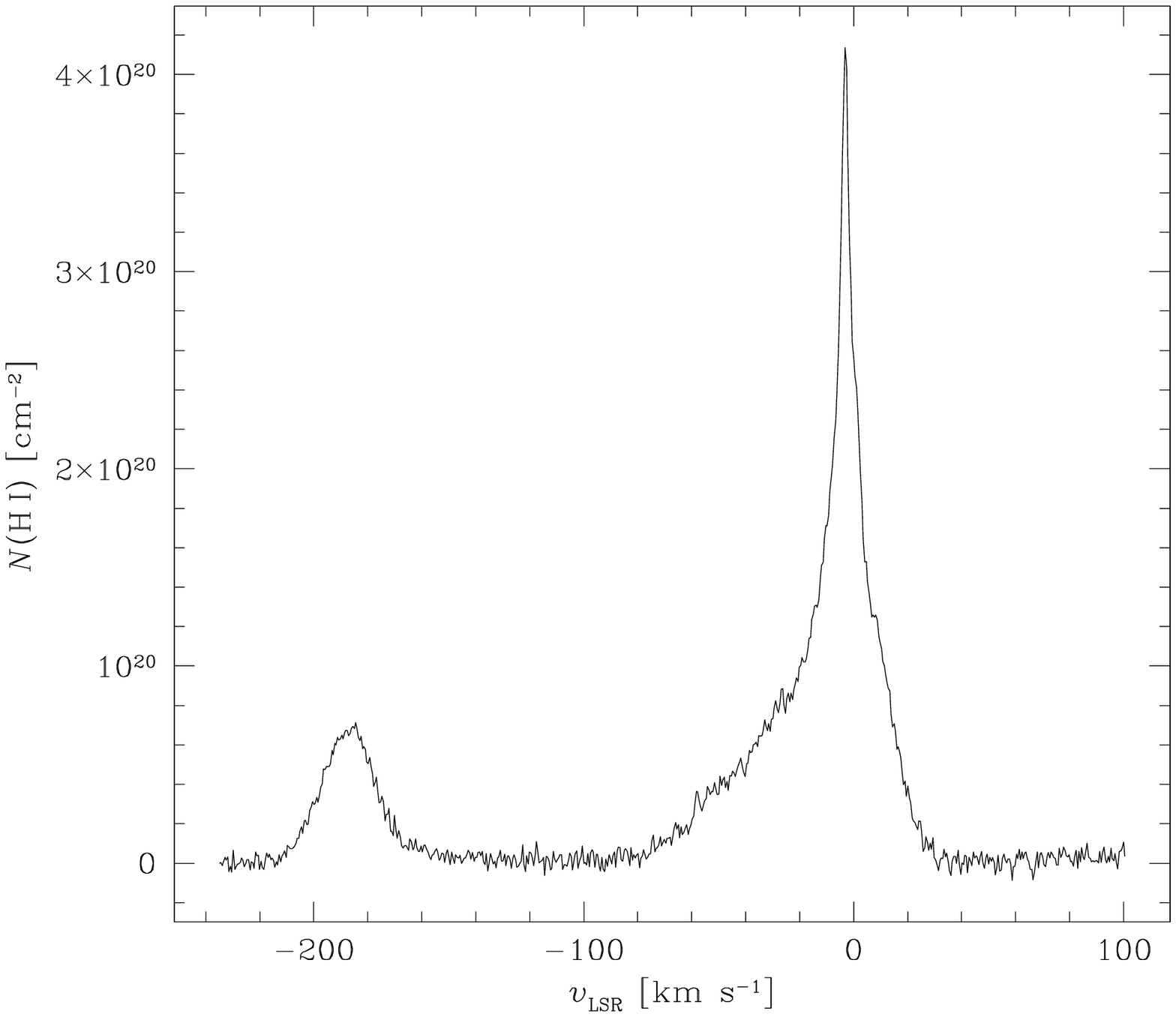}
\caption{{\it Left.} \ion{H}{i} column density map in units of cm$^{-2}$ of the 
HVCs in Complex C within the XFLS field. The map is obtained from the public 
GBT 21~cm data \citep{lockman05}. The HVCs are well isolated in space and their 
integrated emission was computed by adding all the channels between $-210 < 
\upsilon_{\rm LSR} < -132$ km~s$^{-1}$. The selected regions for the CO 
observations are the two framed \ion{H}{i} cores labeled HVC-A and HVC-B, 
respectively. They correspond to two maxima of the HVC \ion{H}{i} emission with 
column densities of $\sim 7\times 10^{19}$ cm$^{-2}$, and are the most 
promising regions for CO detection in these HVCs. {\it Right.} Average
\ion{H}{i} spectrum in the direction of the two selected cores. It clearly 
shows the contribution of the local ISM and of the HVC \ion{H}{i} emissions.}
\label{HVC-HI}
\end{figure*}
%
%________________________________________________________________
 
\section{Observations and data reduction}\label{observations}

The straightforward consequence of the discovery of cold dust emission from 
HVCs in Complex~C by MD05 is a search for molecular gas via the CO emission. 
The Complex~C spans some 1500 degrees squared on the sky. It has an average 
metallicity of $0.1-0.3$ solar \citep{tripp03} and is located at a distance 
greater than 5 kpc from the Sun \citep{wakker01b}. The Spitzer Extragalactic 
First Look Survey (XFLS) field analyzed by MD05 covers only a $3^{\circ}\times 
3^{\circ}$ area on the edge of Complex~C centered at ${\rm RA} = 
17{\rm h}\,18{\rm m}$, ${\rm DEC} = +59^{\circ}\,30{\rm m}$ (J2000).

To select the most promising HVC regions for CO emission detection in this 
$3^{\circ}\times 3^{\circ}$ area, we first carefully re-examined the results 
obtained by MD05 to check whether we could spatially identify the infrared dust 
emission peaks in the HVCs. Using the public IRIS maps \citep{miville05b} and 
the GBT 21~cm \ion{H}{i} data of the XFLS field kindly made available by 
\citet{lockman05}, we extracted, as in MD05, the dust emission component 
corresponding to the HVCs from the decomposition of the infrared signal into 
different \ion{H}{i} components identified in the 21~cm spectrum, namely the 
HVCs, two IVCs, and the local ISM \ion{H}{i} emission (Fig.~4 in MD05). We 
confirm the global infrared-\ion{H}{i} correlation found by MD05, but no 
spatial information on the HVC infrared emission can be derived. Indeed, the 
HVC infrared signal is spatially detected only at $2-3$~$\sigma$ (with a higher 
signal-to-noise ratio at larger scales), which makes reliable spatial 
identifications of infrared emission maxima difficult. 

Therefore, we decided to choose the hopefully suitable HVC regions for our CO 
observations on the basis of \ion{H}{i} core selection, corresponding to a 
\ion{H}{i} column density maxima selection. Such a strategy is supported by the
recent results of \citet{wakker06} which indicate that the fraction of sight 
lines with H$_2$ detections in UV, increases with $N$(\ion{H}{i}) for the IVCs. 
On the left panel of Fig.~\ref{HVC-HI}, we show the 21~cm emission map of the 
$3^{\circ}\times 3^{\circ}$ field analyzed by MD05, integrated over the HVC 
velocity channels $-210 < \upsilon_{\rm LSR} < -132$ km~s$^{-1}$. We focused on 
two HVC \ion{H}{i} cores: the HVC-A core located at (J2000) 
${\rm RA} = 17{\rm h}\,14{\rm m}\,59.055{\rm s}$, 
${\rm DEC} = +60^{\circ}\,08{\rm m}\,45.84{\rm s}$ and the HVC-B core located 
at (J2000) ${\rm RA} = 17{\rm h}\,15{\rm m}\,11.678{\rm s}$, 
${\rm DEC} = +60^{\circ}\,15{\rm m}\,03.56{\rm s}$, both having among the 
larger \ion{H}{i} column densities with $N$(\ion{H}{i}) $\sim 7\times 10^{19}$ 
cm$^{-2}$. The average 21~cm spectrum in the direction of the two cores is 
shown on the right panel of Fig.~\ref{HVC-HI}. We clearly see the contribution 
of the dominant local ISM \ion{H}{i} emission and of the HVC \ion{H}{i} 
emission at the LSR velocity of about $-190$ km~s$^{-1}$.

%
%________________________________________________________________

\begin{figure*}[!]
\centering
\includegraphics[width=2.5cm]{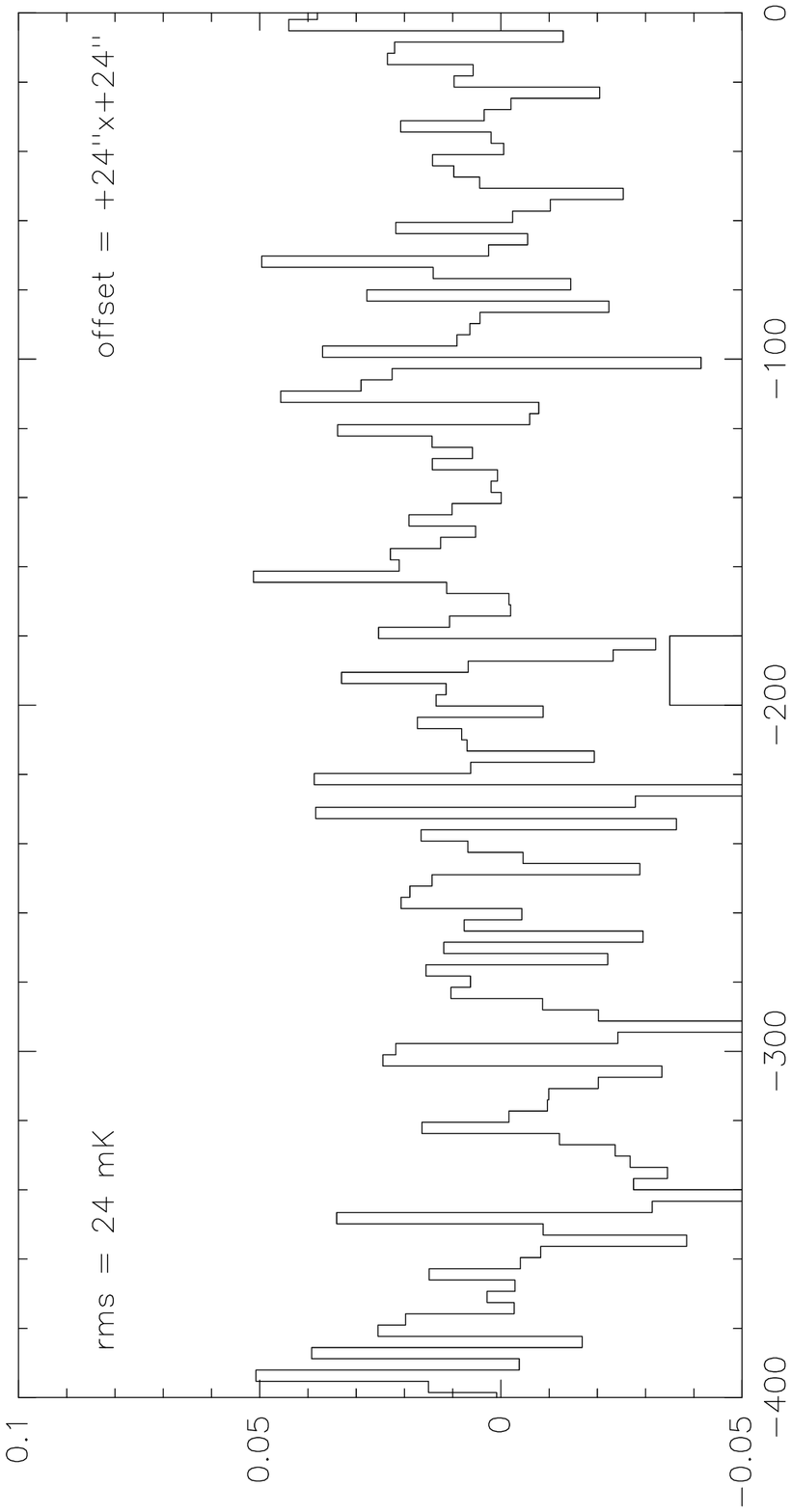}
\includegraphics[width=2.5cm]{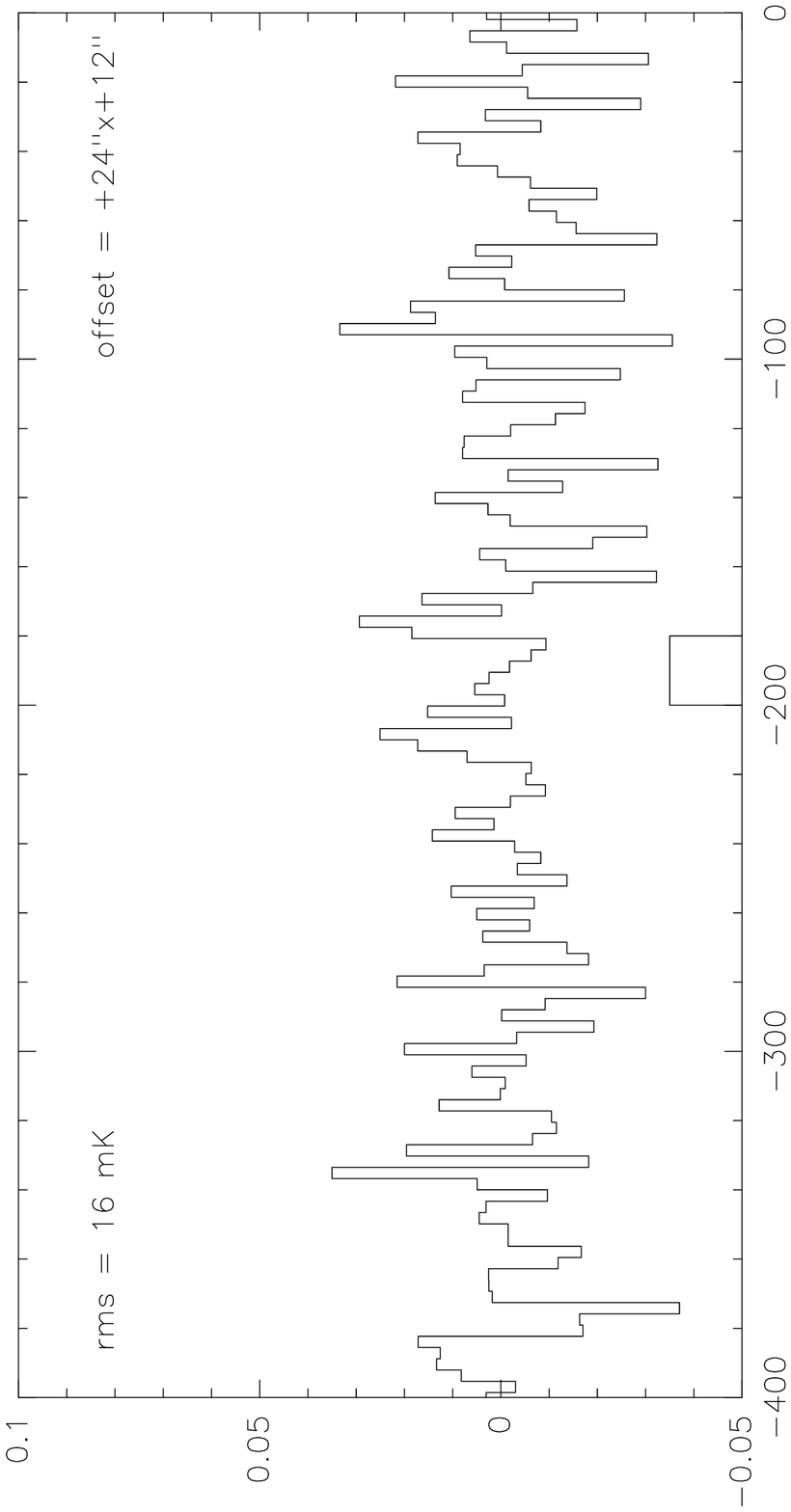}
\includegraphics[width=2.5cm]{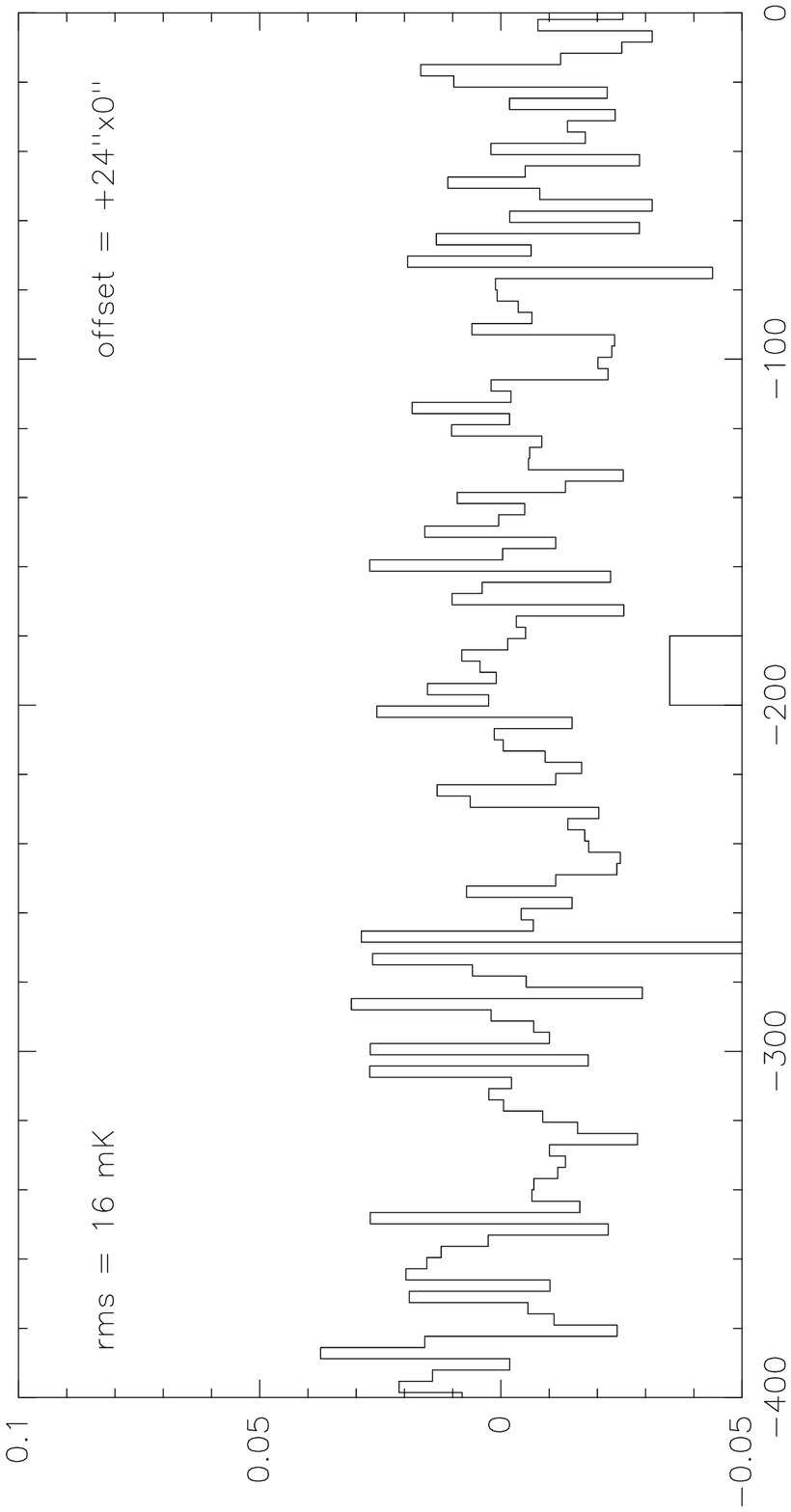}
\includegraphics[width=2.5cm]{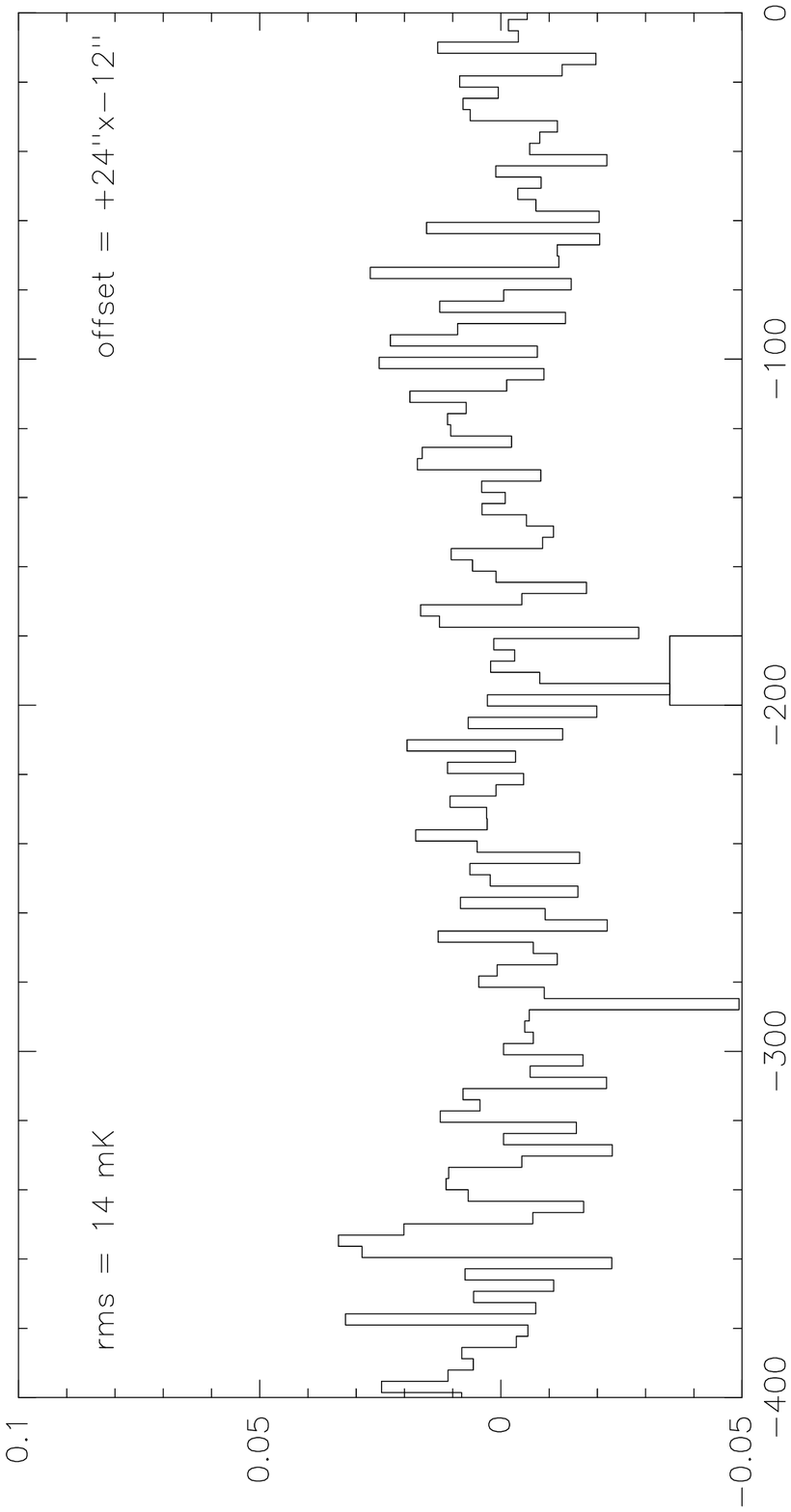}
\includegraphics[width=2.5cm]{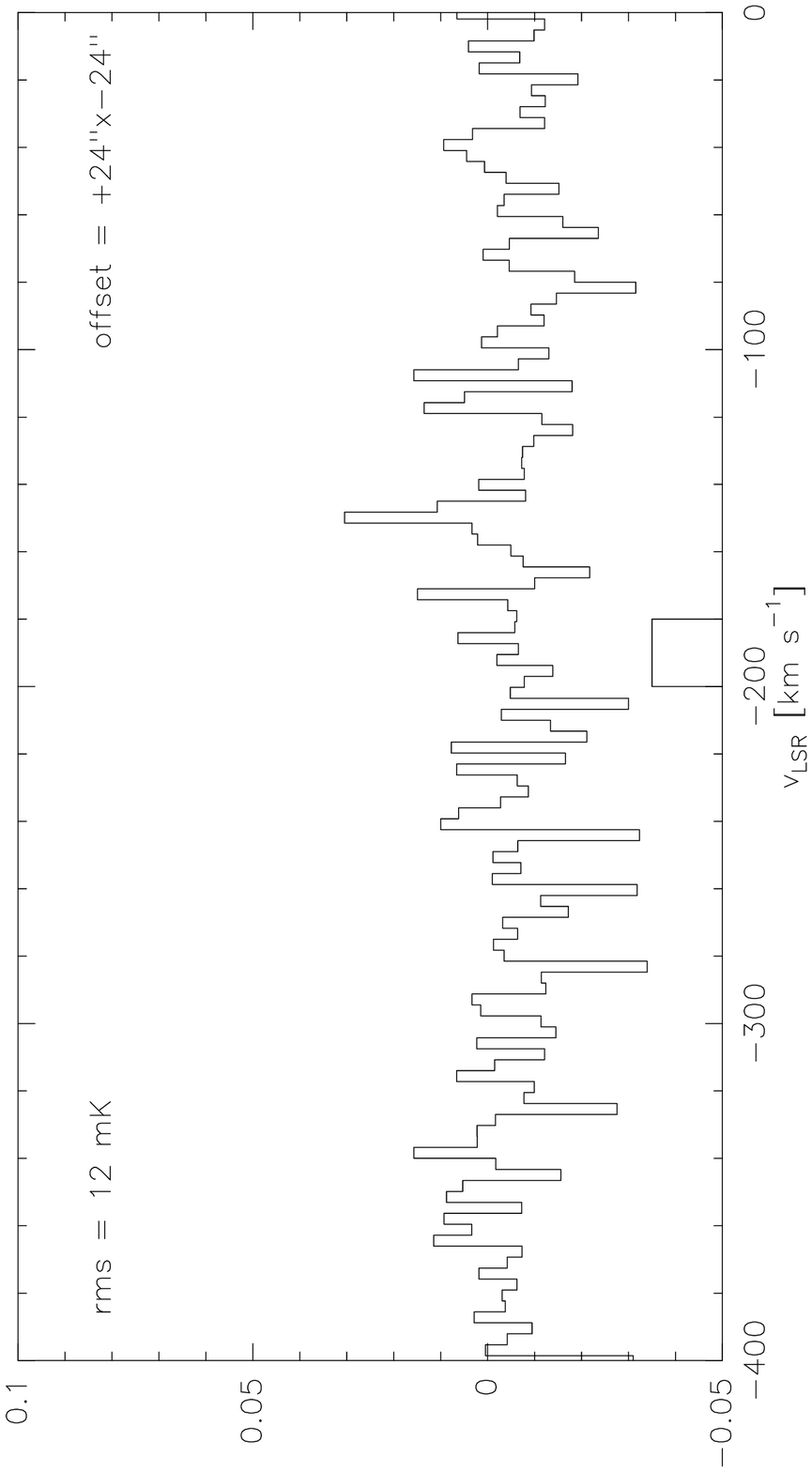}

\includegraphics[width=2.5cm]{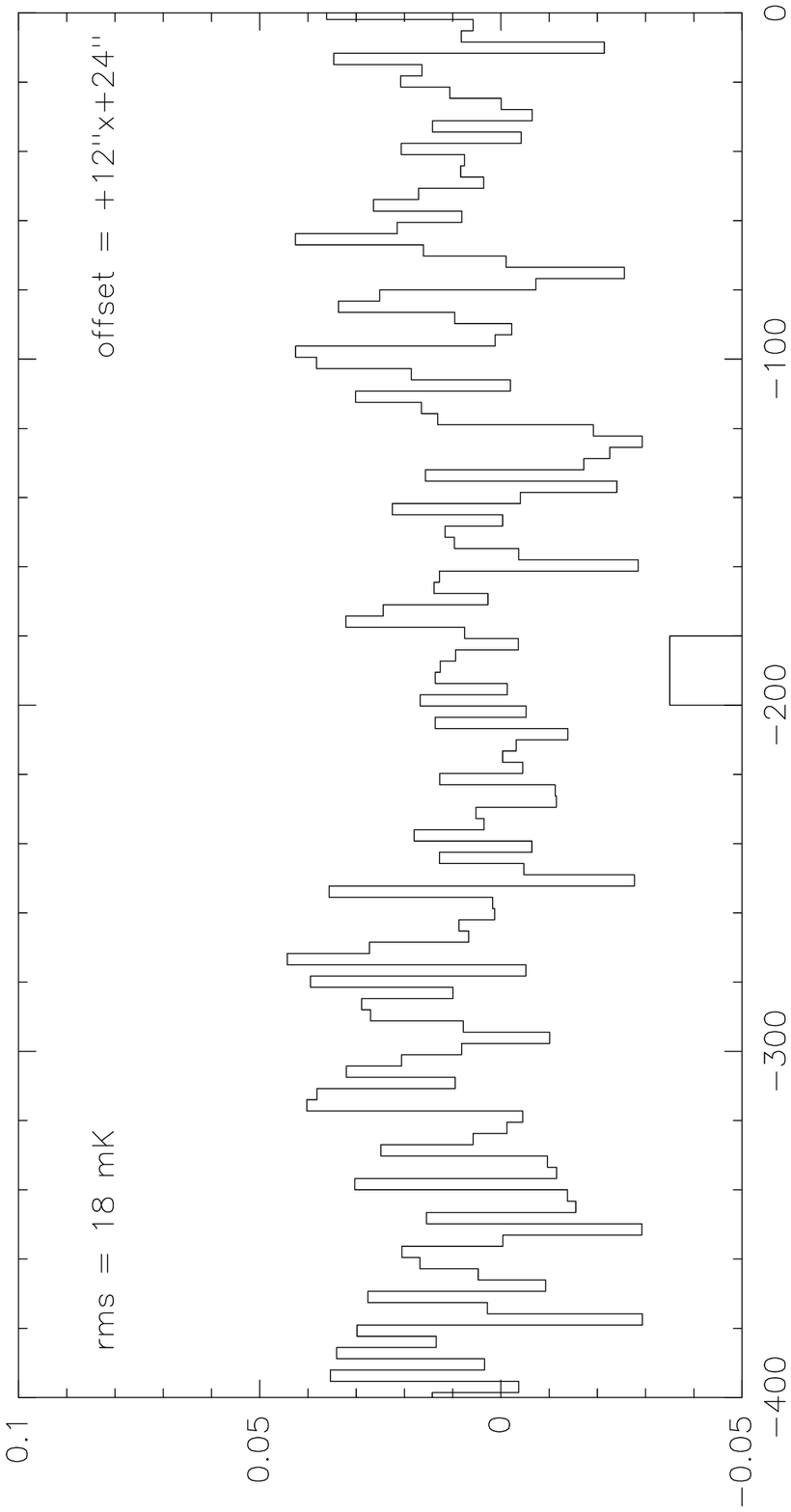}
\includegraphics[width=2.5cm]{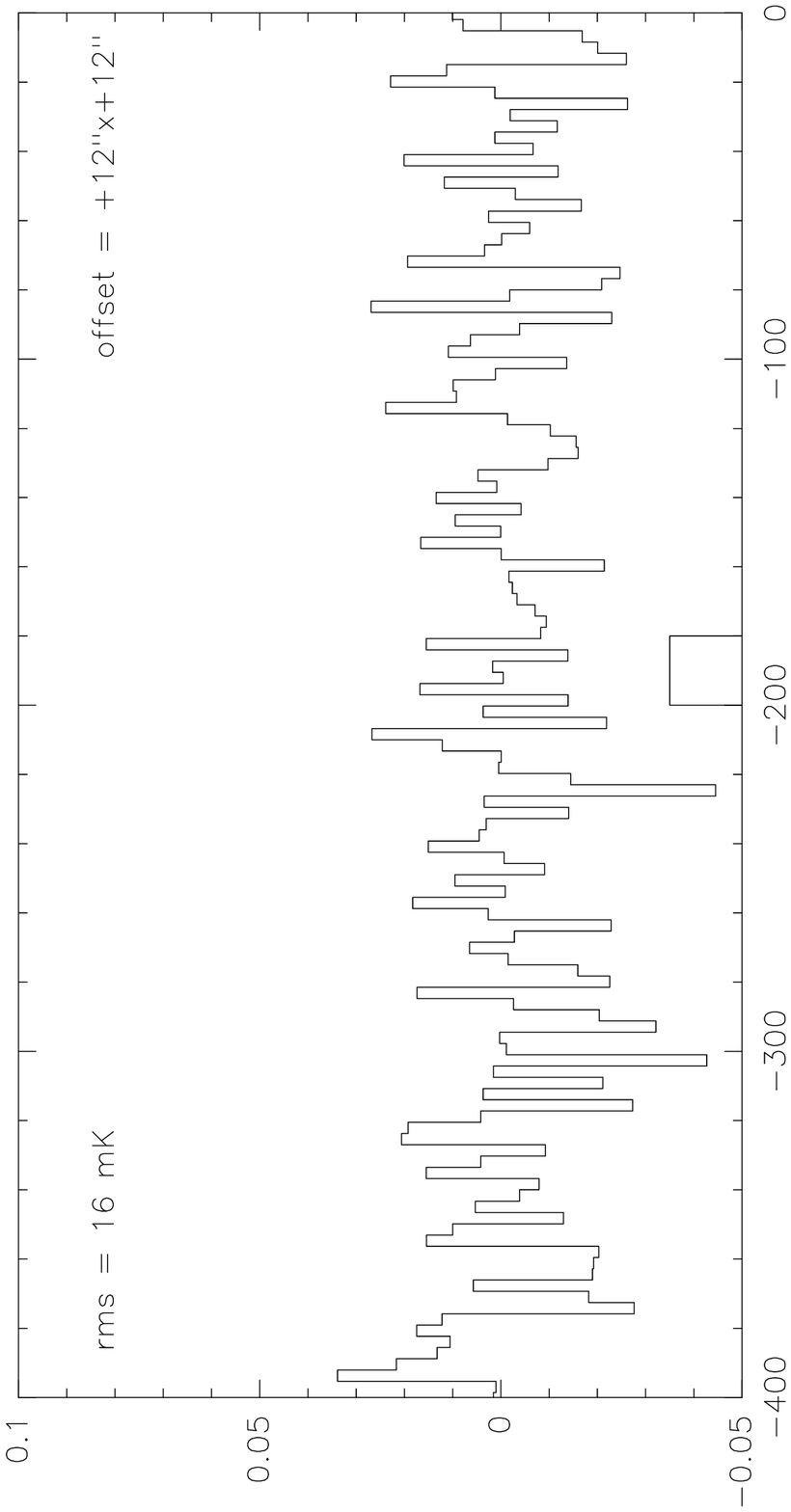}
\includegraphics[width=2.5cm]{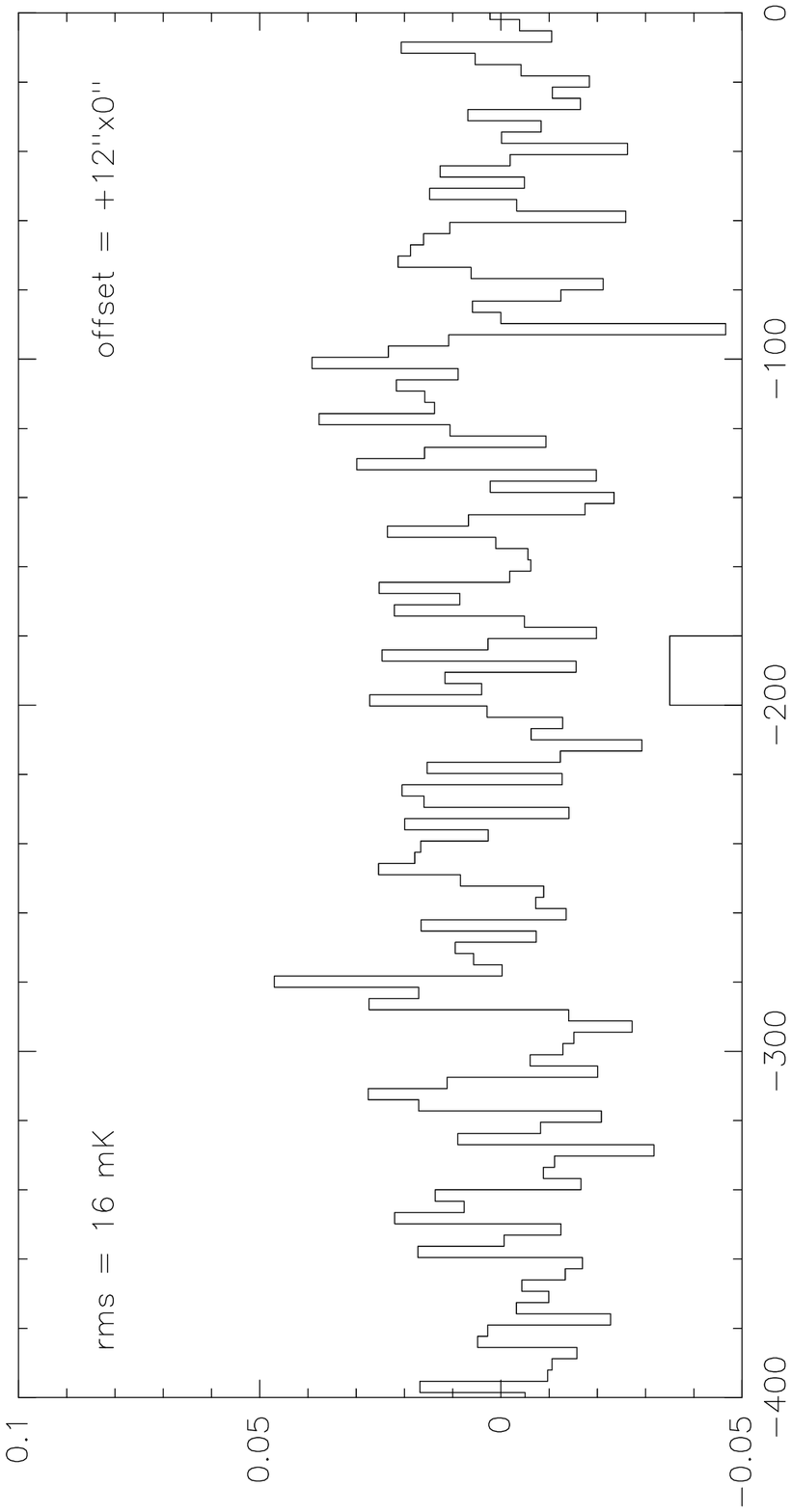}
\includegraphics[width=2.5cm]{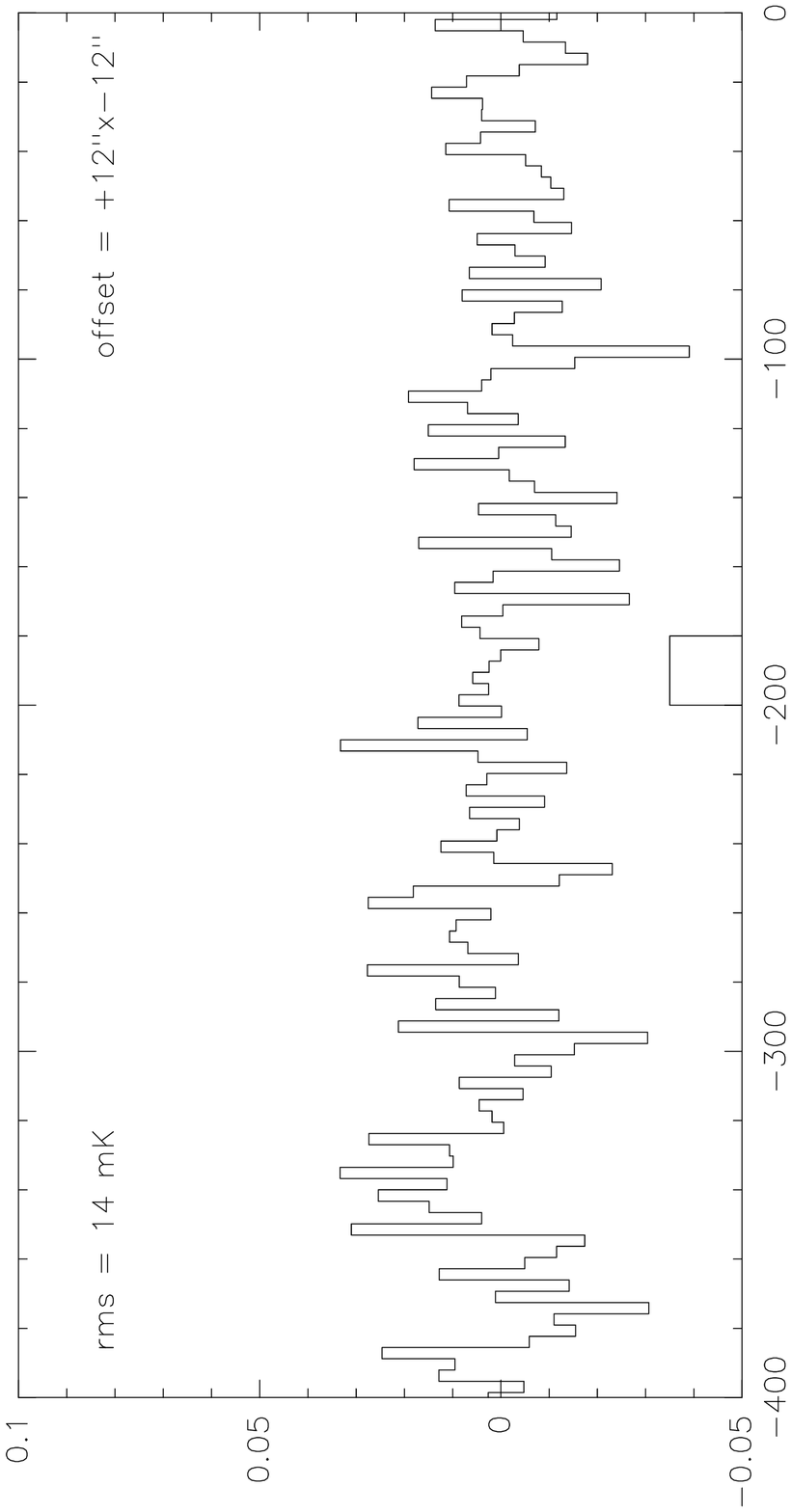}
\includegraphics[width=2.5cm]{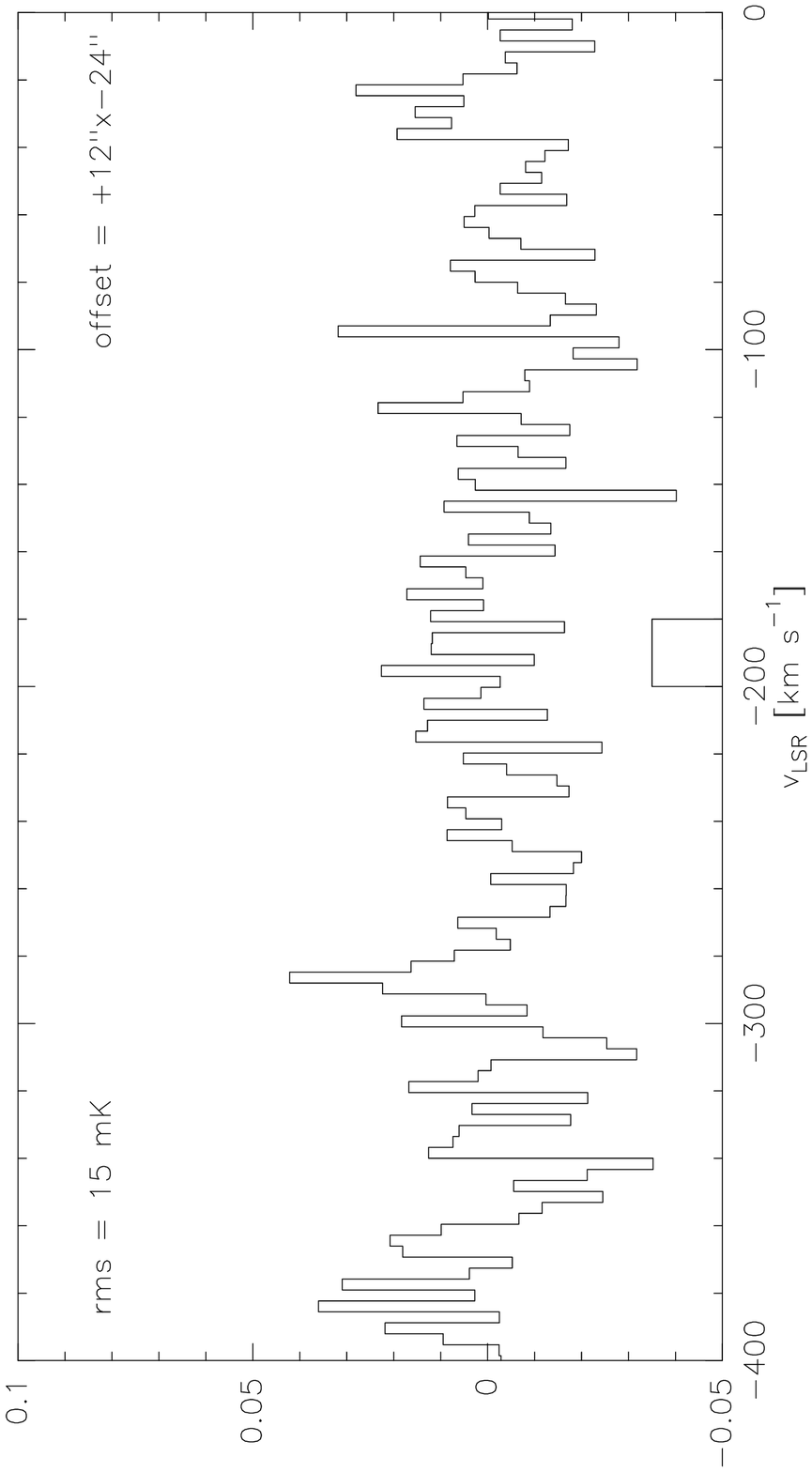}

\includegraphics[width=2.5cm]{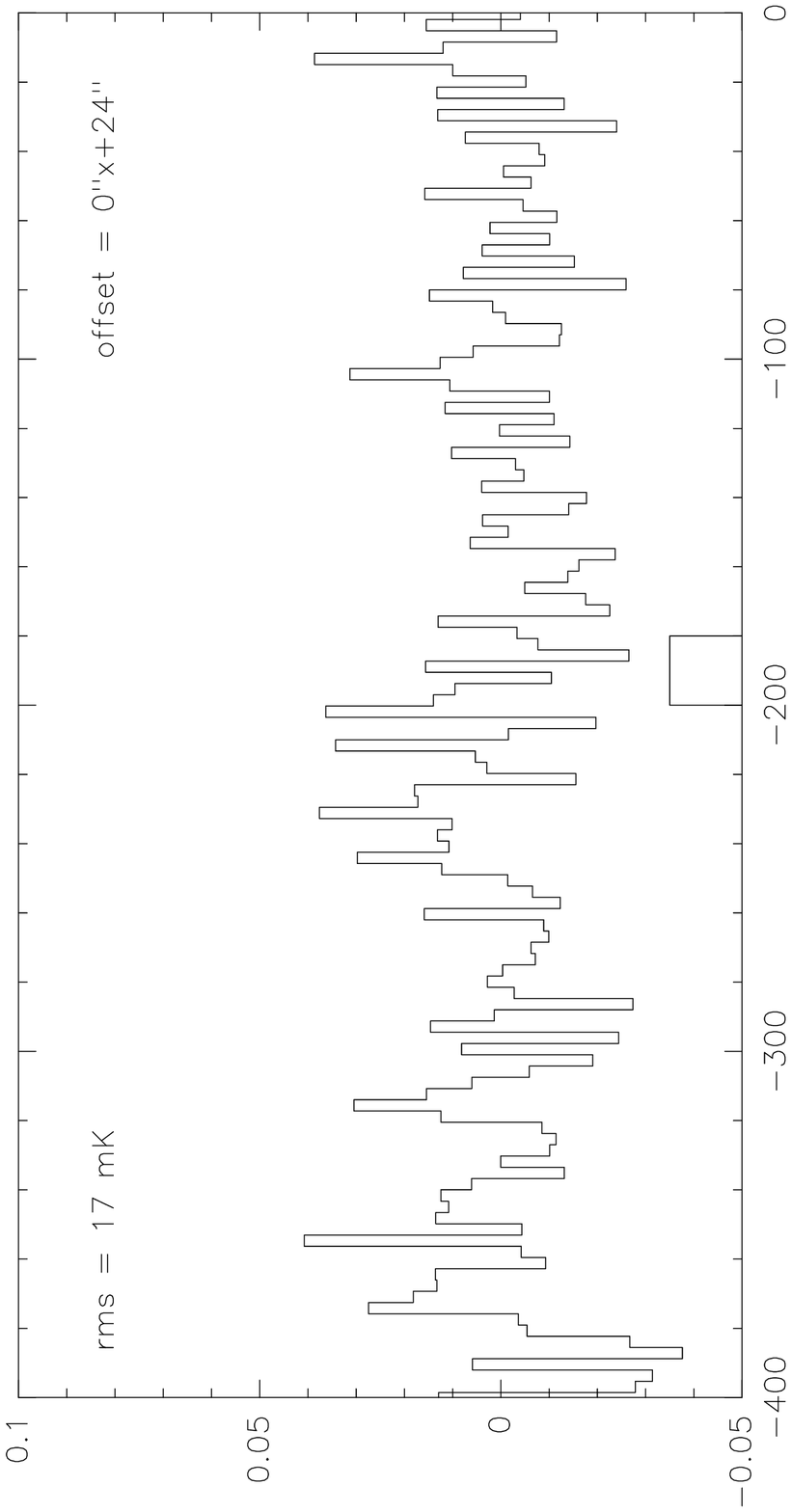}
\includegraphics[width=2.5cm]{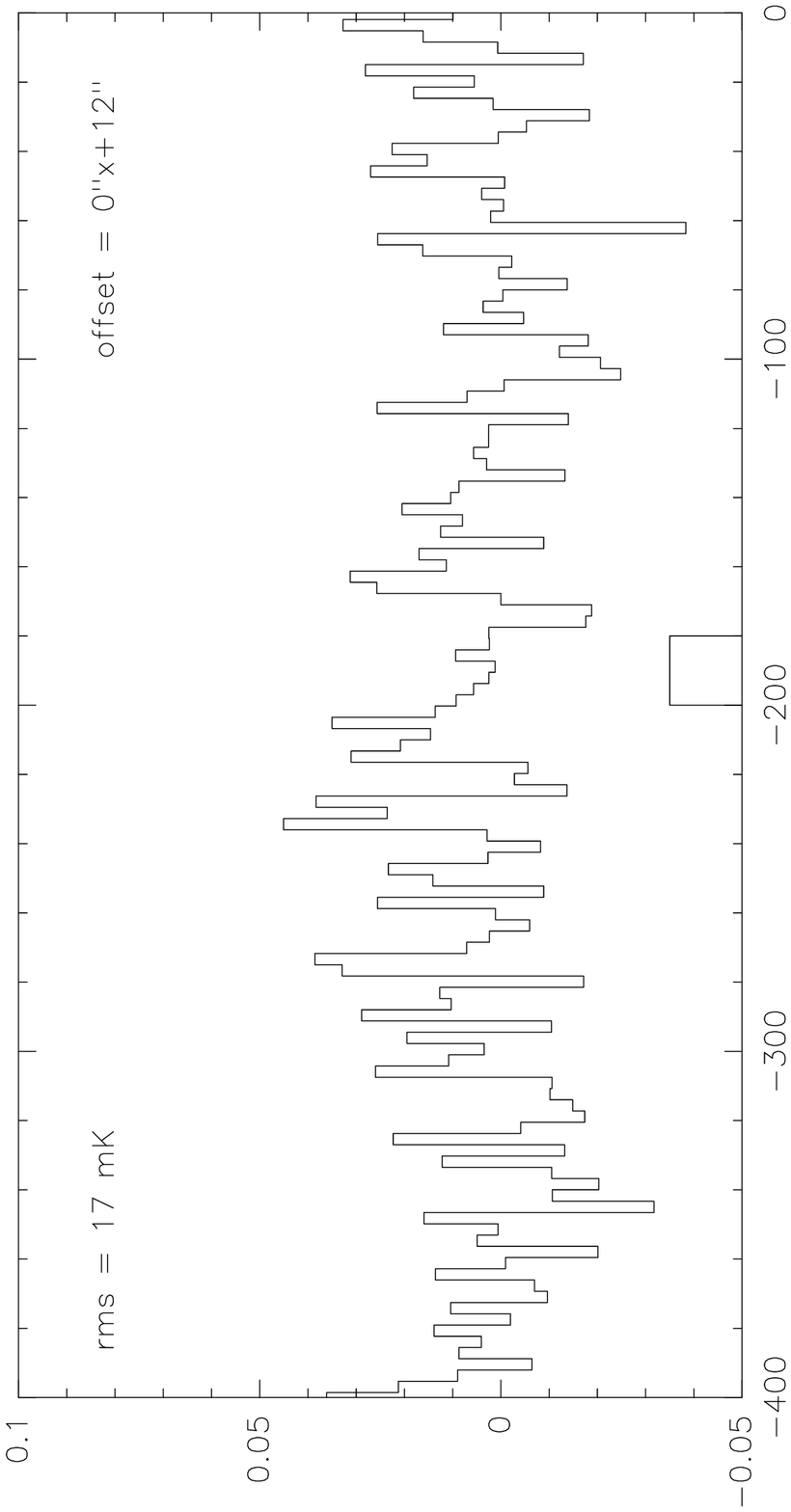}
\includegraphics[width=2.5cm]{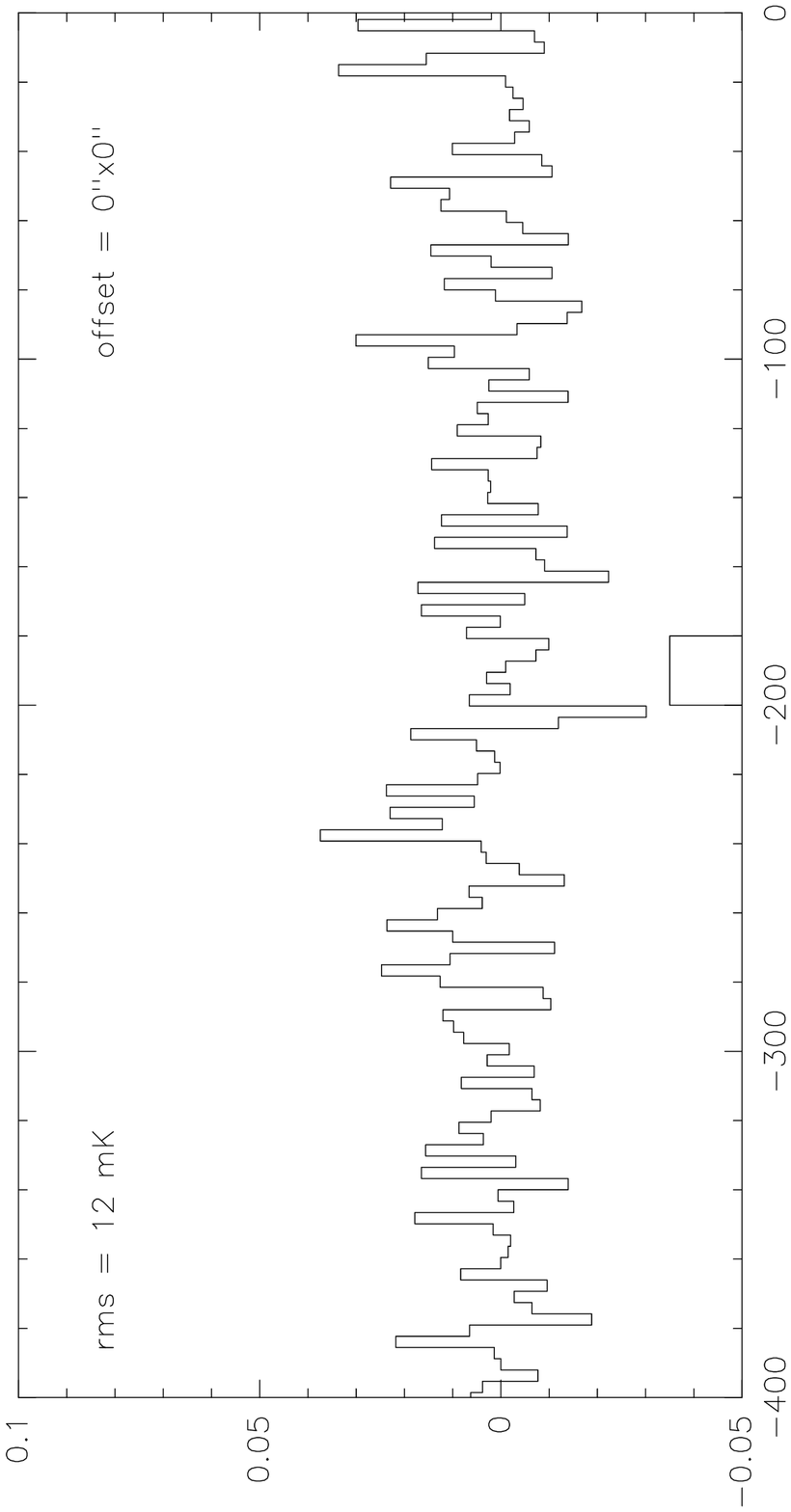}
\includegraphics[width=2.5cm]{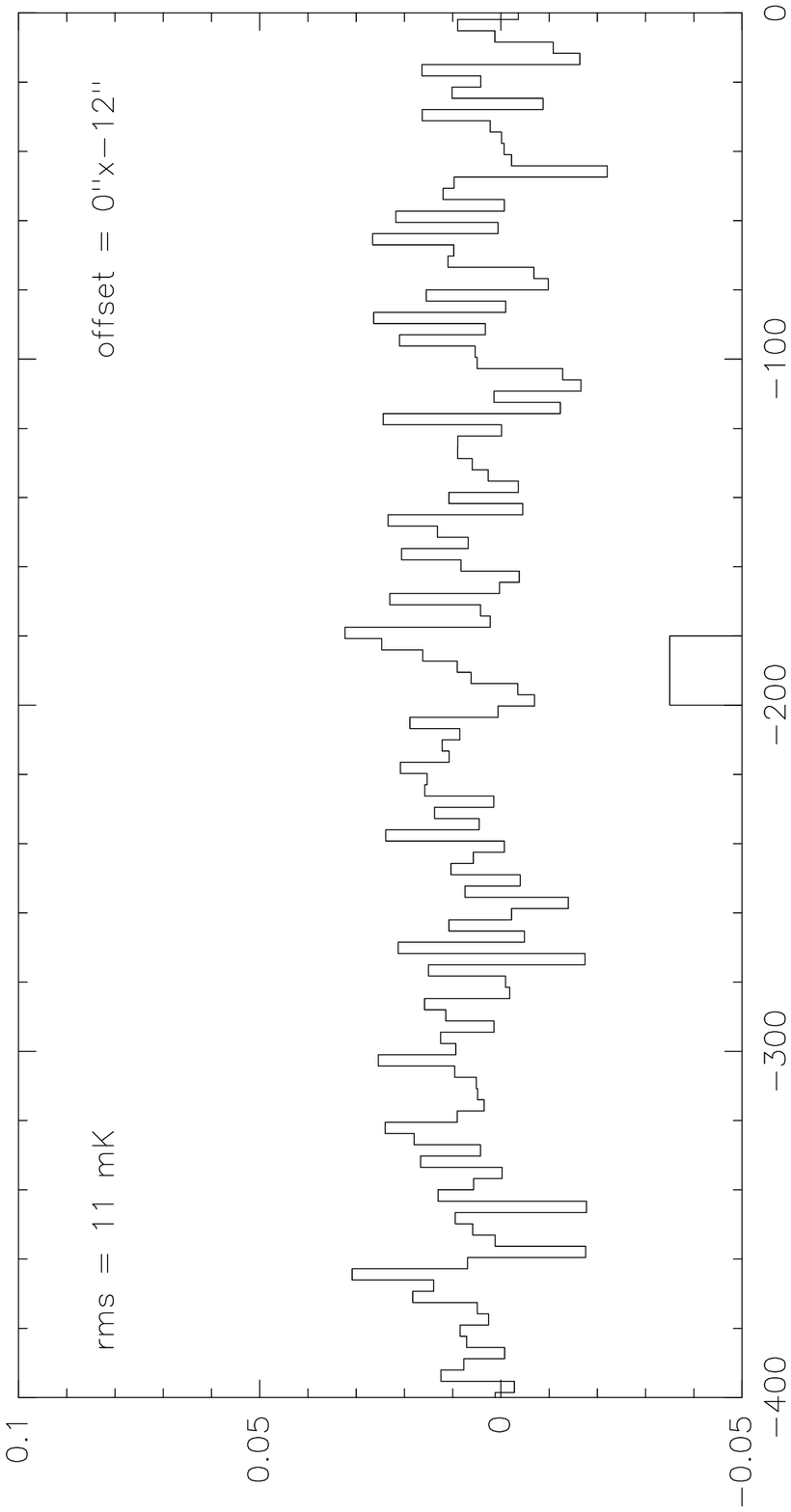}
\includegraphics[width=2.5cm]{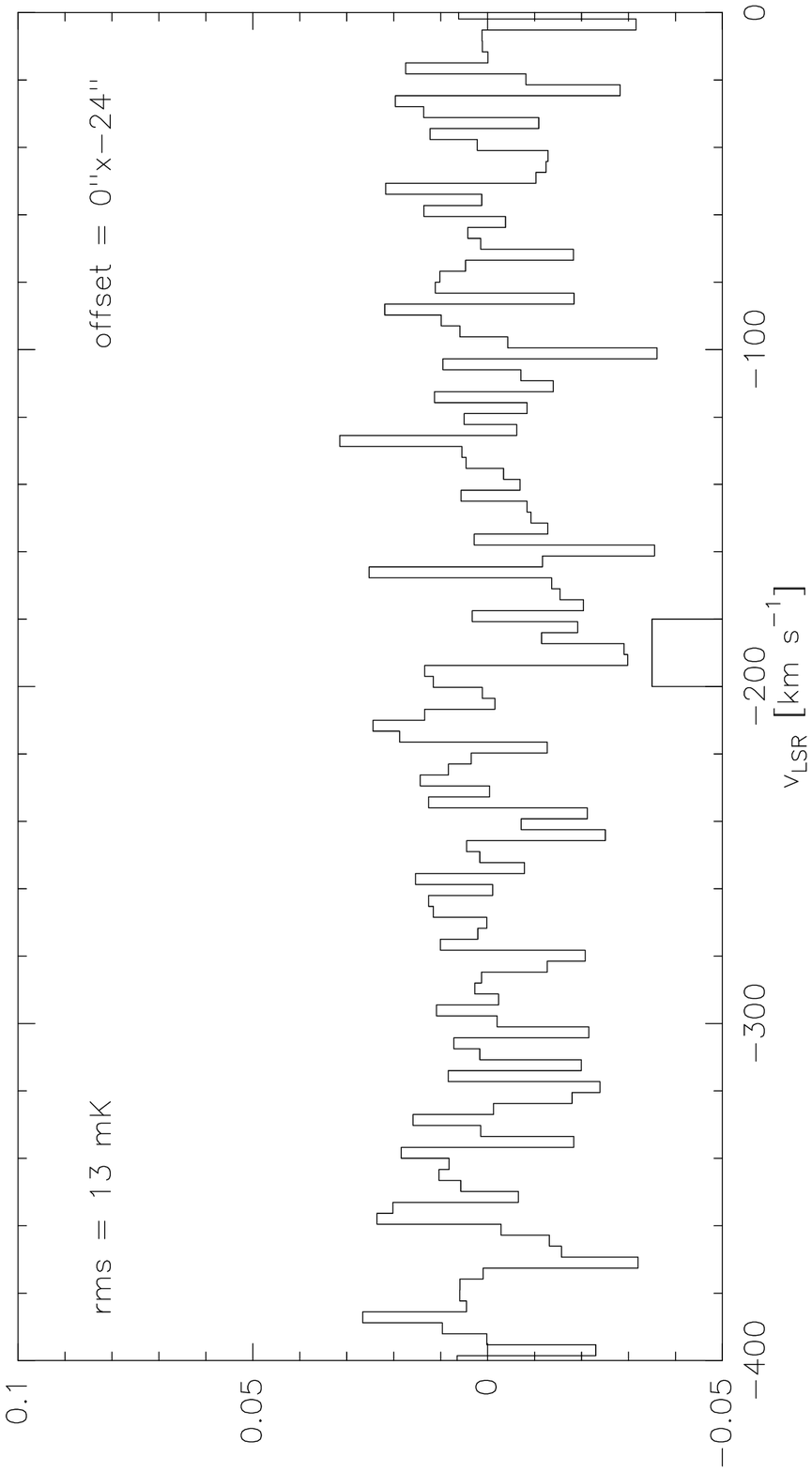}

\includegraphics[width=2.5cm]{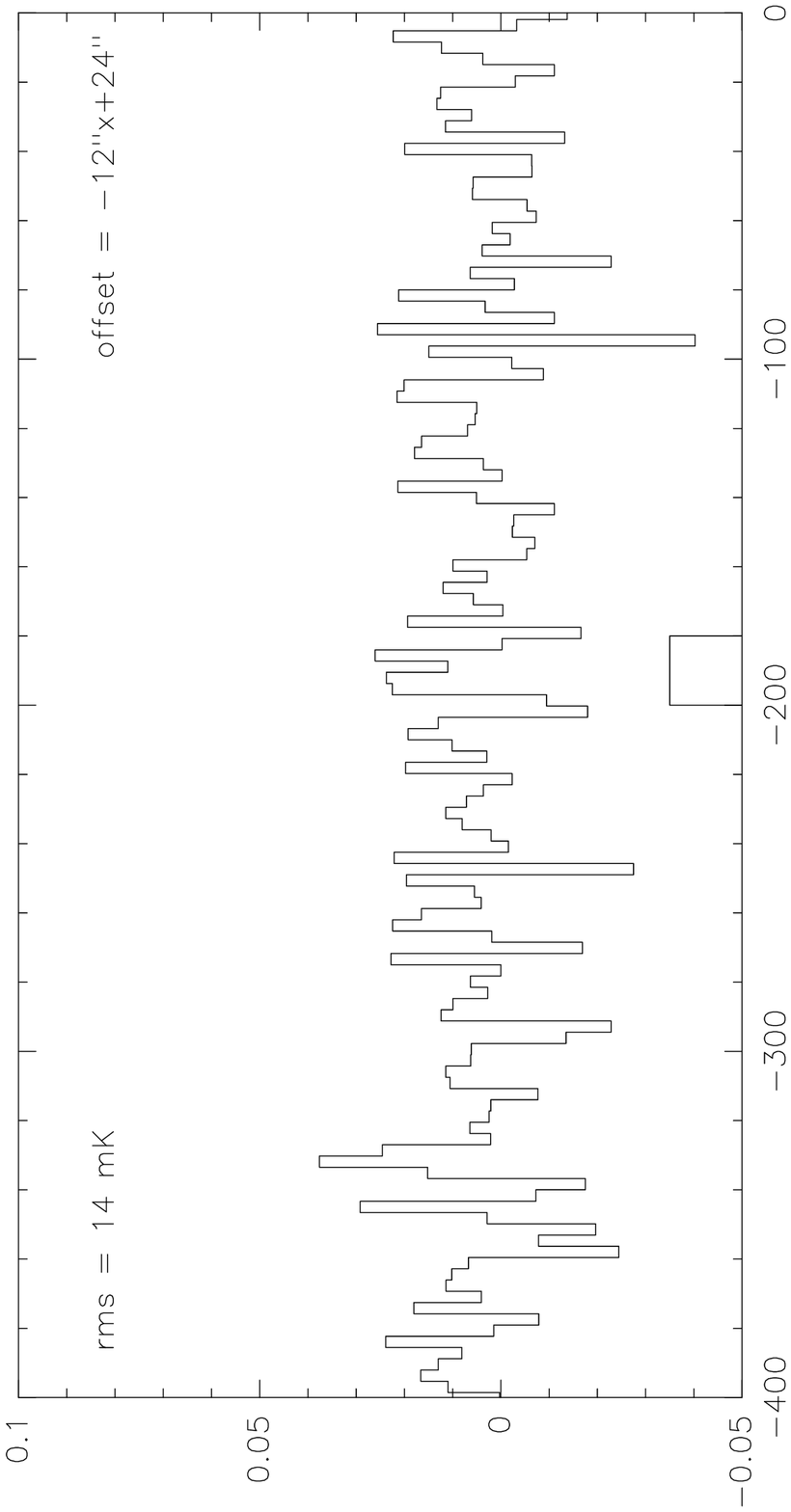}
\includegraphics[width=2.5cm]{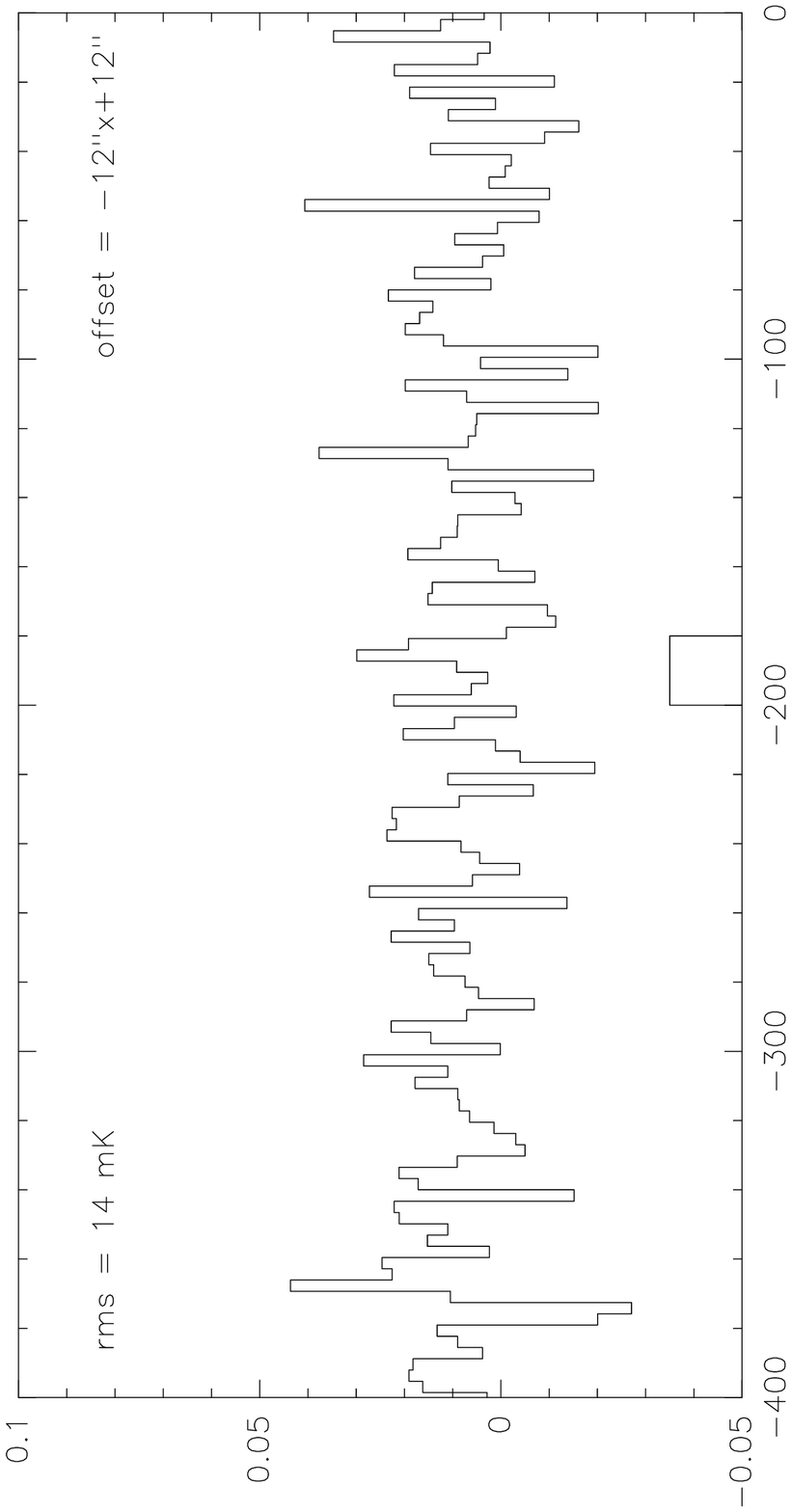}
\includegraphics[width=2.5cm]{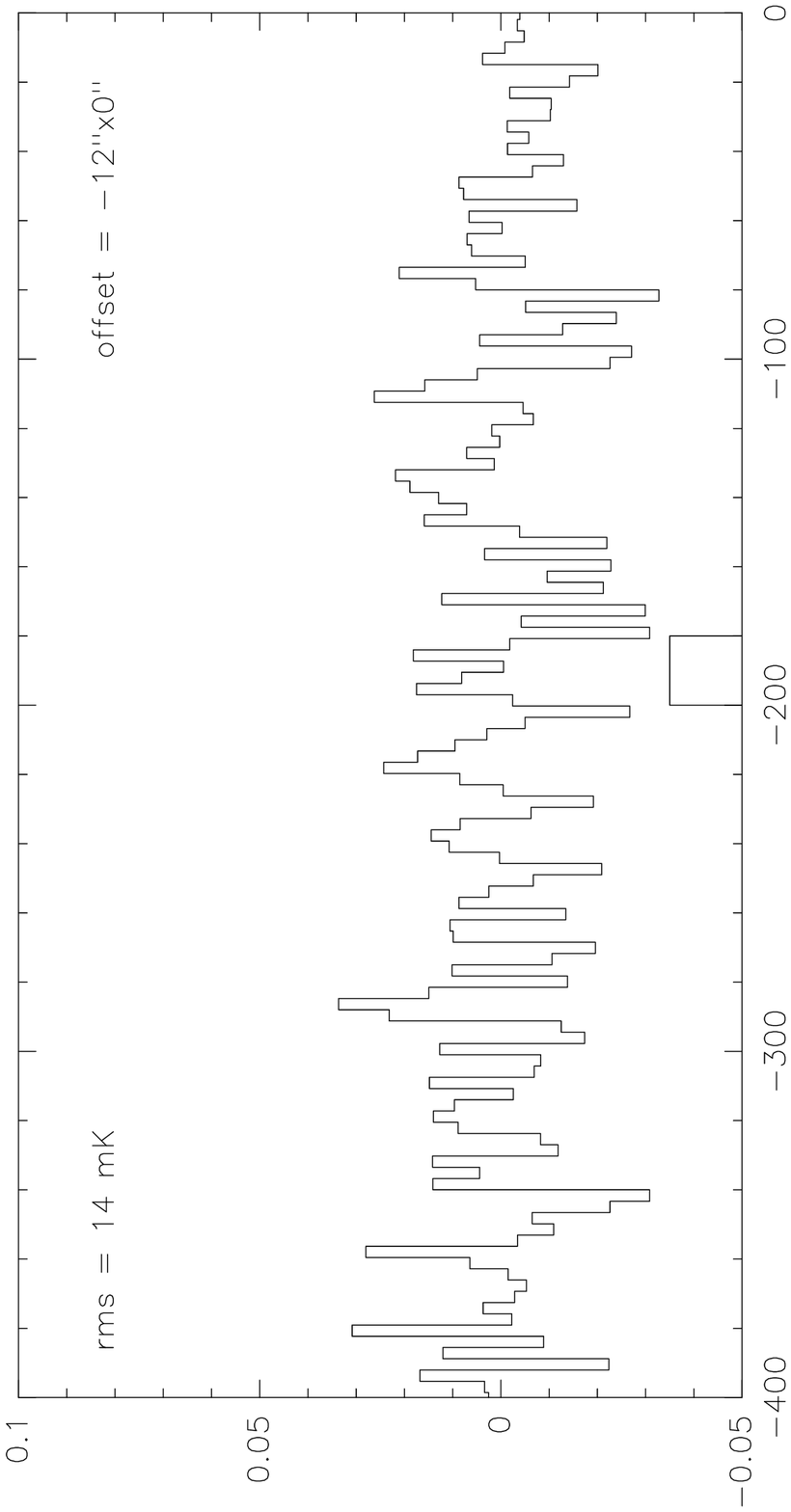}
\includegraphics[width=2.5cm]{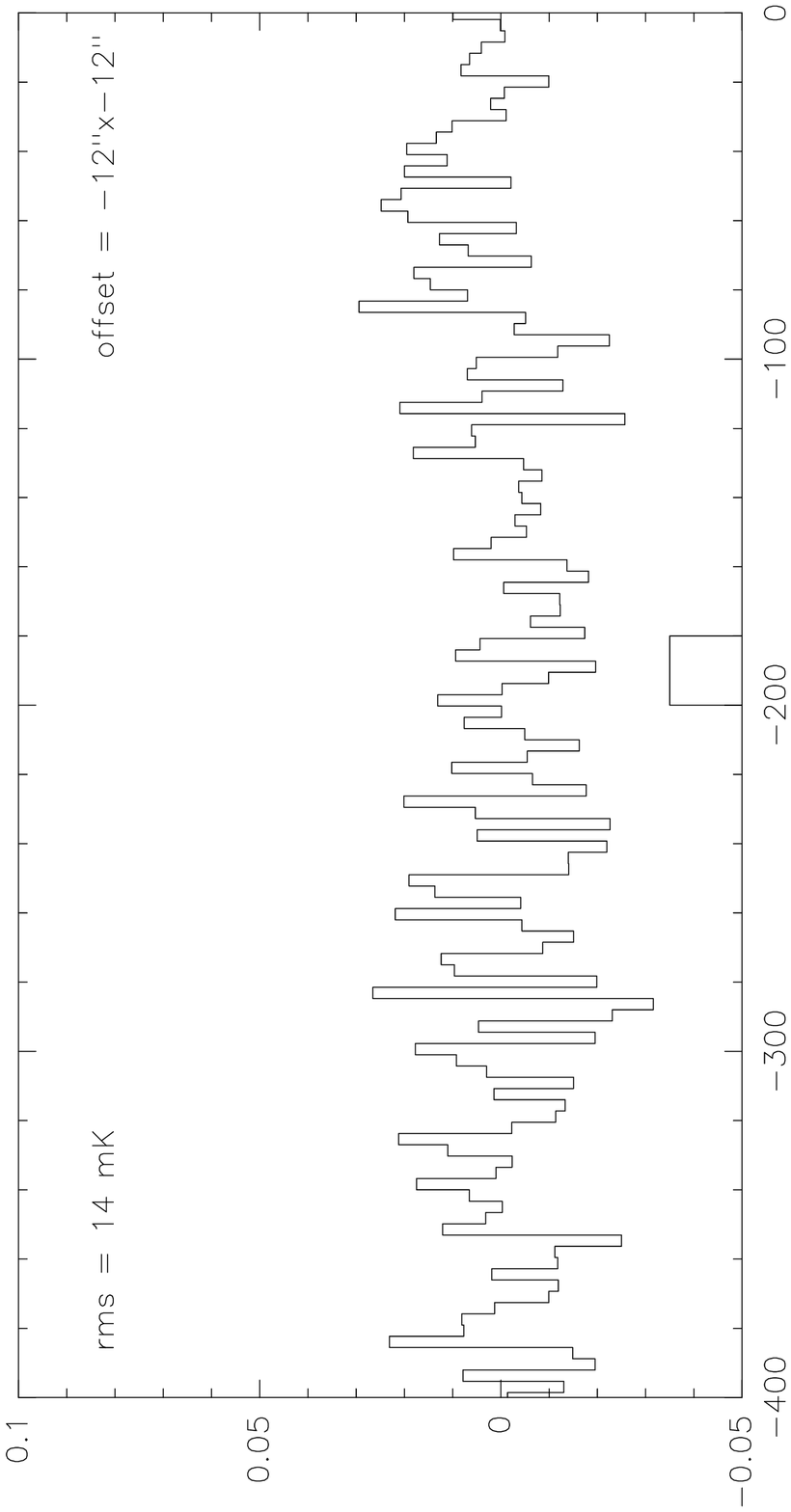}
\includegraphics[width=2.5cm]{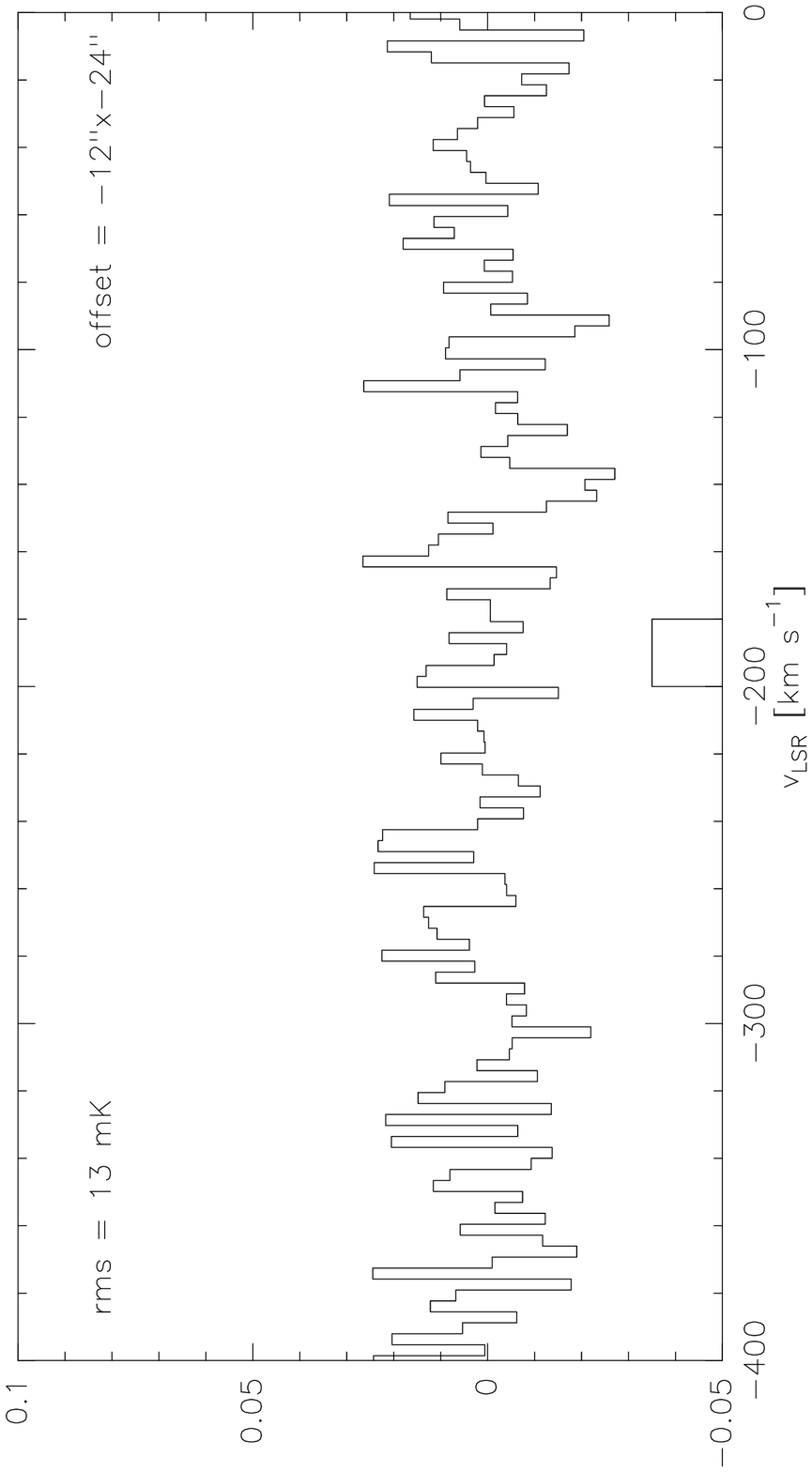}

\includegraphics[width=2.5cm]{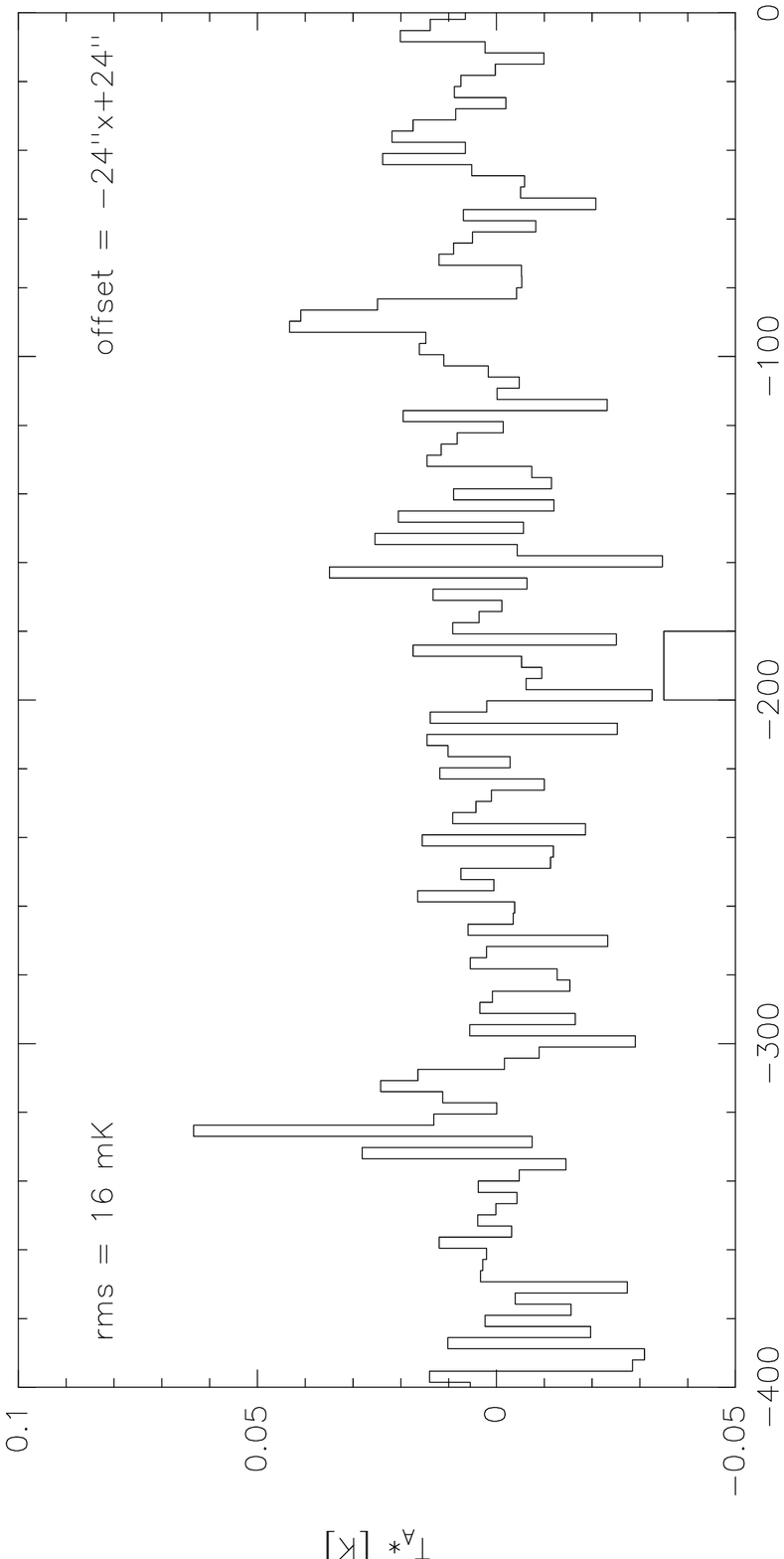}
\includegraphics[width=2.5cm]{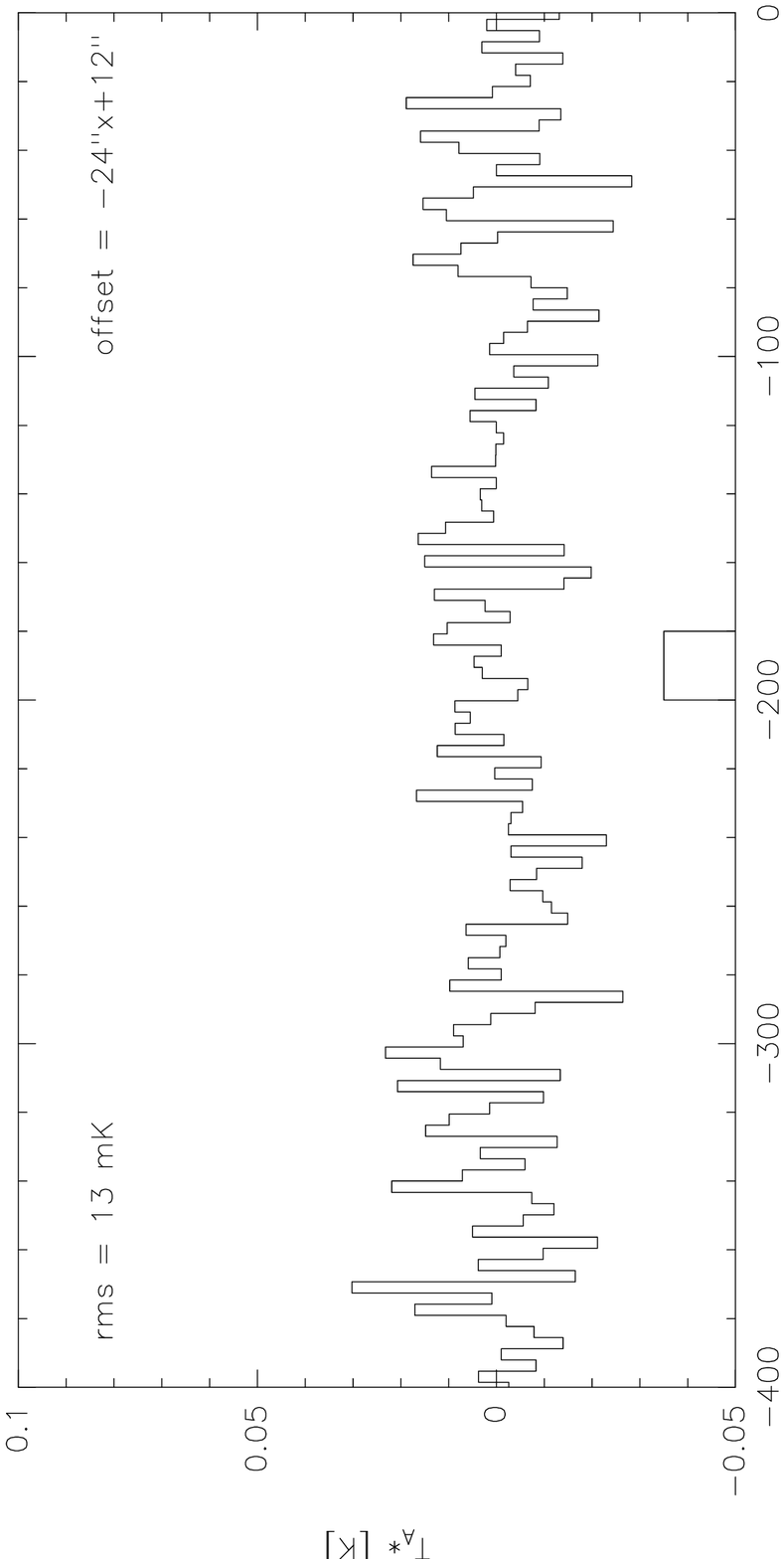}
\includegraphics[width=2.5cm]{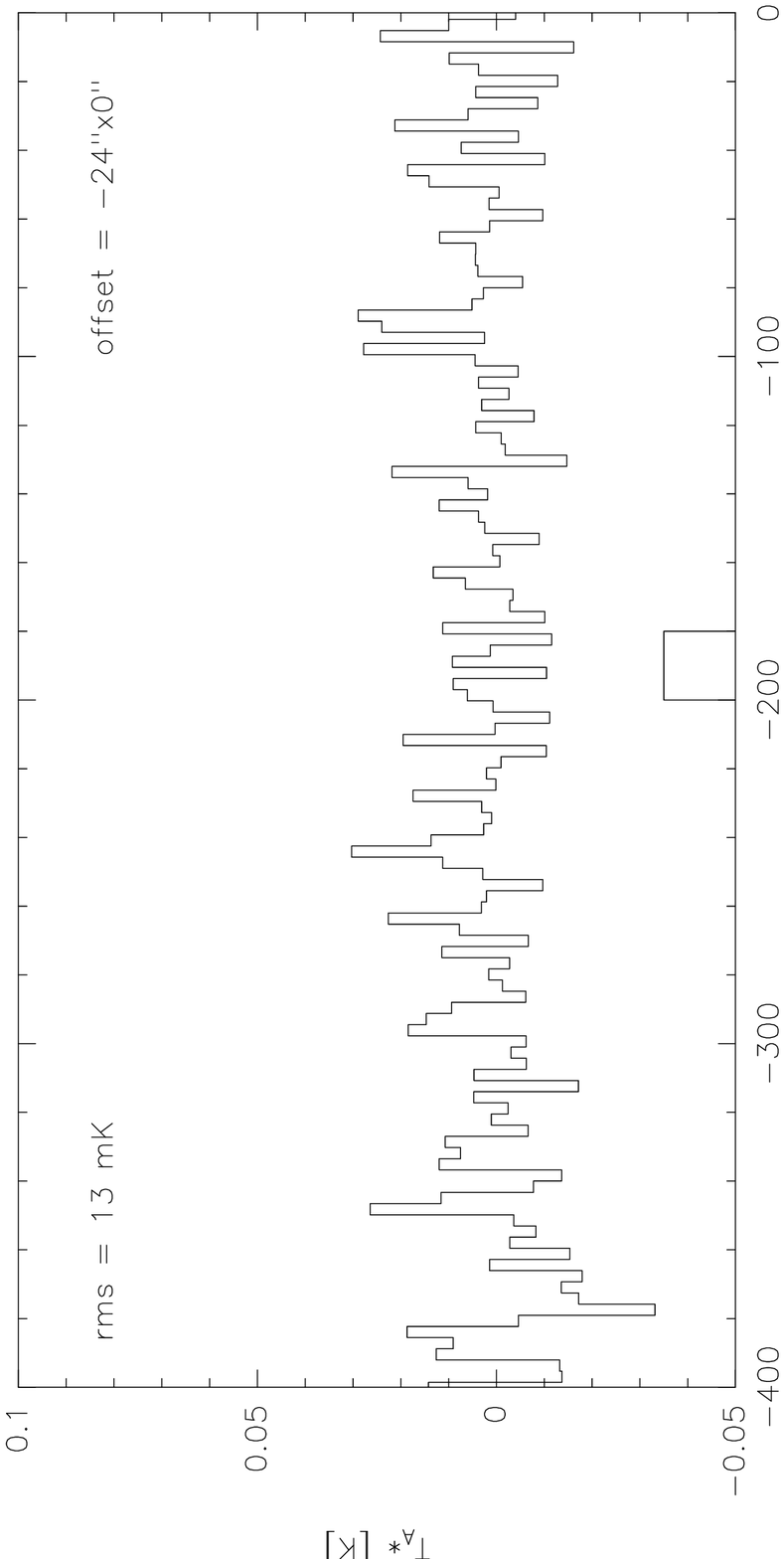}
\includegraphics[width=2.5cm]{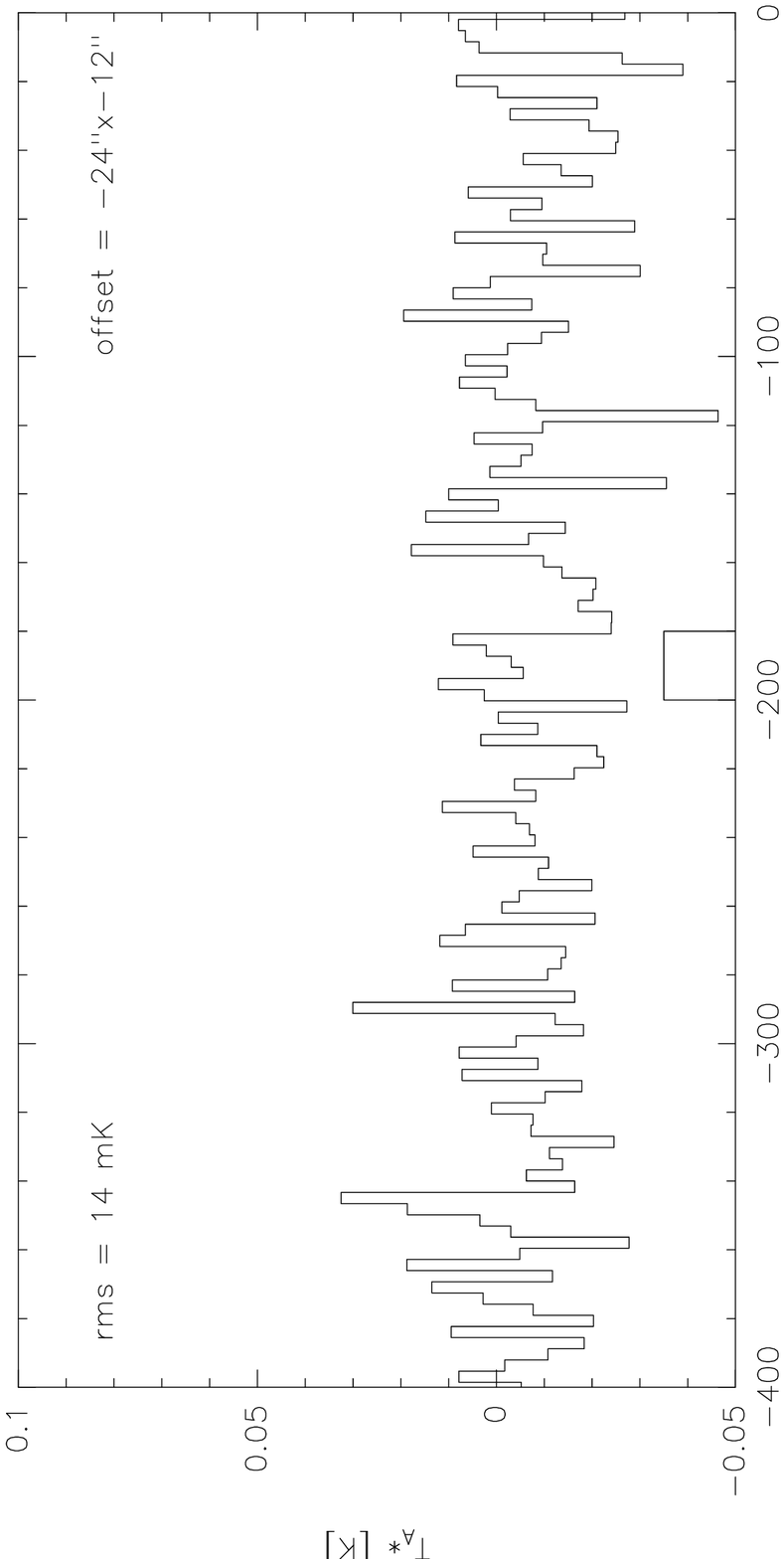}
\includegraphics[width=2.5cm]{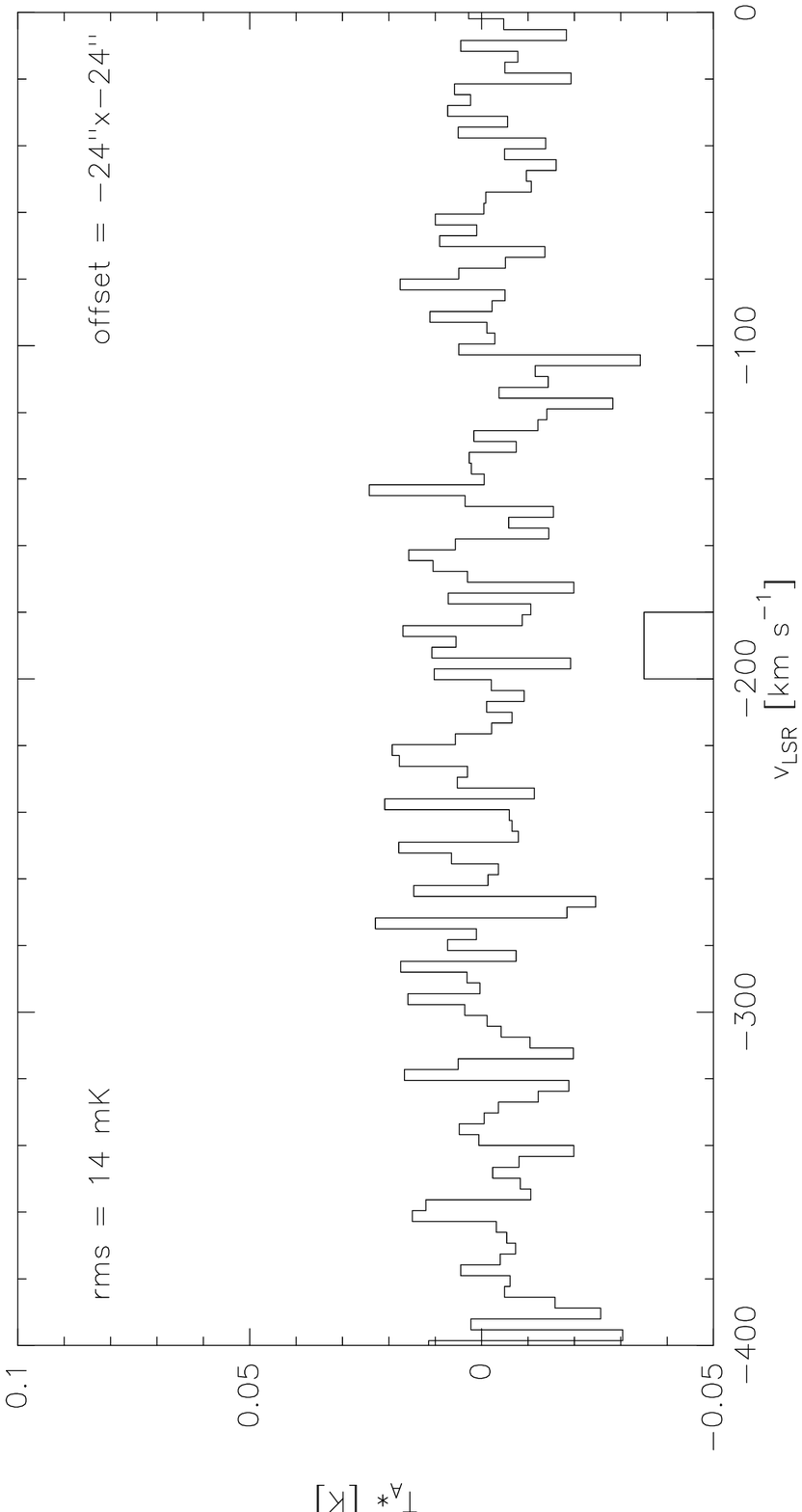}
\caption{Mosaic of 25 $^{12}$CO(1--0) spectra, covering a $60''\times 60''$ 
grid with a $12''$ spacing, centered on the \ion{H}{i} core HVC-A. The spectra 
are smoothed to a resolution of $\sim 3$ km~s$^{-1}$ and are the result of the 
co-addition of all scans obtained at a given position of the mosaic. The box
shows the expected LSR velocity of the CO emission line according to the 21~cm 
data.}
\label{HVC-A-spectra}
\end{figure*}

\begin{figure*}[!]
\centering
\includegraphics[width=2.5cm]{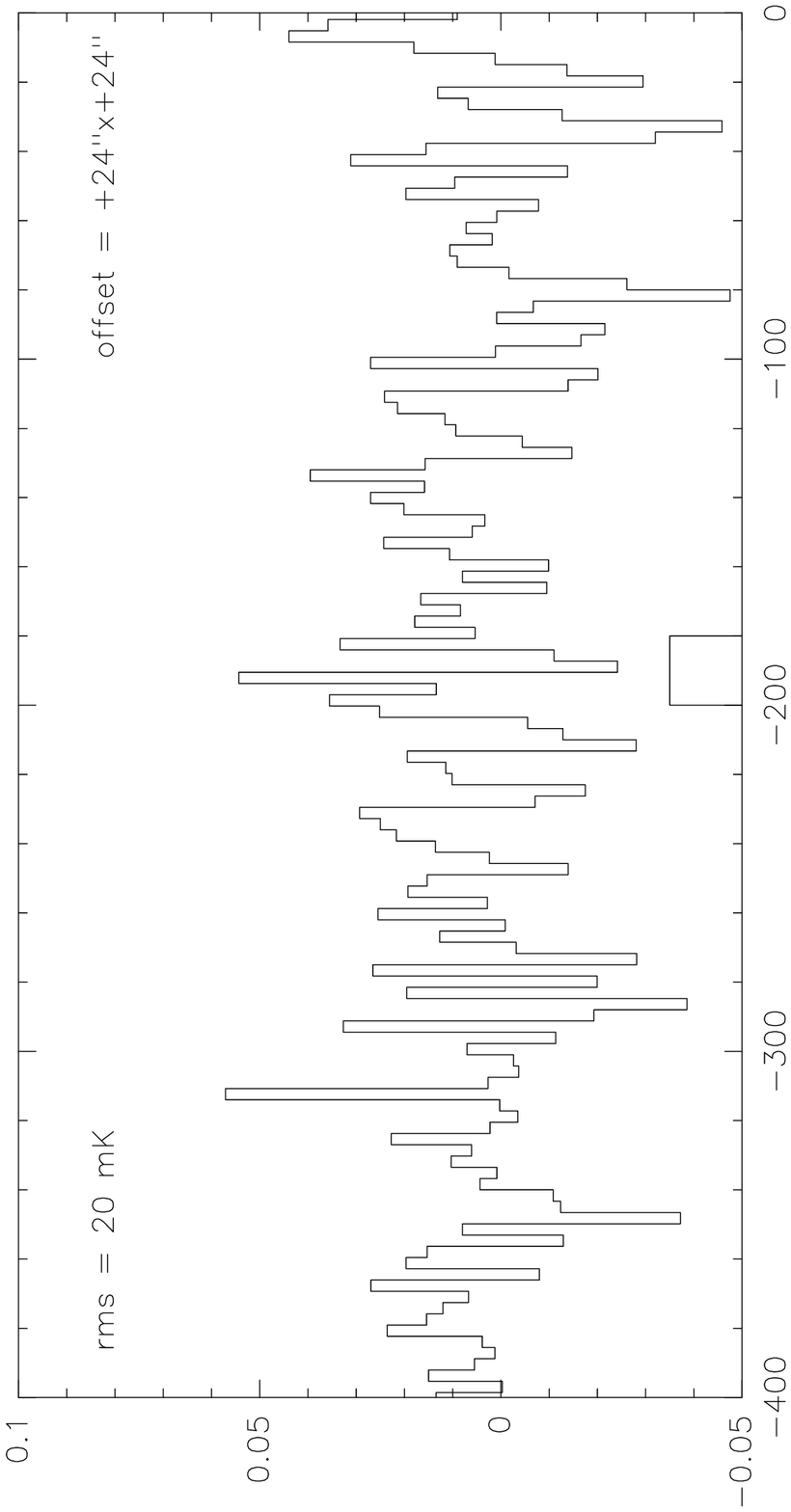}
\includegraphics[width=2.5cm]{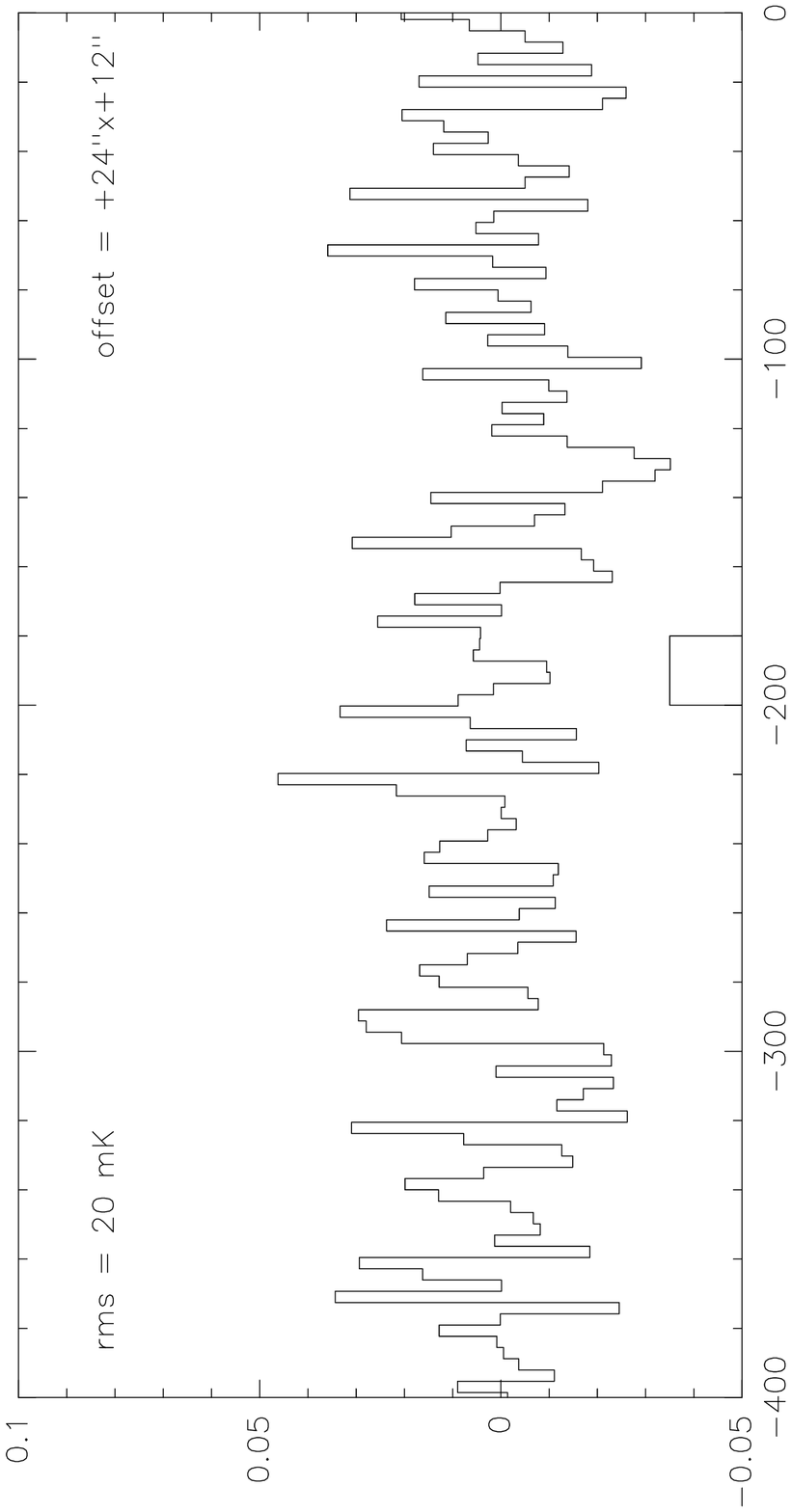}
\includegraphics[width=2.5cm]{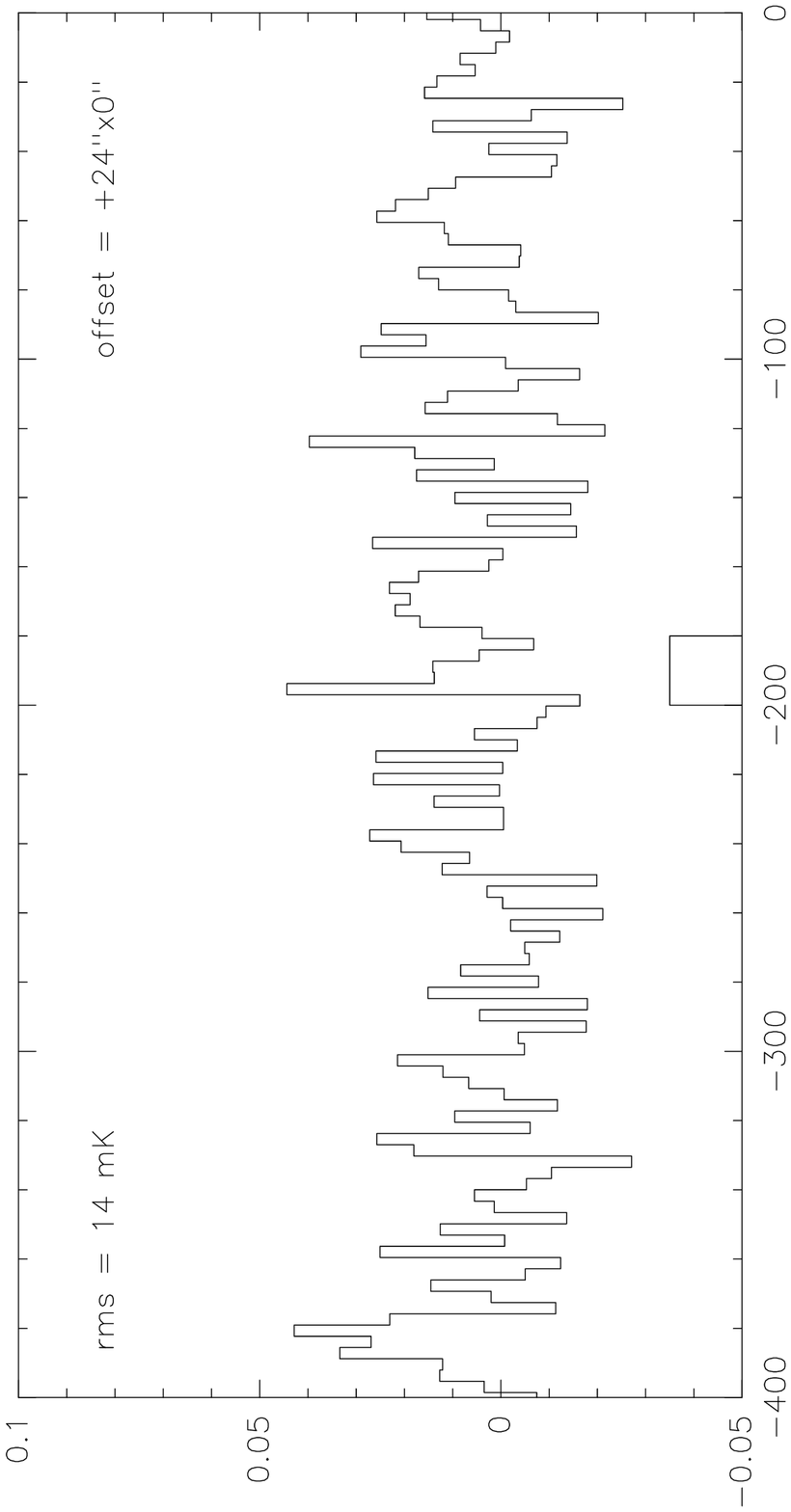}
\includegraphics[width=2.5cm]{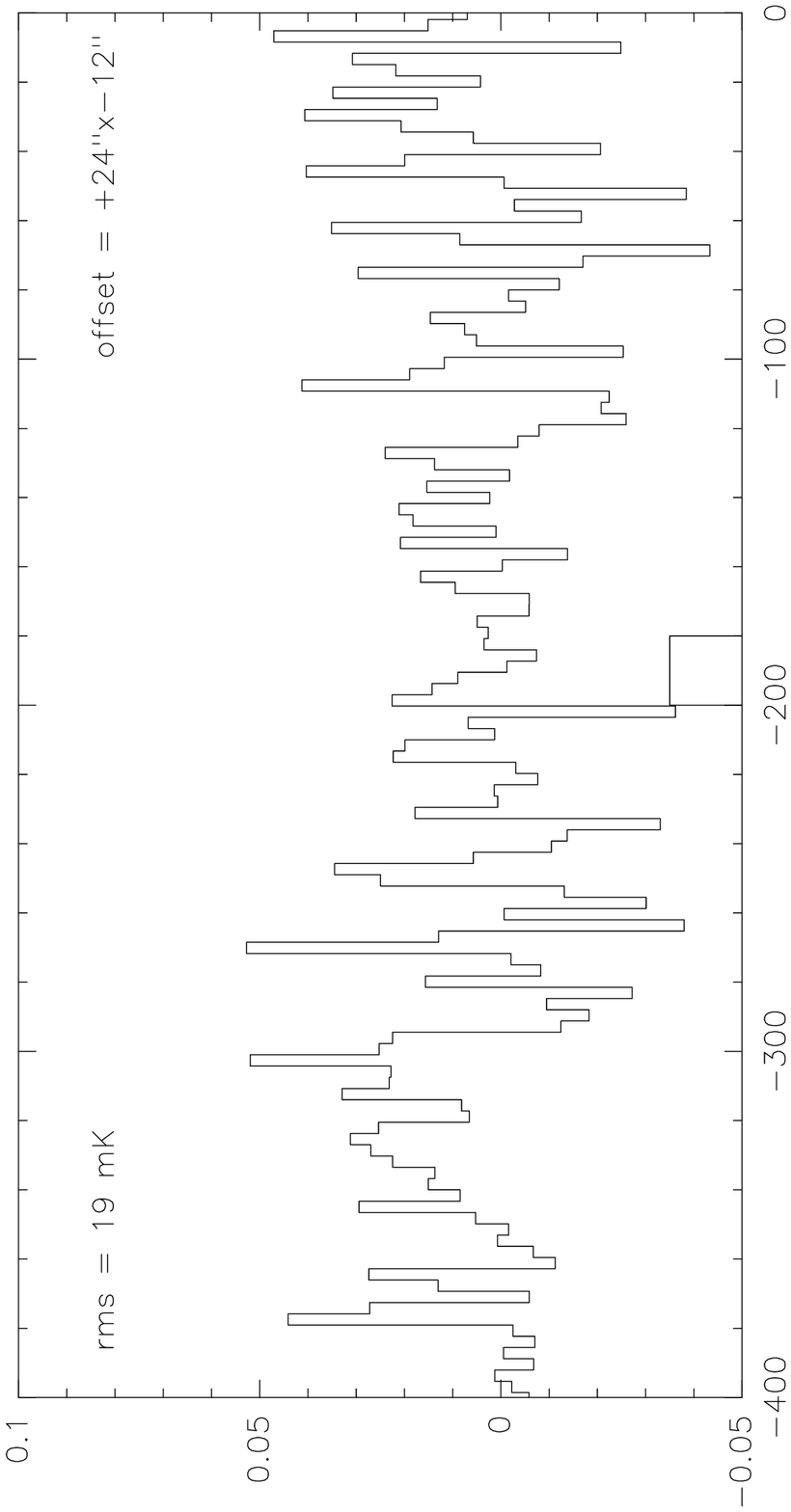}
\includegraphics[width=2.5cm]{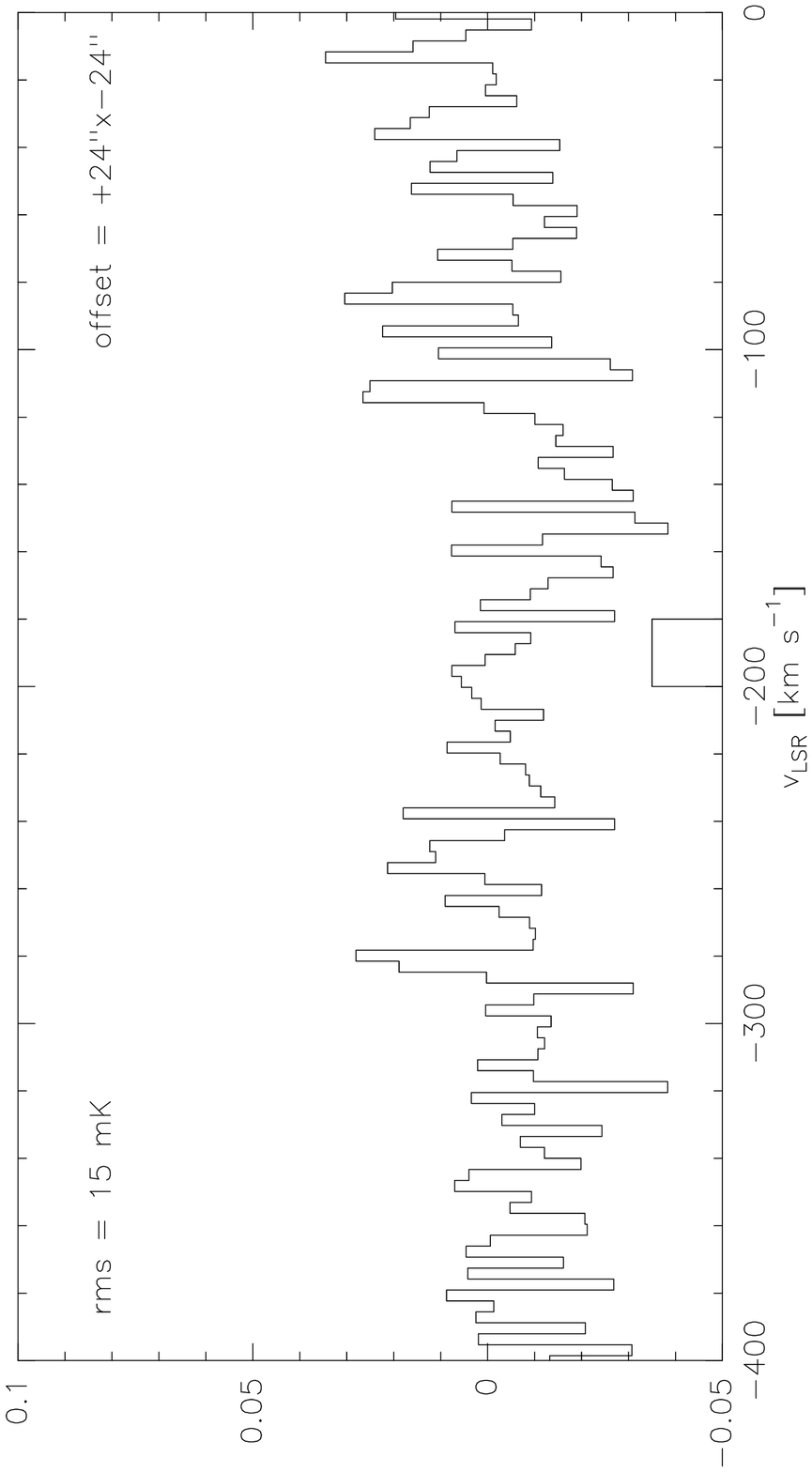}

\includegraphics[width=2.5cm]{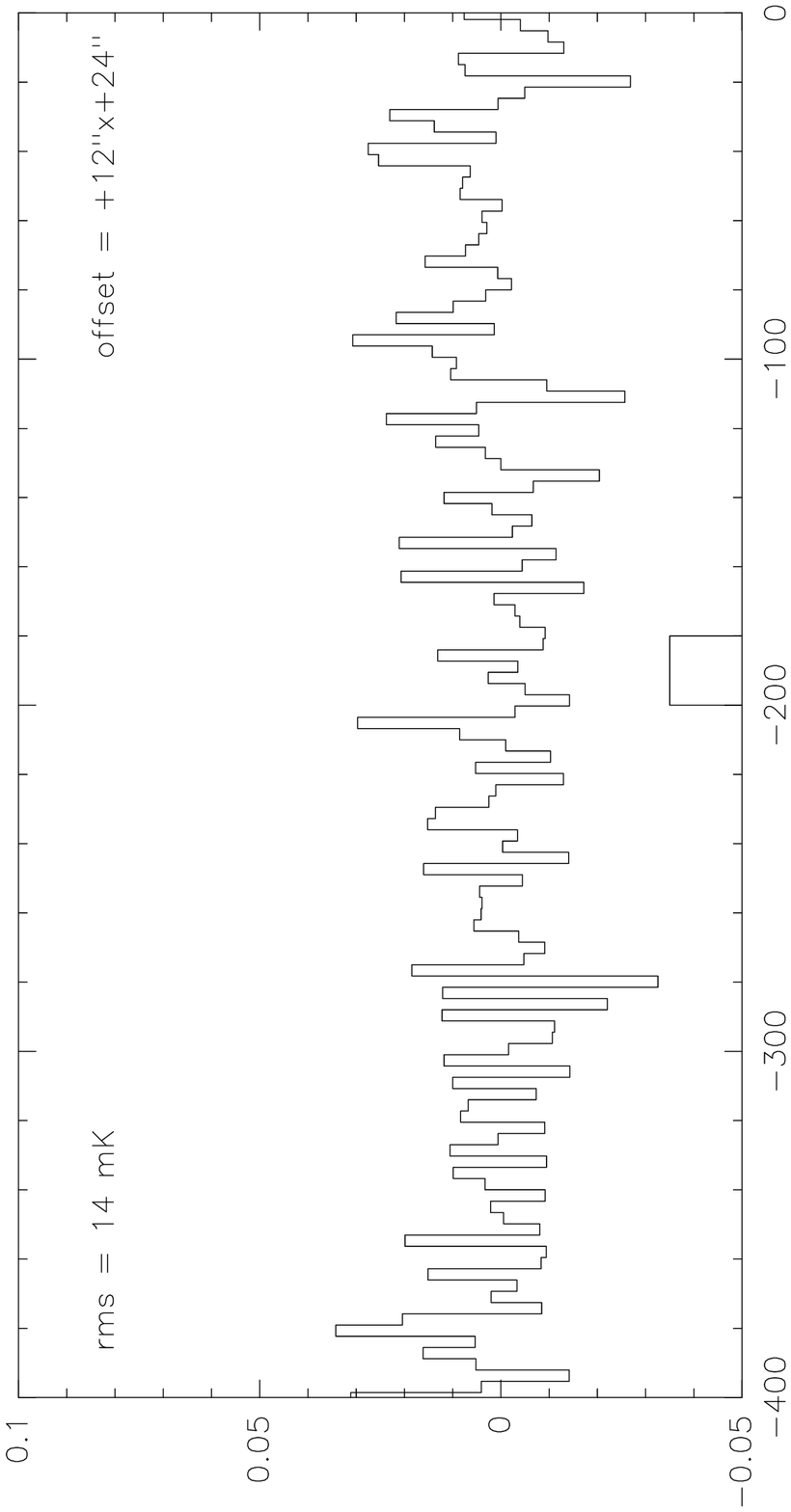}
\includegraphics[width=2.5cm]{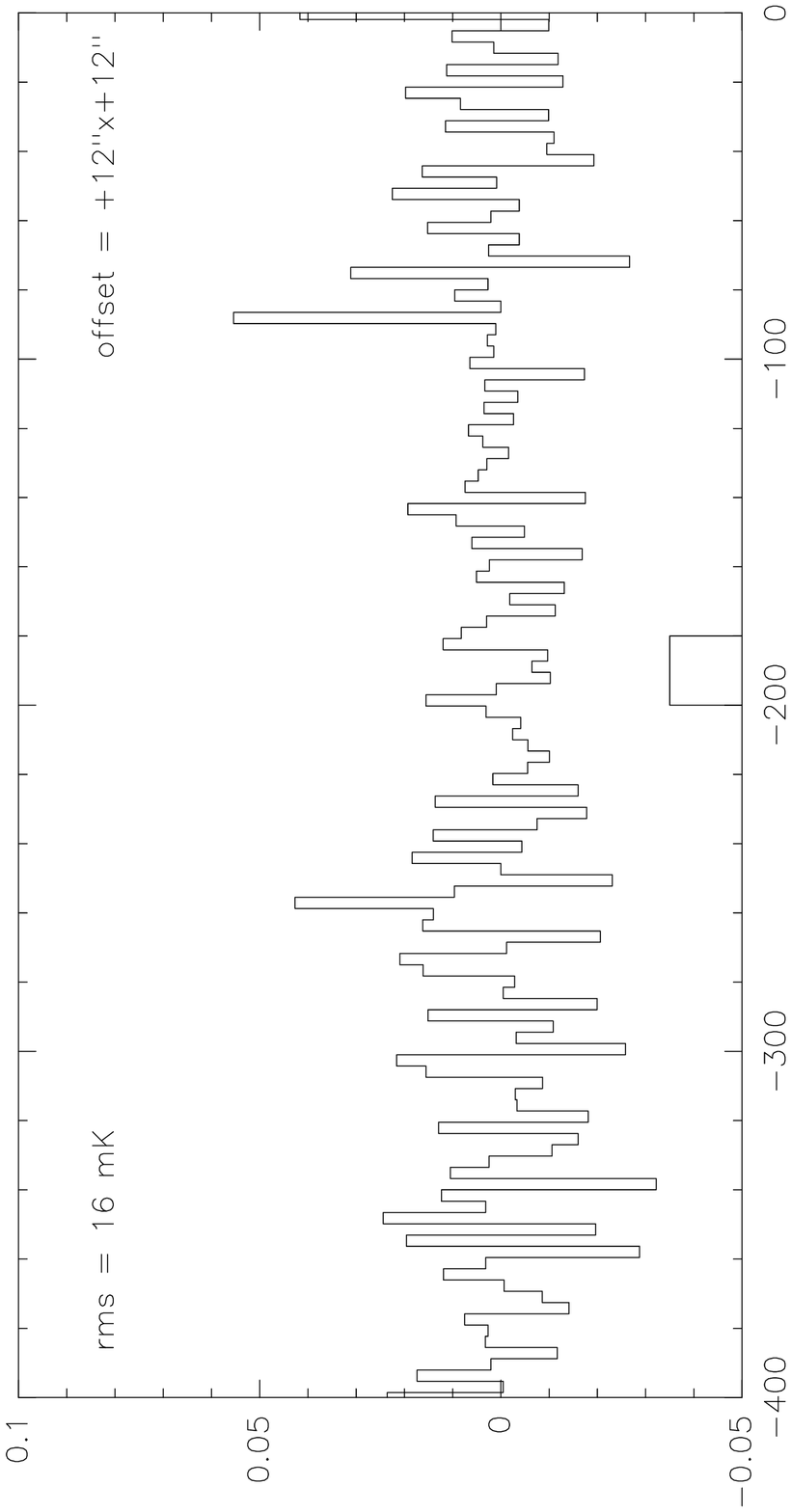}
\includegraphics[width=2.5cm]{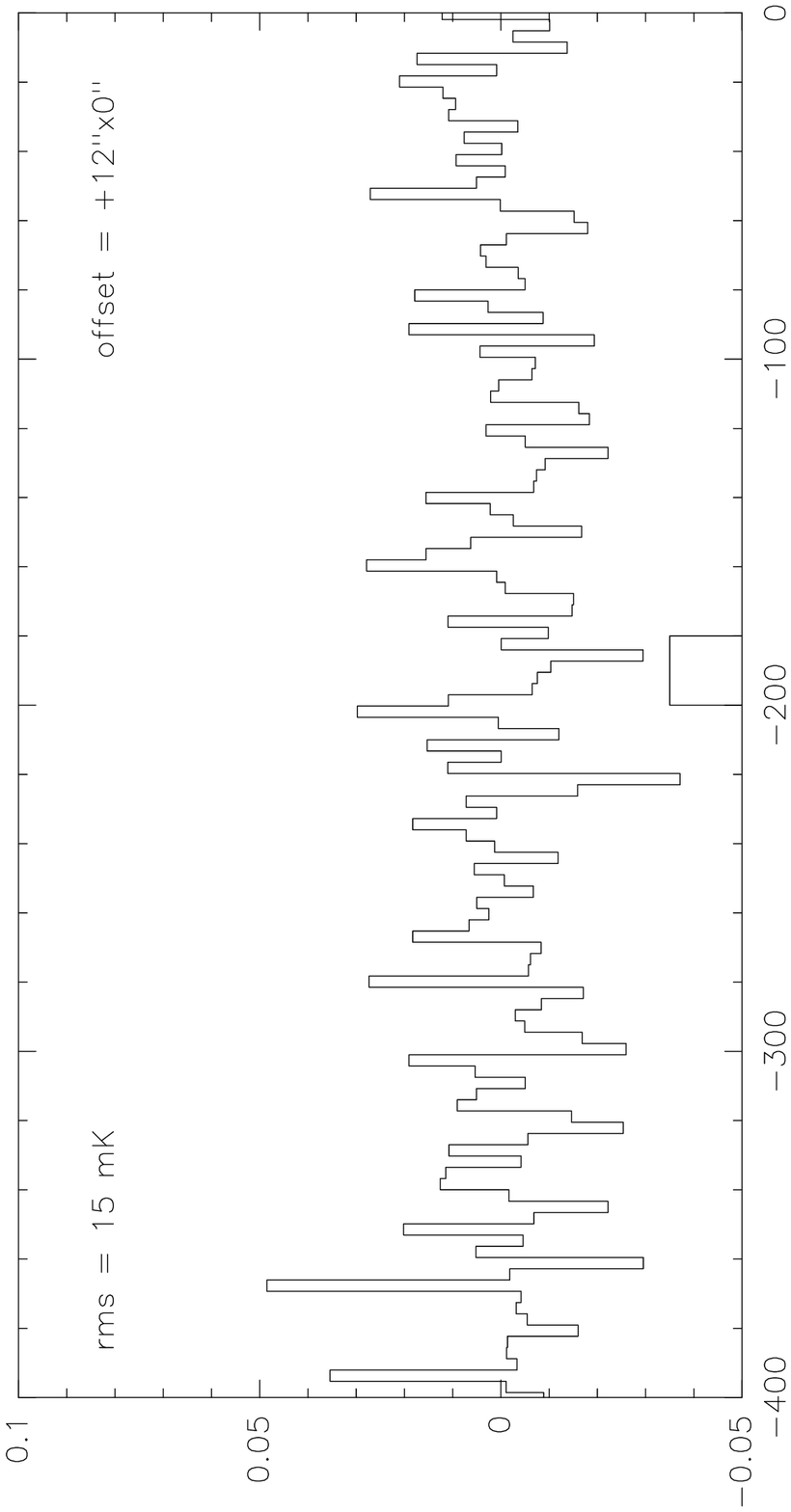}
\includegraphics[width=2.5cm]{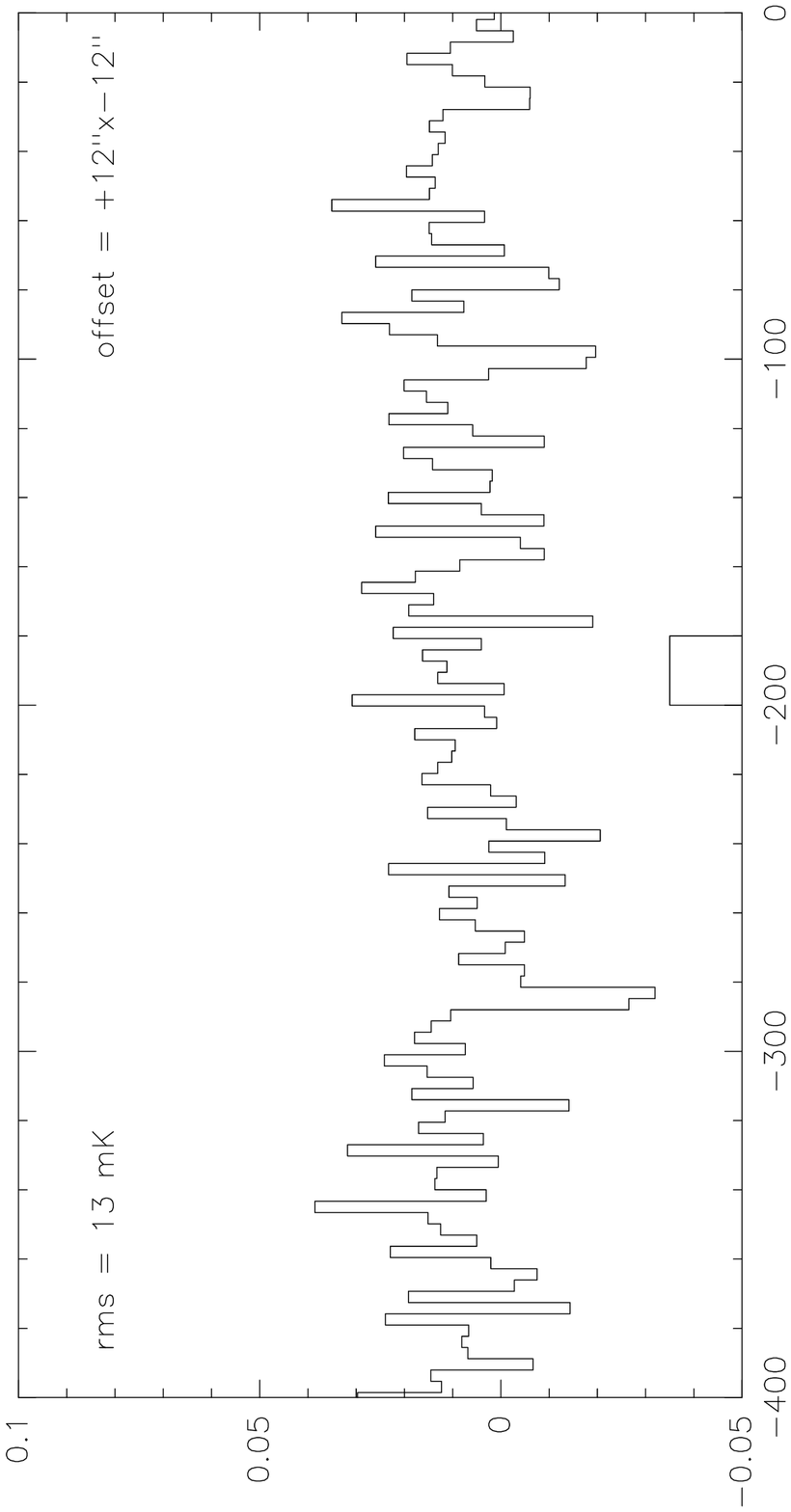}
\includegraphics[width=2.5cm]{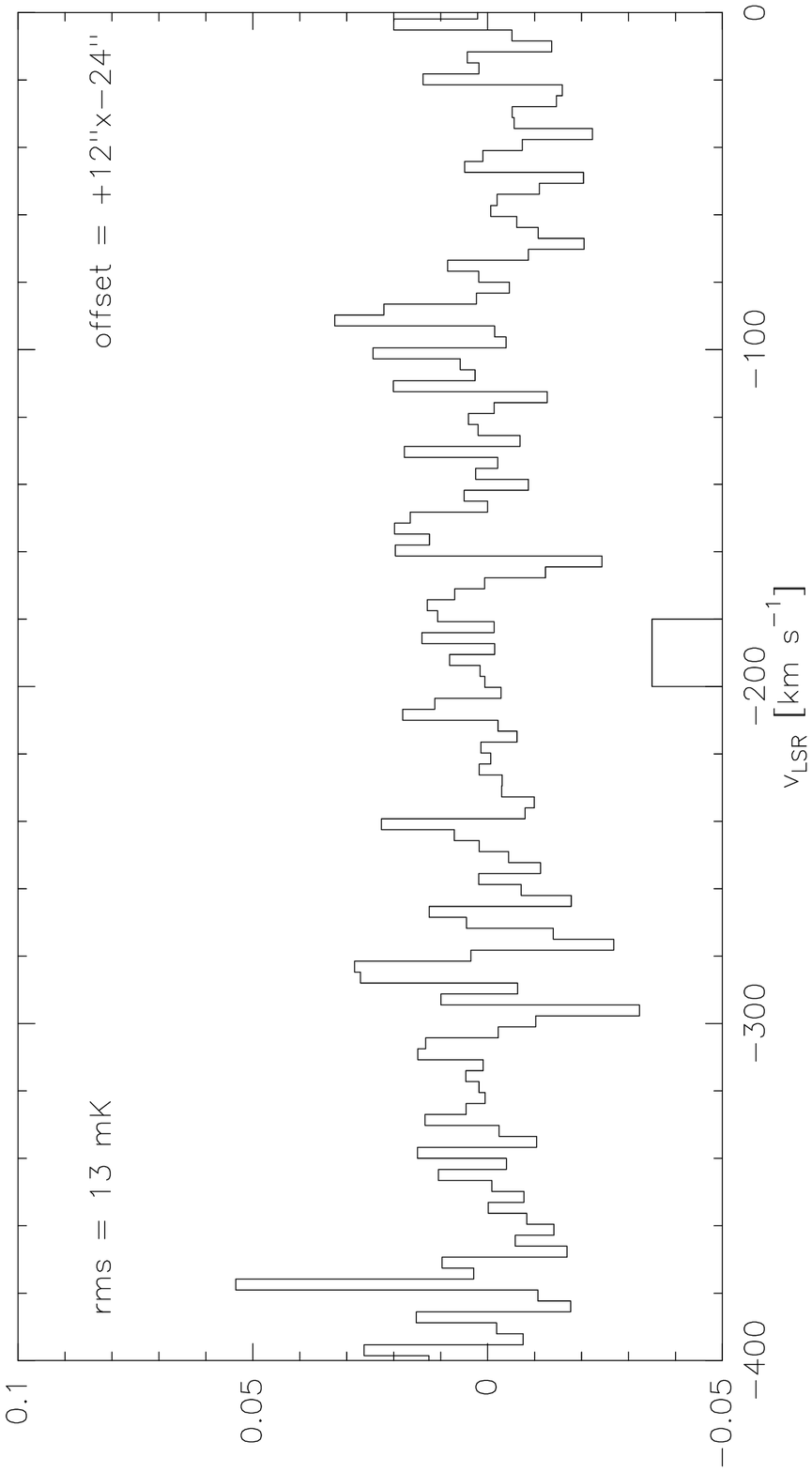}

\includegraphics[width=2.5cm]{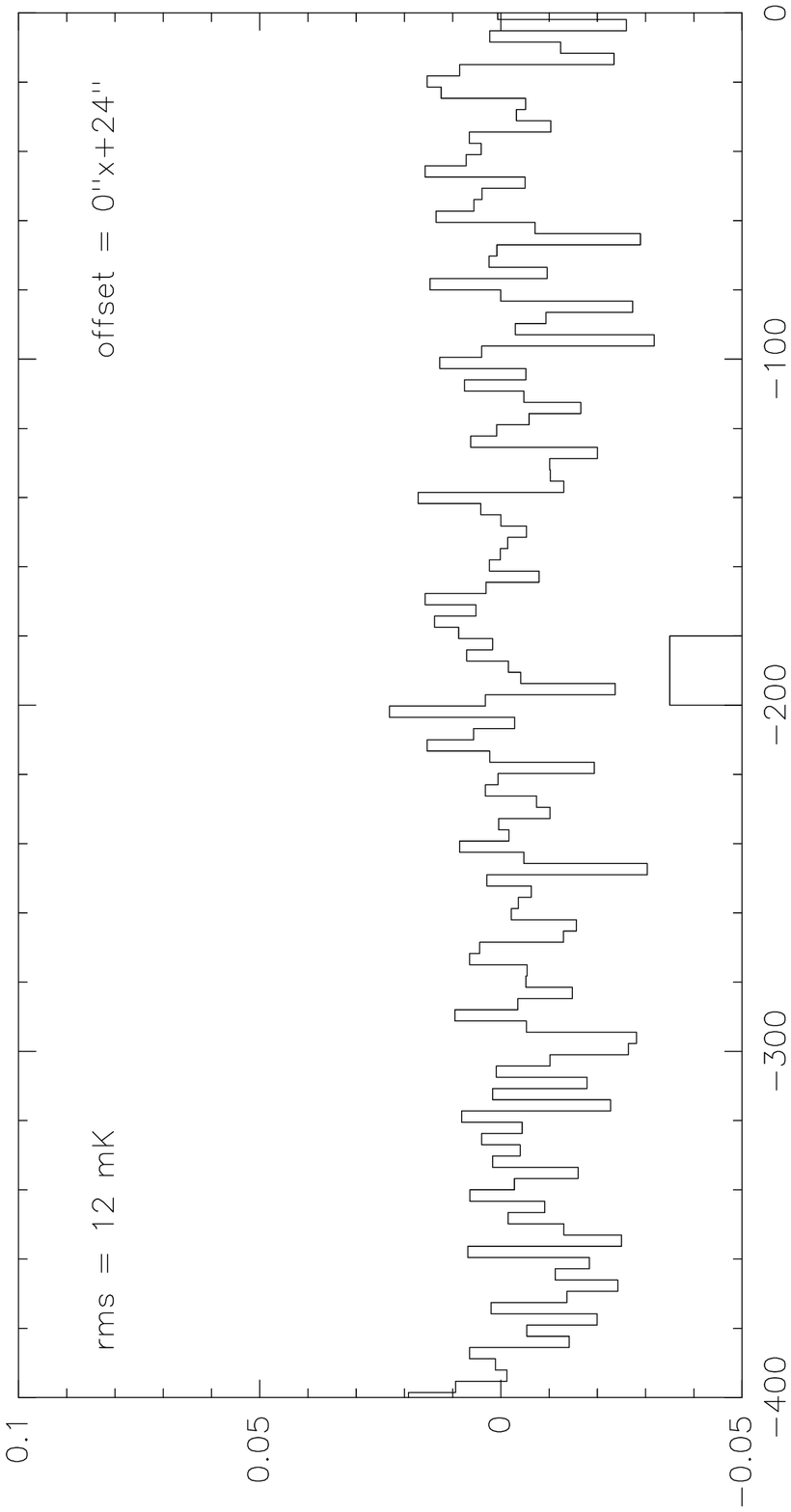}
\includegraphics[width=2.5cm]{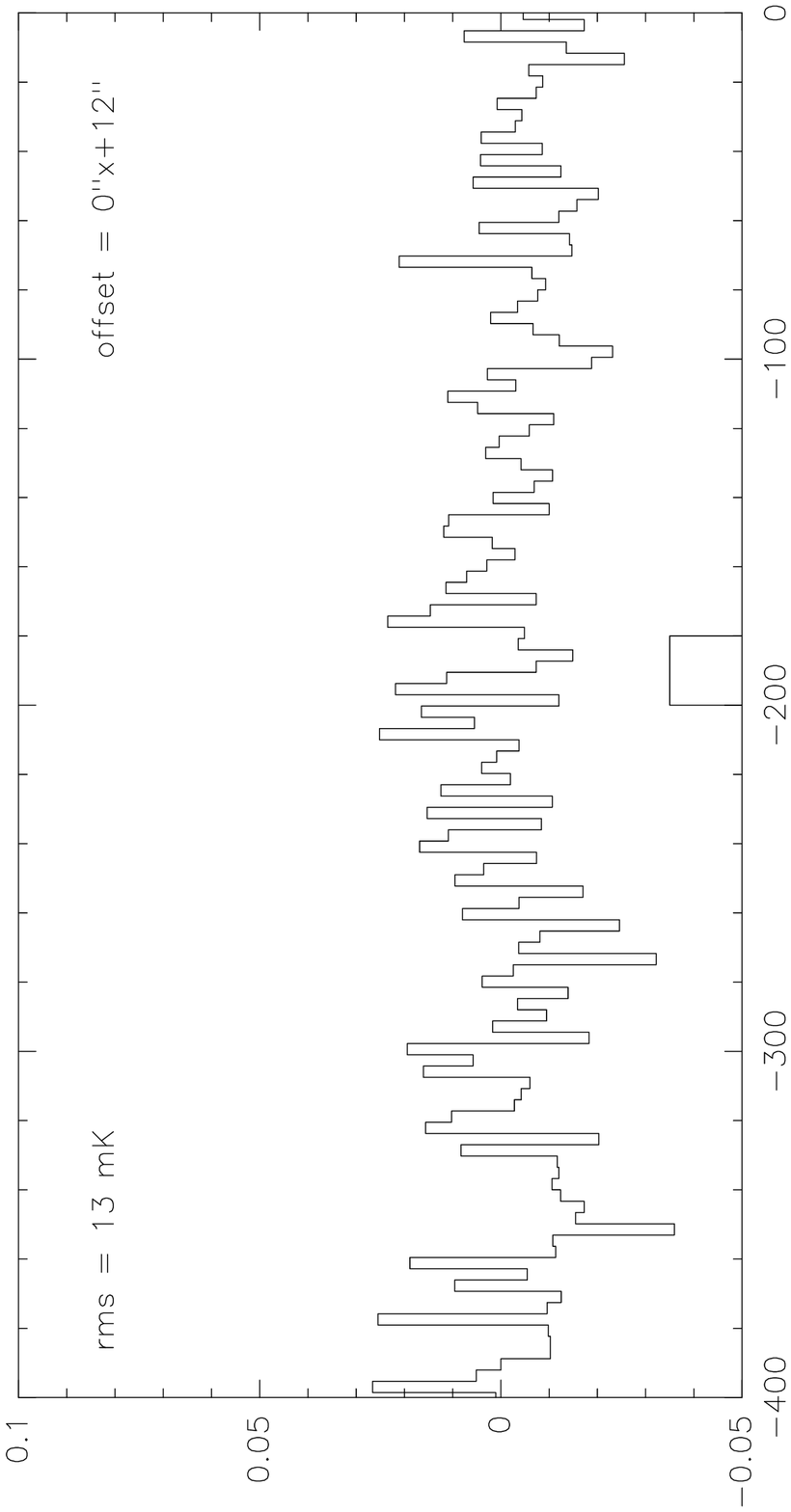}
\includegraphics[width=2.5cm]{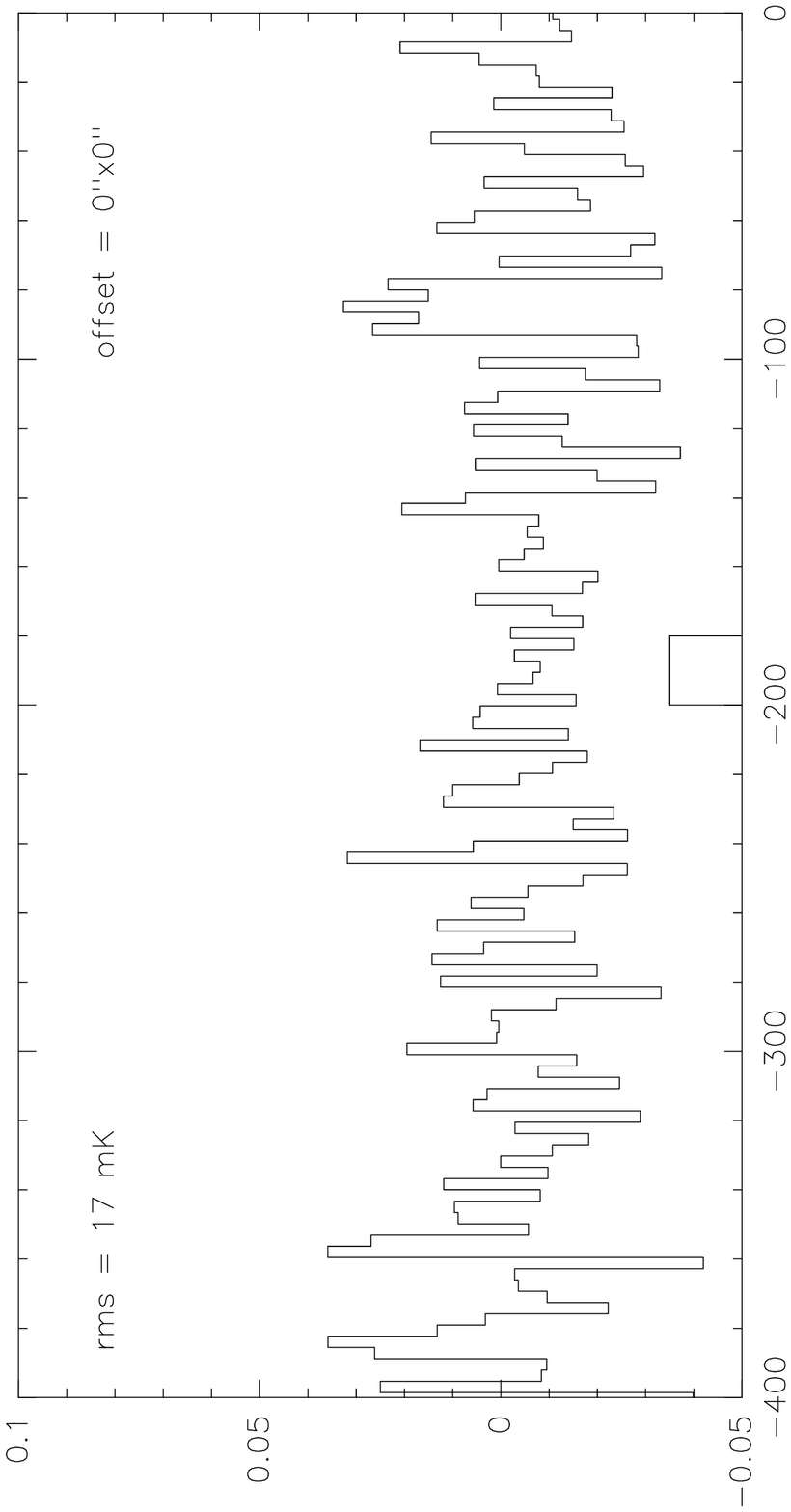}
\includegraphics[width=2.5cm]{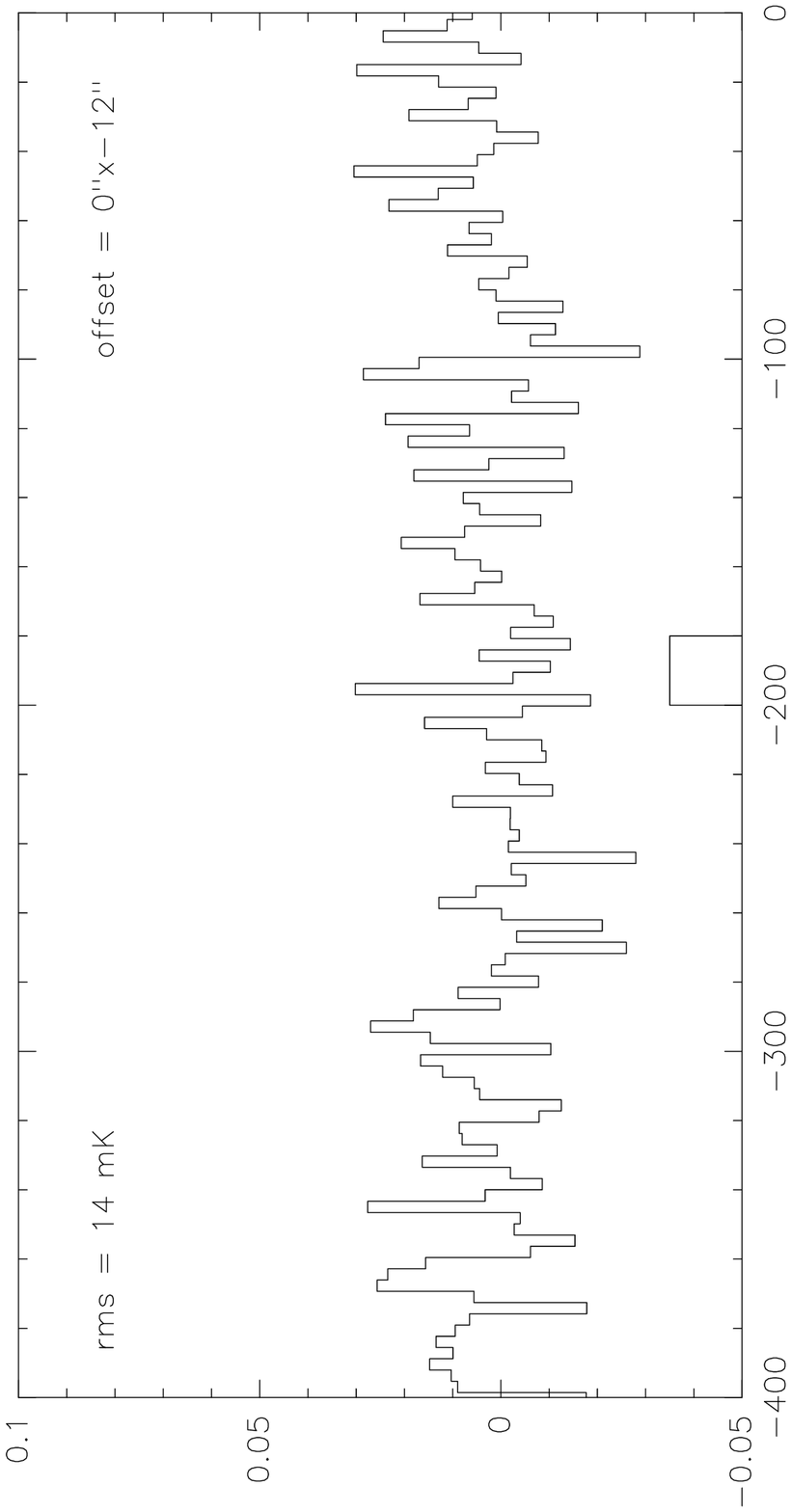}
\includegraphics[width=2.5cm]{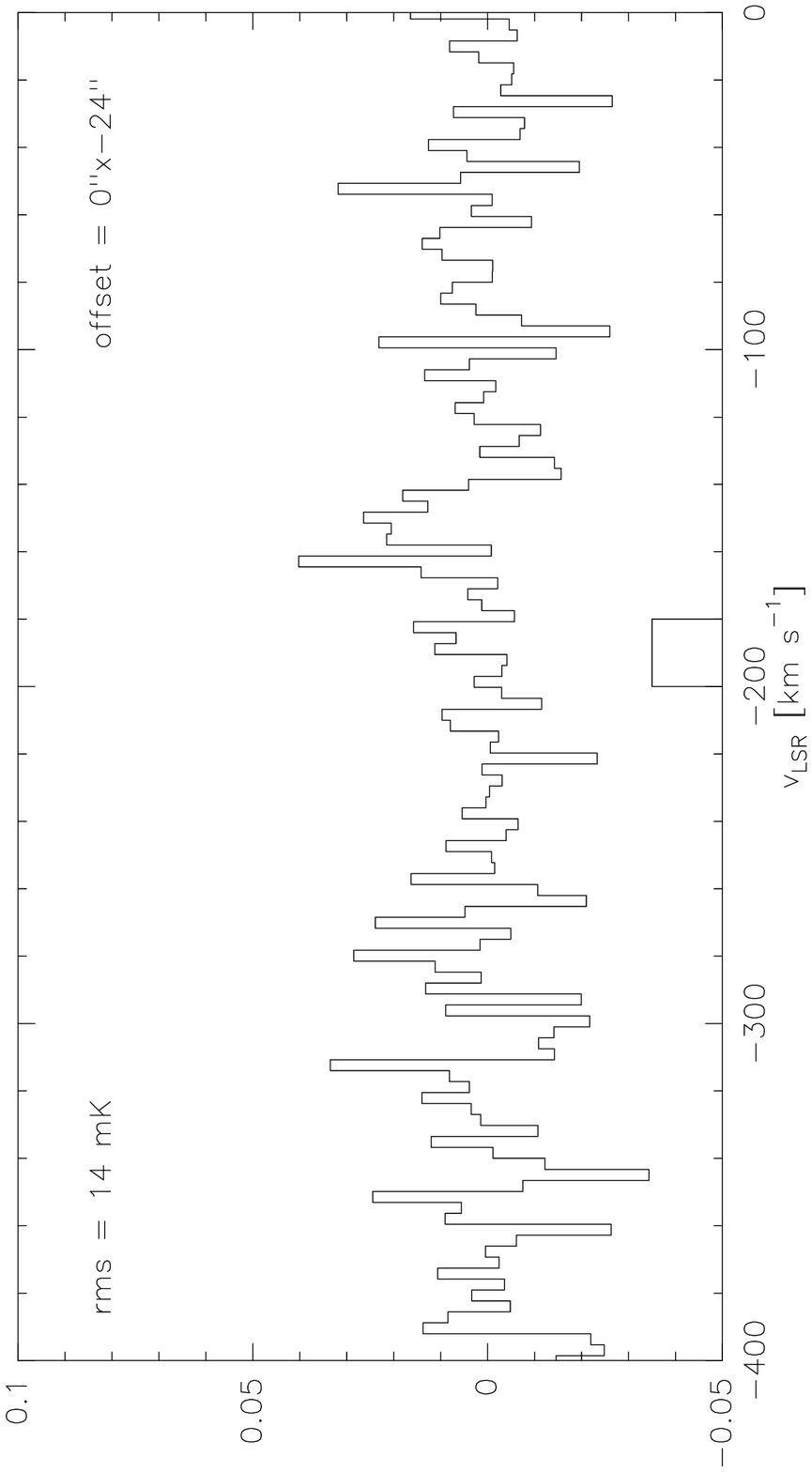}

\includegraphics[width=2.5cm]{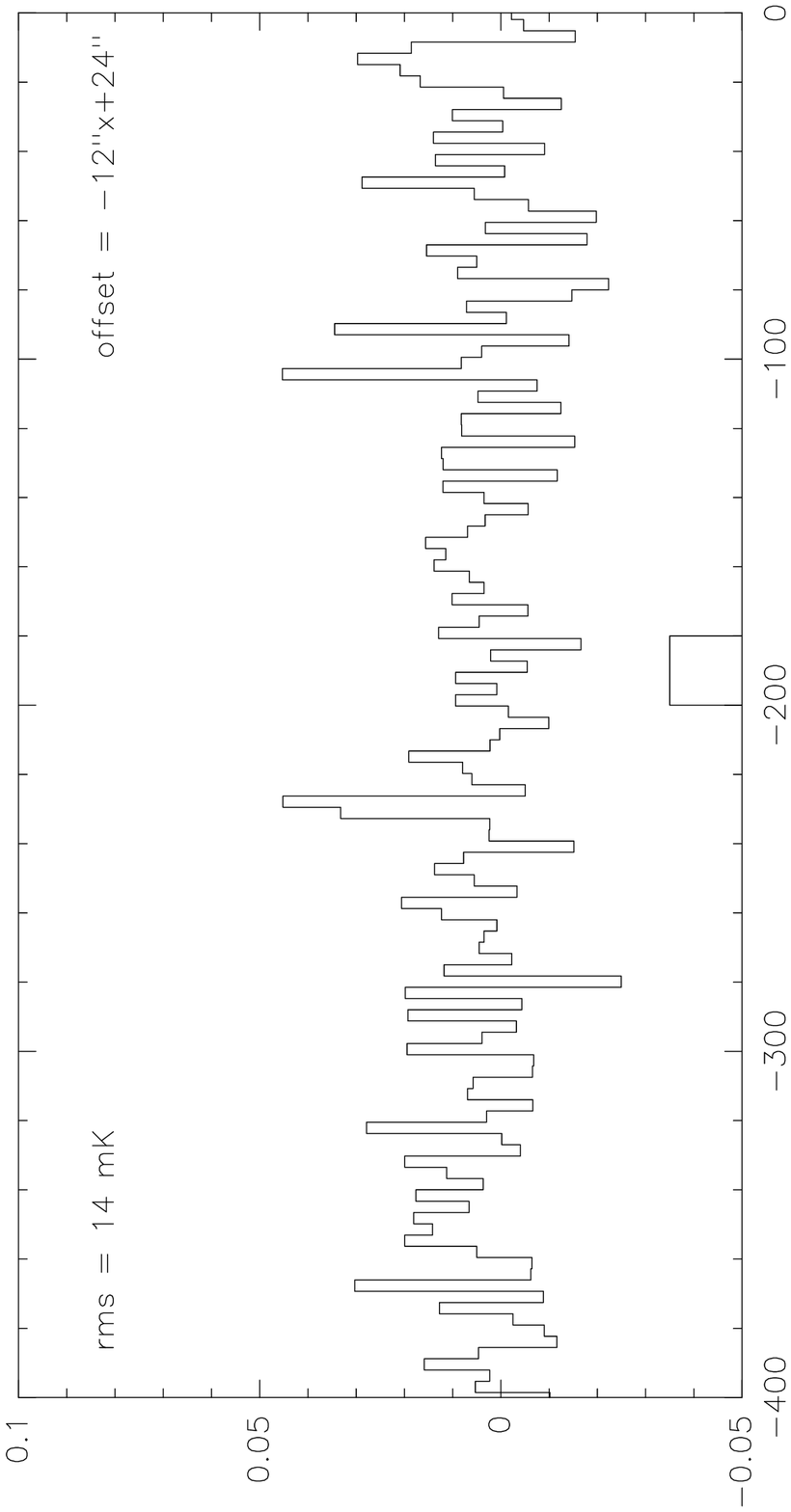}
\includegraphics[width=2.5cm]{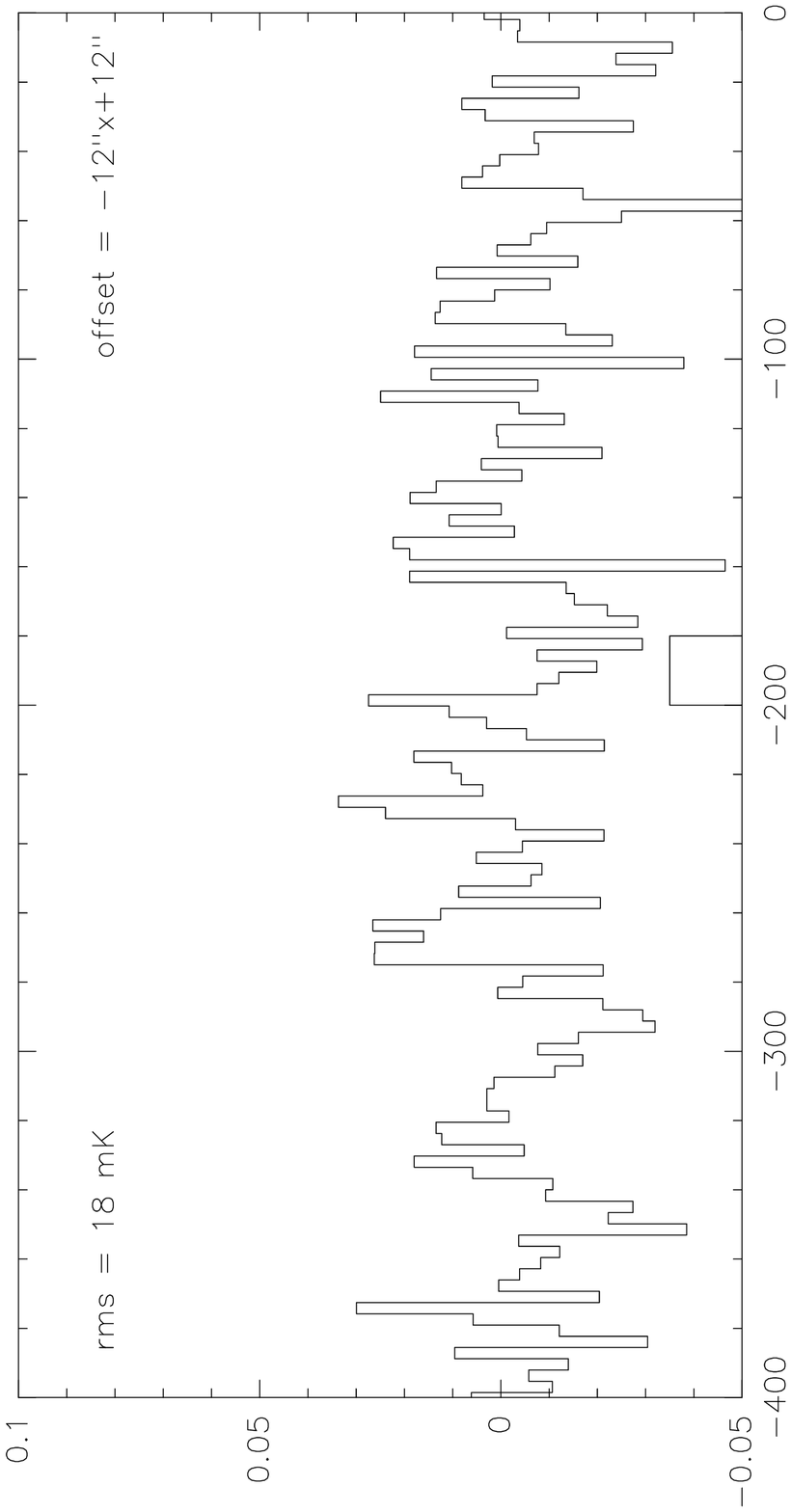}
\includegraphics[width=2.5cm]{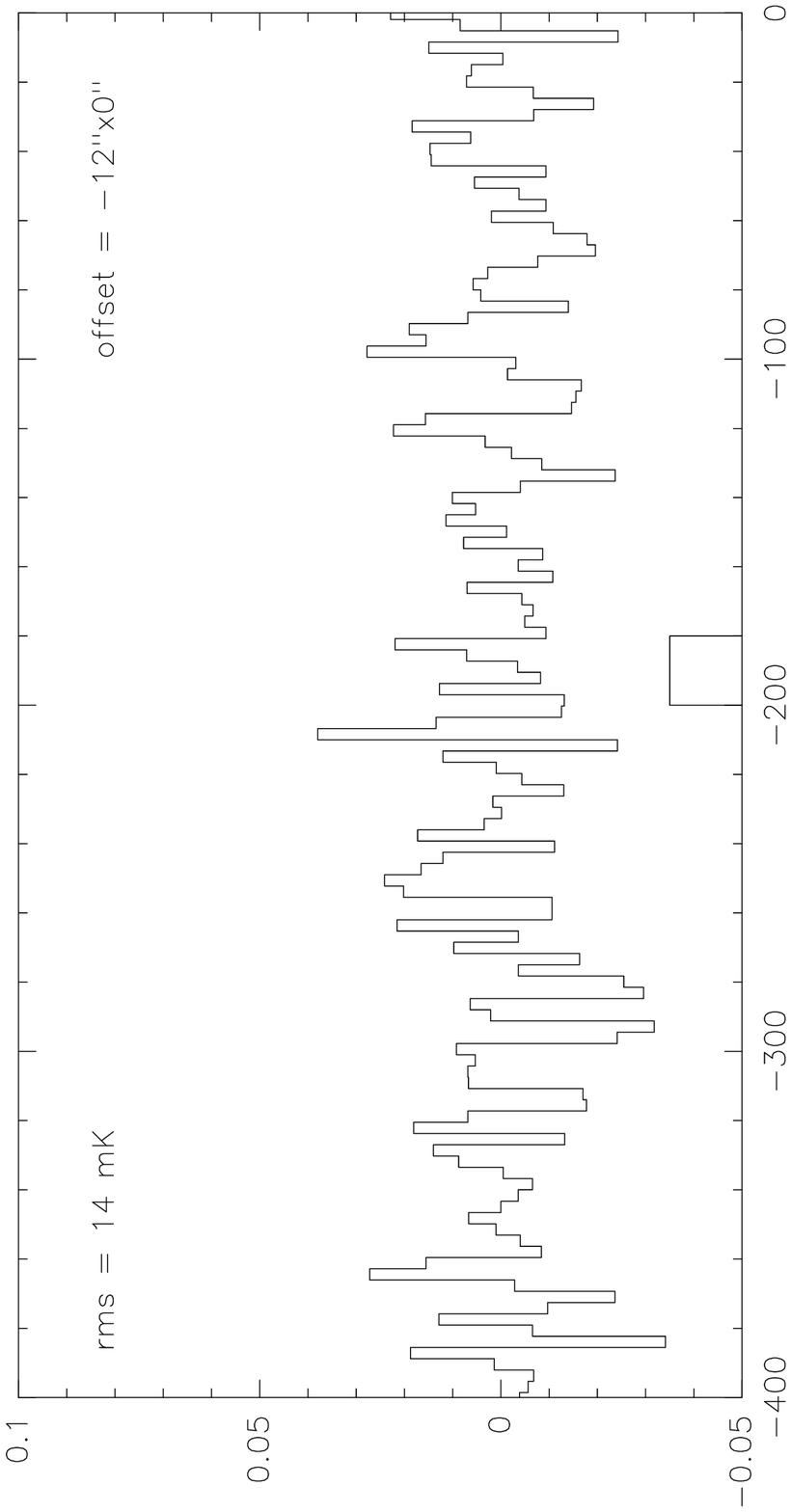}
\includegraphics[width=2.5cm]{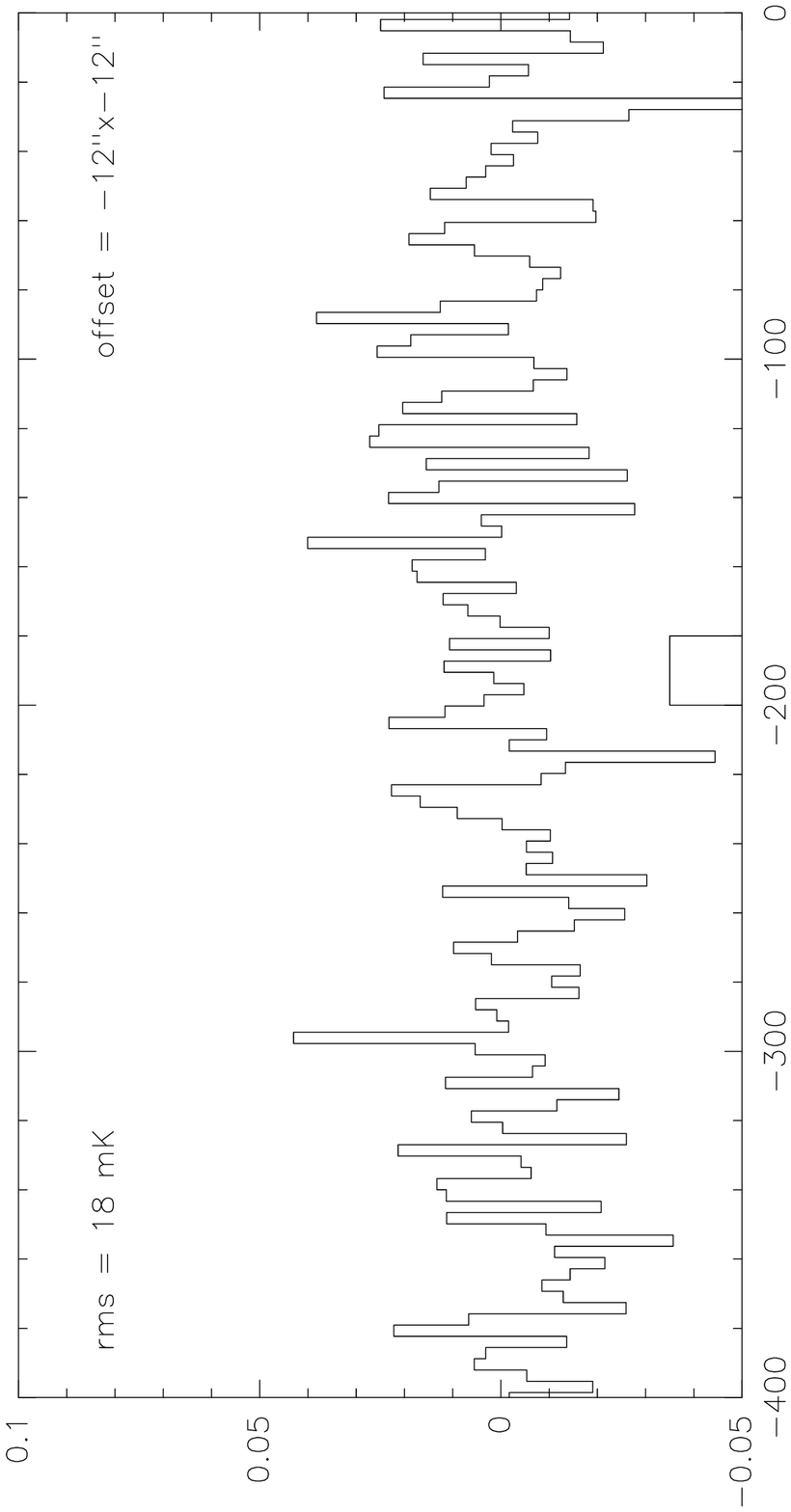}
\includegraphics[width=2.5cm]{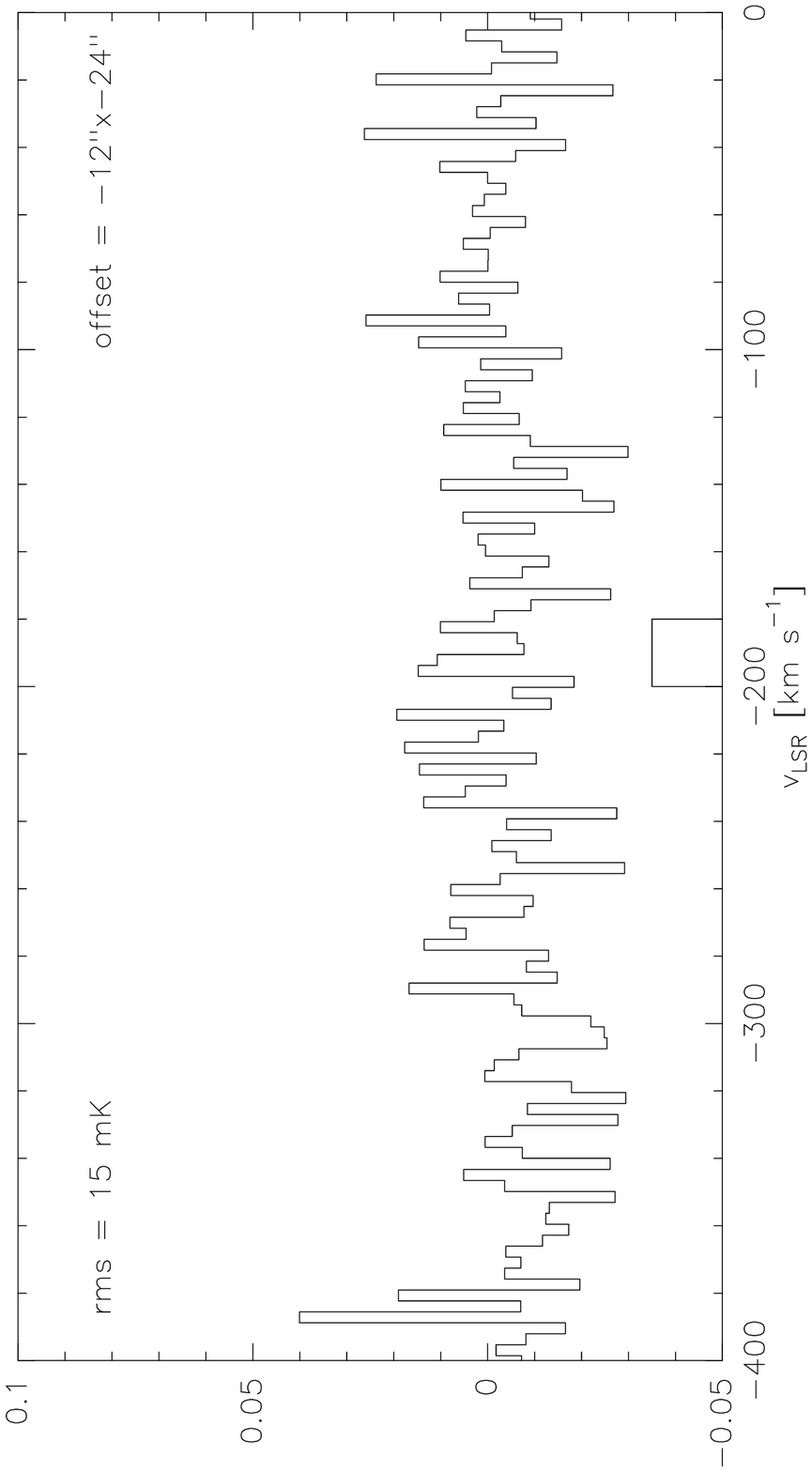}

\includegraphics[width=2.5cm]{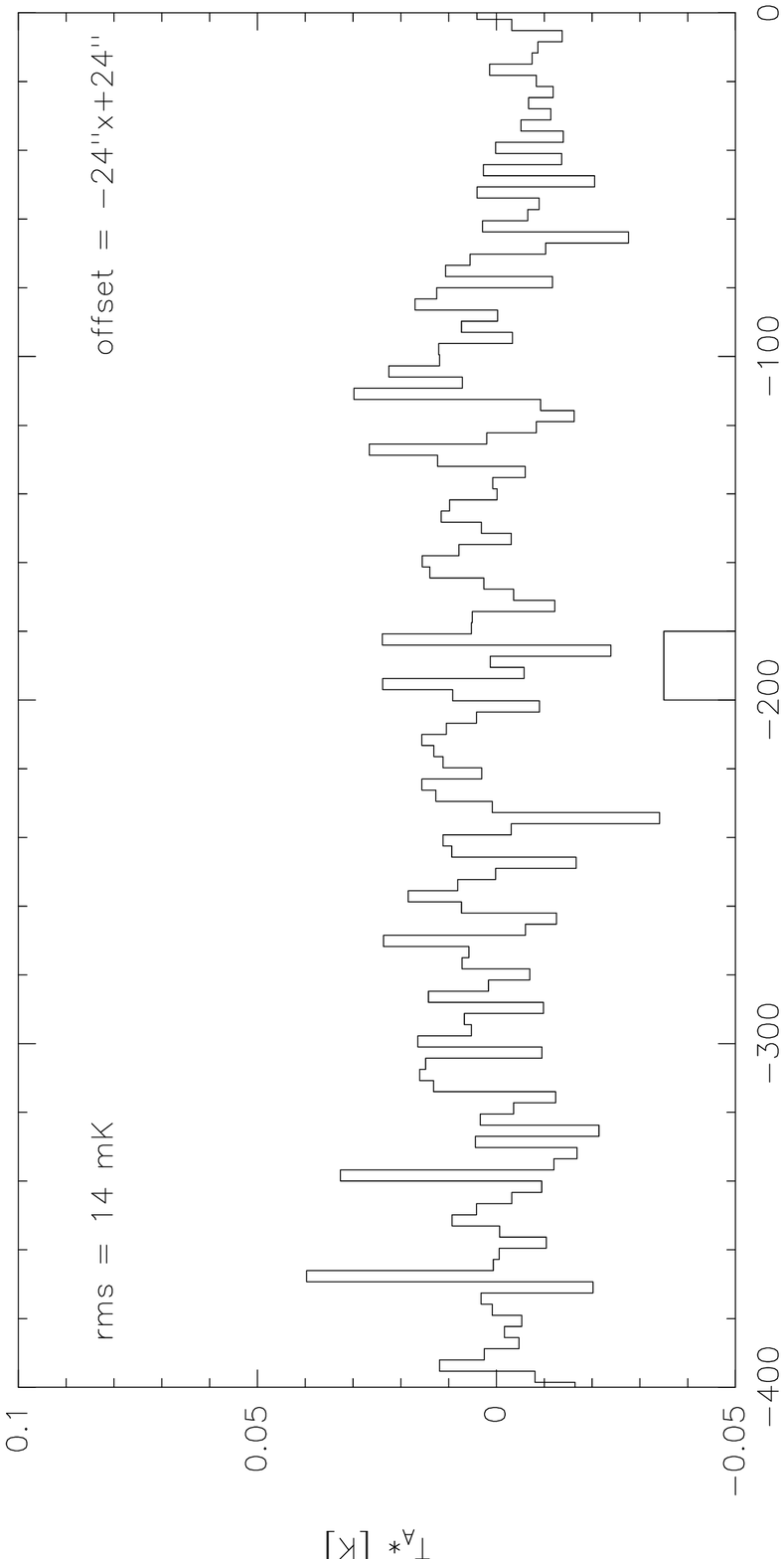}
\includegraphics[width=2.5cm]{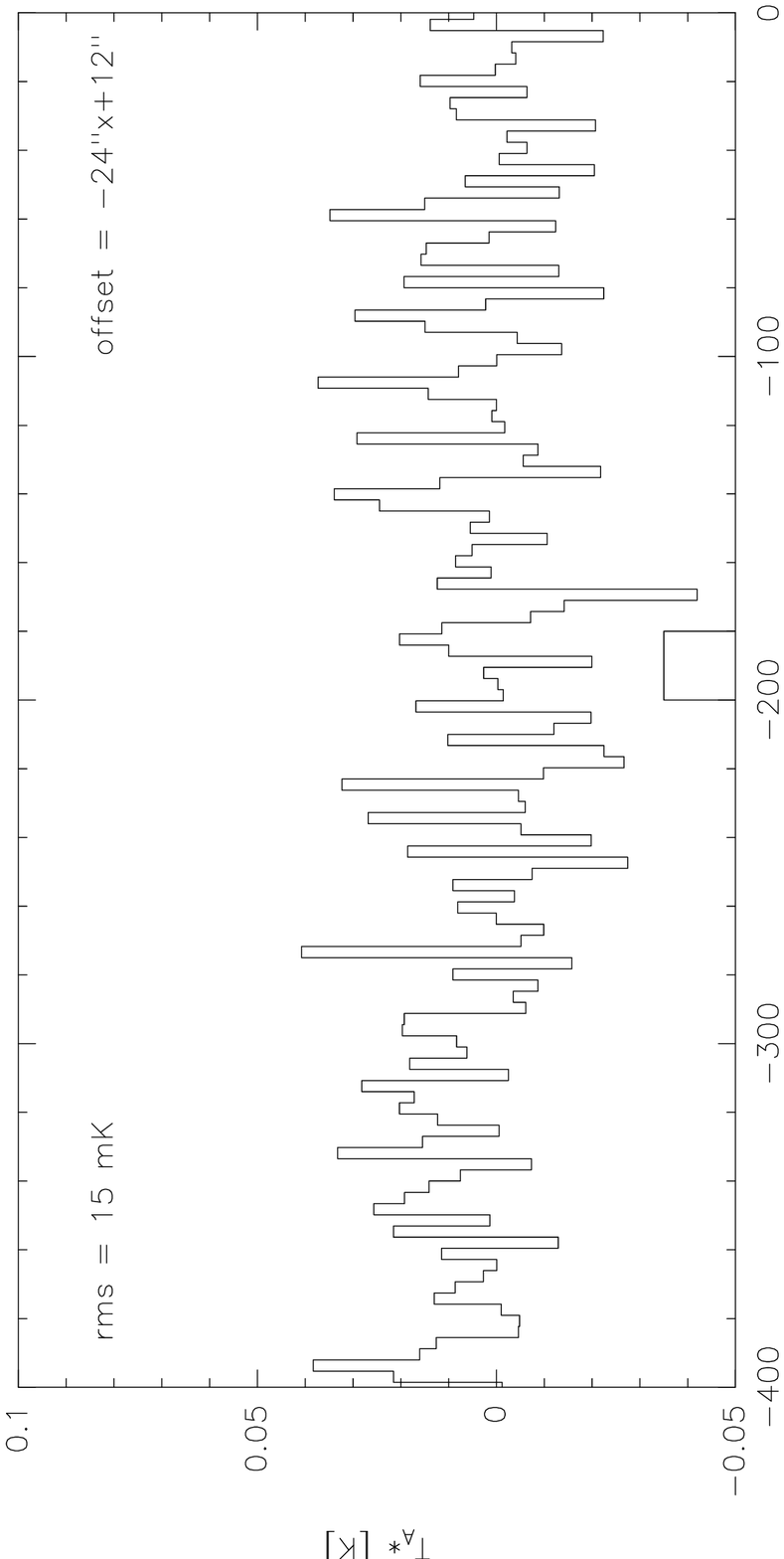}
\includegraphics[width=2.5cm]{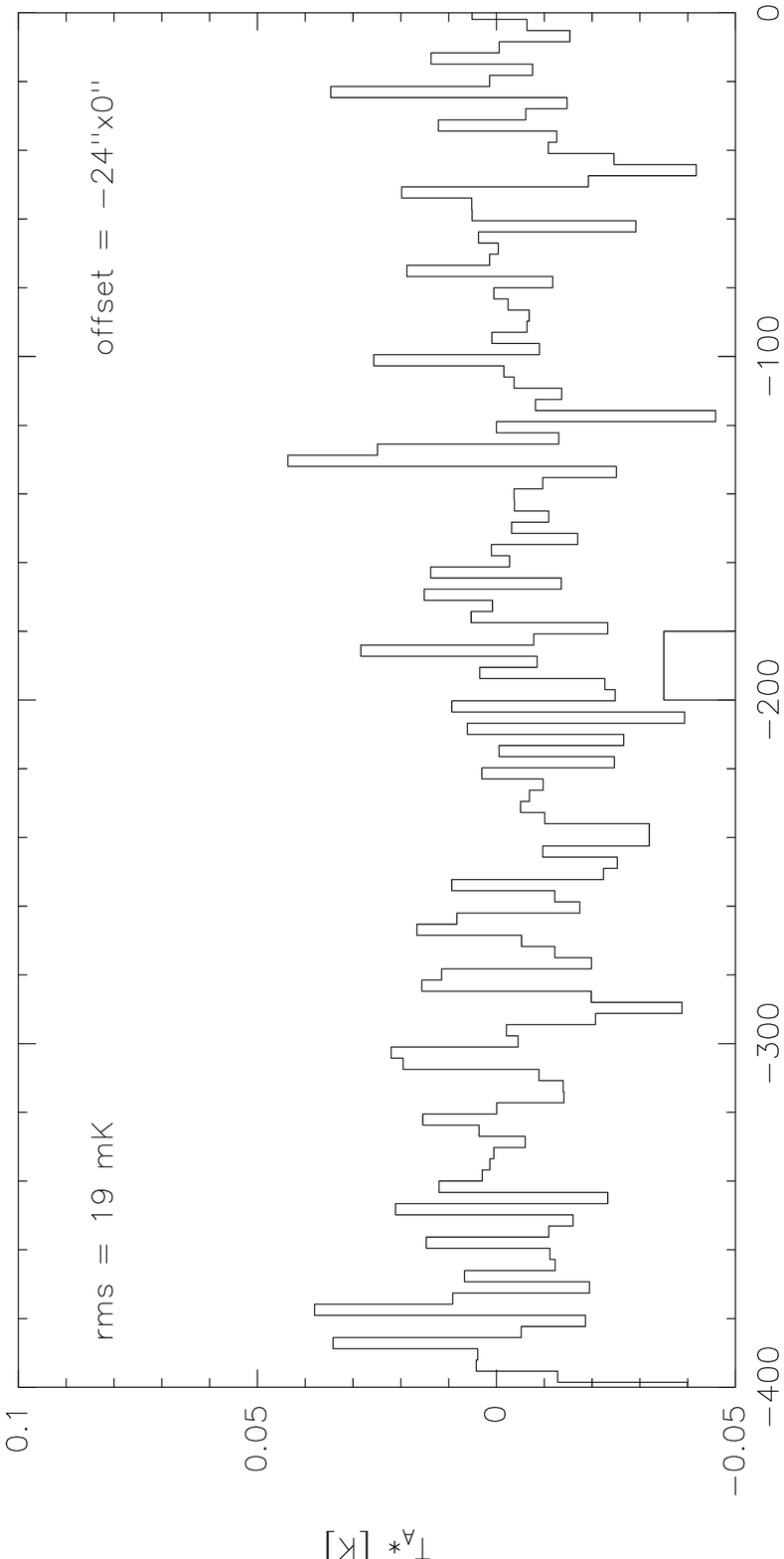}
\includegraphics[width=2.5cm]{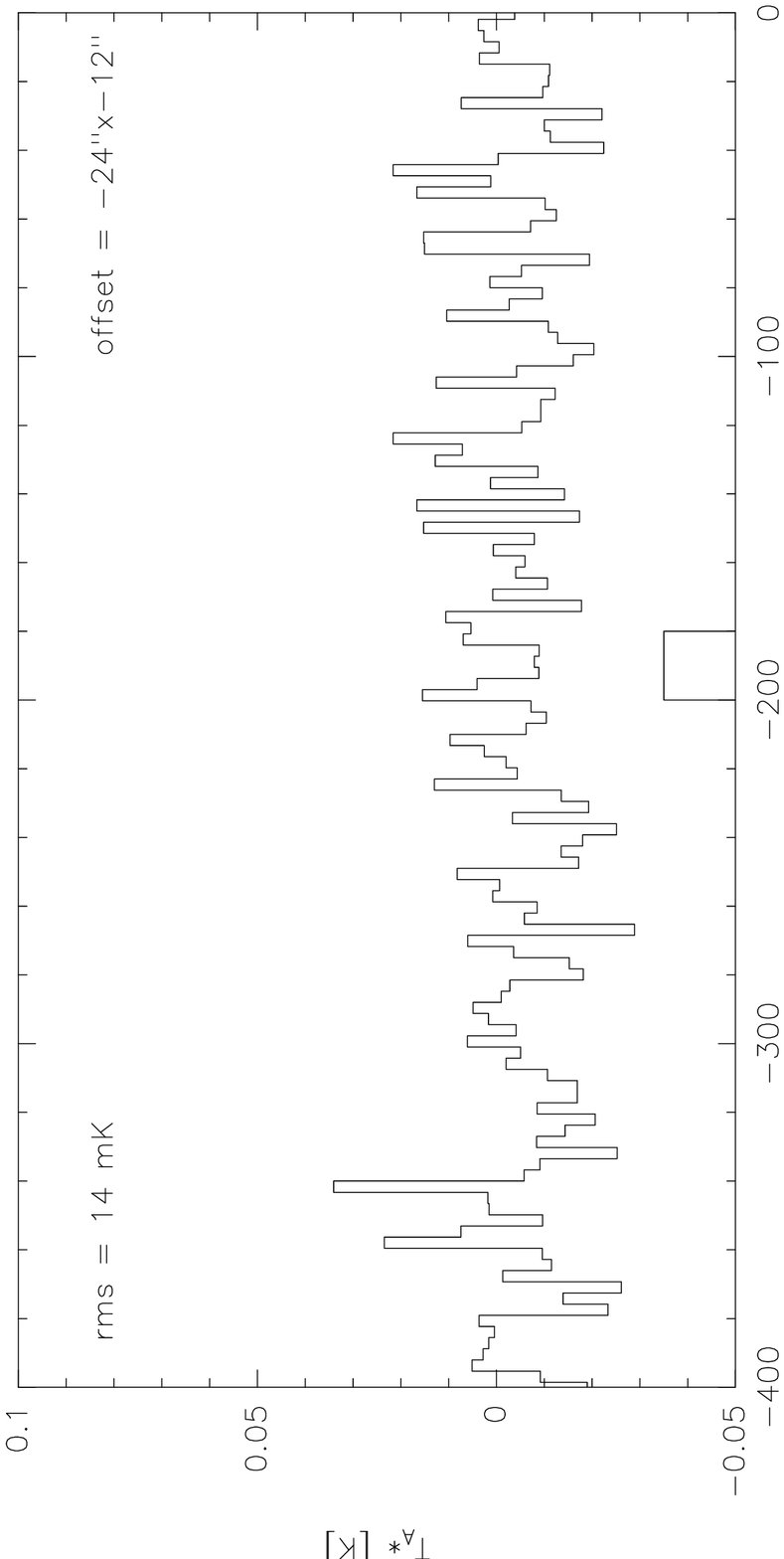}
\includegraphics[width=2.5cm]{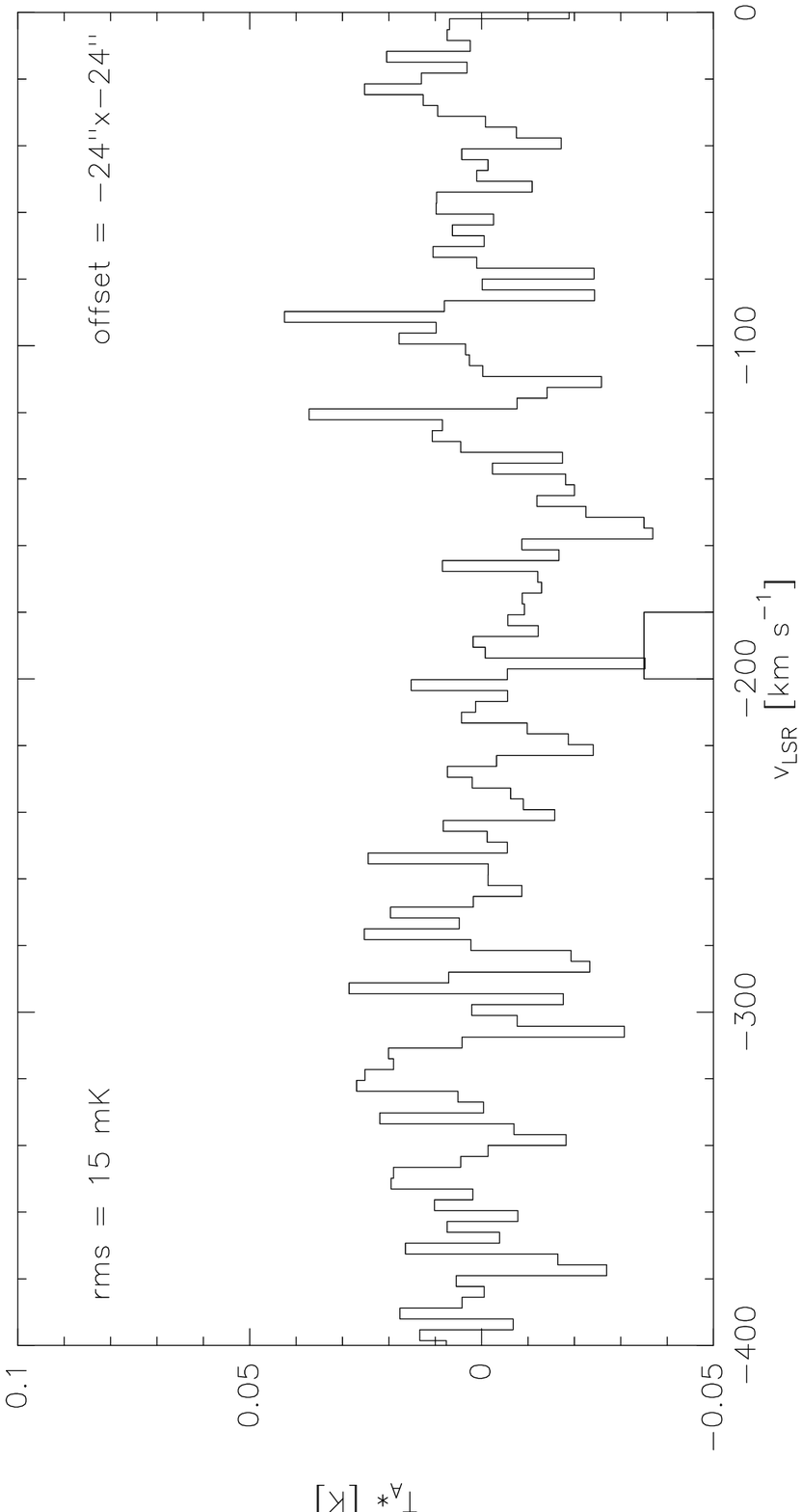}
\caption{Same as Fig.~\ref{HVC-A-spectra}, but the mosaic is centered on the
\ion{H}{i} core HVC-B.}
\label{HVC-B-spectra}
\end{figure*}
%
%________________________________________________________________

The observations were performed with the IRAM 30~m millimeter-wave telescope at 
Pico Veleta, Spain, on June 20--23, 2006 under good weather conditions. We used
four receivers simultaneously, two centered on the $^{12}$CO(1--0) line at 115
GHz and two on the $^{12}$CO(2--1) line at 230 GHz. The beam width of the 
telescope at these two frequencies is $22''$ and $11''$, respectively. The data
were recorded using the VESPA autocorrelator with 480 MHz bandwidth and 0.32 
MHz resolution at 3~mm and two 1 MHz filter banks (512 channels each) at 1~mm.
The resulting velocity coverage at 115 GHz is 1250 km~s$^{-1}$ with 0.8 
km~s$^{-1}$ resolution. The corresponding values at 230 GHz are 660 km~s$^{-1}$ 
and 1.3 km~s$^{-1}$. To sample significant probability to find the CO molecular 
emission, we made $5\times 5 = 25$ points maps centered on each of the two 
selected HVC regions. The observations followed the sequence from $-24''$ to 
$+24''$ offsets relative to the central coordinates of the HVC \ion{H}{i} cores 
with a $12''$ spatial sampling in the right ascension direction and same in the 
declination direction, leading to $1'\times 1'$ maps. Spectra were obtained 
using the wobbler switch technique with a wobbler throw of $90''$. The total 
integration time on each point of the HVC-A and HVC-B maps varied between 9 and 
26 minutes, with a final observing time of $\sim 13$ hours on the sources.

The data were reduced with the CLASS software from the GILDAS package. The 
expected average CO line width is $\sim 10$ km~s$^{-1}$, thus we first Hanning 
smoothed all the raw spectra at 3~mm to a resolution of $\sim 3$ km~s$^{-1}$, 
assuming that a good line sampling requires at least 3 points. This leads to an 
increase of the signal-to-noise ratio per spectrum by a factor of 1.6. We then 
co-added all the smoothed spectra obtained with the two receivers tuned on the 
$^{12}$CO(1--0) line and corresponding to the scans obtained at a given point 
of the HVC-A and HVC-B maps. We finally get a mosaic of 25 spectra centered on 
each of the selected HVC \ion{H}{i} cores (Figs.~\ref{HVC-A-spectra} and 
\ref{HVC-B-spectra}). No baseline subtraction was done on individual spectra 
before the co-addition, since we perform a linear sum.

%
%________________________________________________________________

\section{Results}\label{results}

In Figs.~\ref{HVC-A-spectra} and \ref{HVC-B-spectra} we show the constructed 
mosaics of 25 $^{12}$CO(1--0) spectra, centered on the two selected HVC 
\ion{H}{i} cores for which cold dust emission was recently detected by MD05.
These spectra show clearly that reliable molecular emission is not detected 
toward any position of the maps. We do observe a few tentative CO line 
detections with a significance level of $2-3$~$\sigma$: for instance, the 
spectra at the offset positions $(-24''\times +24'')$, $(0''\times-12'')$,
$(+12''\times -24'')$ and $(+24''\times -24'')$ in the HVC-A region, and 
$(-12''\times 0'')$ and $(0''\times -24'')$ in the HVC-B region. However, such 
spikes are very common in millimeter observations, occur randomly and result
only from noise effects. Moreover, the LSR velocities of these tentative 
detections differ by 10 to 90 km~s$^{-1}$ from the expected LSR velocity of the 
two selected HVC \ion{H}{i} cores as measured from 21~cm data. 
%Only lines observed above $4-5$ times the rms level can be considered as 
%detections.

Despite the lack of firm CO emission detection in our data, we can at least 
derive an upper limit on the molecular content in these HVCs which already 
provides interesting constraints. The baseline rms of our $^{12}$CO(1--0) 
spectra vary between 11 and 24 mK according to the different number of scans 
obtained at the various positions of the maps (see Figs.~\ref{HVC-A-spectra} 
and \ref{HVC-B-spectra} for more details), with a mean rms value of 14.8 mK for 
the HVC-A region and 15.4 mK for the HVC-B region. Assuming a Gaussian line 
with FWHM of 3 km~s$^{-1}$ 
\citep[similar to][ and see also Sect.~\ref{indirect}]{wakker97b}, the mean 
rms values lead to average 5~$\sigma$ CO intensity detection limits of 
$I({\rm CO}) < 0.22$ K~km~s$^{-1}$ and 0.23 K~km~s$^{-1}$, respectively, in 
$22''$ beams. The most sensitive limit is derived toward the position 
$(0'',-12'')$ of the HVC-A map with $I({\rm CO}) < 0.16$ K~km~s$^{-1}$ 
(baseline ${\rm rms} = 11$ mK). Using the CO-to-H$_2$ conversion factor 
$X({\rm CO}) = 1.9\times 10^{20}$ cm$^{-2}$~(K~km~s$^{-1}$)$^{-1}$ determined 
by \citet{strong96} from the diffuse Galactic $\gamma$-ray emission, we get 
from the most sensitive observations the 5~$\sigma$ H$_2$ column density 
detection limit of $N({\rm H}_2) < 3\times 10^{19}$ cm$^{-2}$. 
%This upper limit is, however, very uncertain given the controversy going 
%around the $X({\rm CO})$ factor. Higher conversion factors seem indeed to be 
%required in presence of low metallicity and low excitation clouds, like the 
%HVCs in the Complex~C (see Sect.~\ref{discussion}).

In the case that the molecular gas is diffuse rather than clumpy, the
co-addition of all the $^{12}$CO(1--0) spectra of each HVC map is justified in 
order to obtain a CO intensity measurement over a larger beam hence equivalent 
to a $70''$ beam. The co-addition of spectra of the 25 positions leads to very 
good baseline rms of 2.9 and 3.0 mK for the HVC-A and HVC-B regions, 
respectively, which correspond to a 5~$\sigma$ CO intensity detection limit of 
$I({\rm CO}) < 0.04$ K~km~s$^{-1}$. This limit can then be directly compared 
with the upper limit $I({\rm CO}) < 0.077$ K~km~s$^{-1}$ derived by 
\citet{wakker97b}, since it was determined over a similar beam size. We 
improved the sensitivity of the CO observations by a factor of 2, but this is 
still not sufficient for detecting CO emission in HVCs. 
%At a first glance, this does not favor the presence of diffuse molecular gas 
%in the HVCs. 

%As a general result, either the molecular gas in HVCs has CO emission lines 
%much weaker than those reached by the sensitivity of the current millimeter 
%instruments, or the molecular gas in HVCs has a clumpy structure with clump 
%sizes much smaller than those reached by the angular resolution of the IRAM 
%30~m telescope. This will be discussed in the next section.

%
%________________________________________________________________

\section{Discussion}\label{discussion}

We collected a new set of CO observations toward two \ion{H}{i} cores of HVCs 
in Complex~C for which cold dust emission was recently detected by MD05. 
Although our data have a three times better angular resolution and a two times 
better sensitivity than those of \citet{wakker97b}, no firm CO emission is 
detected. How can we reconcile this persistent non-detection of CO emission in 
HVCs with the possible presence of cold dust as suggested by MD05 and with the 
recent FUSE results in IVCs and HVCs~?

%
%________________________________________________________________

\subsection{Direct implications of the CO non-detection}\label{direct}

Half of the investigated IVCs with FUSE show the presence of molecular hydrogen 
in absorption \citep{richter03b,wakker06,gillmon06}. The total H$_2$ column 
densities measured are $\log N({\rm H}_2) = 14.1-16.4$. This widespread 
existence of H$_2$ with low molecular fractions, $\log f$, varying between 
$-1.4$ (upper limits) and $-5.3$ (measured values) and low volume densities 
estimated to be $n_{\rm H} = 10-50$ cm$^{-3}$ from the H$_2$ 
formation-dissociation equilibrium, very likely traces the cold, {\it diffuse} 
medium of the halo clouds, because of its large filling factor 
\citep{richter03b}. Indeed, the chance of detecting such a diffuse H$_2$ 
absorber by way of absorption spectroscopy toward a limited number of 
background sources is much higher than that of finding a small, dense clump 
which would have a small angular extent: for a 10 times higher volume density, 
the linear size of the molecular clouds is 10 times smaller, if the \ion{H}{i} 
column density stays constant; and thus, the filling factor is already 100 
times lower when assuming a spherical geometry.

The sightlines passing through HVCs do not seem to show the same ubiquity of 
H$_2$ as the IVCs. Indeed, H$_2$ was found toward the Magellanic Stream with 
$\log N({\rm H}_2) = 16.4-18.2$ \citep{sembach01,richter01b,wakker06}, but no 
H$_2$ is for instance detected in Complex~C down to limits of 
$\log N({\rm H}_2) < 13.8$ \citep{richter01b,wakker06}. This shows that the low 
FUV radiation field, at least as weak as in the IVCs reducing the UV 
photodissociation of H$_2$, and the relatively high \ion{H}{i} column densities 
sampled of up to $10^{20}$ cm$^{-2}$ are not sufficient to allow diffuse H$_2$ 
to exist at moderate densities. The $\sim 0.1$ sub-solar metallicity and the 
abundance patterns measured in Complex~C \citep{tripp03} indicating a low dust 
content led \citet{richter03b} to conclude that the H$_2$ formation on the 
surface of dust grains working as catalysts must be suppressed and that the 
alternative gas phase H$_2$ formation processes \citep{black78} are not 
efficient enough to enable diffuse molecular gas.  

The non-detection of diffuse H$_2$ together with the non-detection of CO 
emission despite a two-fold better sensitivity reached in our observations in 
comparison with the \citet{wakker97b} data do not provide at a first glance any 
evidence that the HVCs in Complex~C contain large amounts of molecular gas. 
Hence these results do not confirm the findings of MD05 and bring into question 
the reliability of their cold dust emission detection.

A close inspection of the MD05 analysis indicates that the derived contribution 
of the HVCs to the infrared emission is at the level of the residual term they 
find from the decomposition of the measured infrared signal into the different 
\ion{H}{i} components identified in the 21~cm observations, namely the HVCs, 
two IVCs, and the local Galactic ISM (see Table~1 in MD05). The determination 
of the dust component of the HVCs is thus strongly dependent on the fit of the 
total infrared emission. In particular, the {\it cold} nature of the dust in 
the HVCs is crucially dependent on the fit of the dust emission at 100 $\mu$m, 
with a derived HVC emissivity 10 times smaller than that of the local ISM gas, 
against only a factor of 2 smaller at 160 $\mu$m. The interpretation of 
Complex~C infrared data by MD05 raises incertainties. Their results thus 
require confirmation by additional measurements.

%
%________________________________________________________________

\subsection{The complexity of the physical conditions at play}\label{indirect}

The situation, however, is even more complex and different configurations can 
explain the non-detection of both the CO emission and the diffuse H$_2$ without 
implying the absence of molecular gas in the HVCs of Complex~C. The presence of 
cold dust emission as supported by MD05 and the lack of CO emission may thus 
well be compatible.

Molecular clumps with high volume densities certainly exist in HVCs, but they
are not traced by the diffuse molecular gas component. These clumps should 
contain molecular hydrogen at a high fractional abundance, since the 
dissociating FUV field in the halo clouds is not intense and the H$_2$ 
formation timescale becomes shorter at high densities ($t_{\rm form}\propto 
n_{\rm H}^{-1}$). Our CO observations with a detection limit 5 orders of 
magnitude higher than that of H$_2$ absorption studies, precisely aim to detect 
the emission peaks of these clumps. It has, indeed, been demonstrated over the 
last few years that the cold neutral component of the ISM consists of 
substantial small-scale structures at AU scales, from 1 pc down to 10 AU (the 
resolution limit of the current observations), indicating ubiquitous tiny 
\ion{H}{i} gaseous clumps with temperatures as low as 50~K and very high 
densities, $n_{\rm H}\sim 10^3-10^5$ cm$^{-3}$ 
\citep[e.g.][]{faison98,lauroesch00}. Recently, the CO emission of such tiny 
molecular clumps with sizes of a few hundred AU, temperatures between 7 and 
18~K, volume densities of a few $10^4$ cm$^{-3}$, and column densities of  
$\log N({\rm H}_2) = 19-21$ have been observed in cirrus clouds and have 
revealed that these clumps are fractally structured 
\citep{heithausen04,heithausen06}. 

Such AU-scale molecular clumps could have easily been diluted in 
\citet{wakker97b} observations made at an angular resolution of $1'$ in
contrast to the $3''$ resolution data of Heithausen, and have therefore been 
undetected. Our new survey for CO emission in HVCs benefits from a three times 
better angular resolution. The size of the clumps resolved in our $22''$ beam 
width hence is three times lower, namely 0.5 pc when assuming that the HVCs in 
Complex~C are about 5 kpc away from the Galactic plane\footnote{If the HVCs in 
Complex~C are at a greater distance than 5 kpc from the Galactic plane, the 
resolved size of the clumps is less than 0.5 pc.} \citep{wakker01b}. A 0.5 pc 
resolution sampled by the observations is within the range of reported sizes 
of molecular clouds, and hence our observations should be less affected by the 
dilution effect. However, we still do not detect the CO emission in HVCs. This 
suggests that the dilution problem remains for the very dense clumps. In the 
absence of agitation and heating by star formation, the fraction of dense and
thus small clumps can indeed be higher. Assuming that the HVCs have a similar 
size distribution of molecular clumps as the cirrus clouds 
\citep{heithausen04}, %Indeed, when taking at face the \citet{heithausen04} 
%results, the molecular clumps with 
H$_2$ column densities higher than our detection limit, $N({\rm H}_2) = 3\times 
10^{19}$ cm$^{-2}$, are hence expected to be confined in very small clumps with 
20 times smaller sizes than our 0.5 pc resolution. Although present, they thus 
are still highly diluted in our observations made with the IRAM 30~m telescope. 

Moreover, assuming the same fractal structure for molecular clouds in the HVCs 
as that traced by the size-line width relation \citep[e.g.][]{solomon87}, we 
expect a series of clumps of all sizes in our beam. The global ensemble is 
composed of molecular clumps of the order of 1 pc size (the resolution scale) 
and with a dispersion in velocity $\sigma_\upsilon \sim 1$ km~s$^{-1}$ or a 
FWHM of about 3 km~s$^{-1}$. It is only at that size scale that we detect 
resolved clumps which hence have an average surface density lower than $3\times 
10^{19}$ cm$^{-2}$ as derived from the CO emission detection limit, or a volume 
density lower than 20 cm$^{-3}$, at 5~$\sigma$. This is much lower than the 
minimal density required to excite the CO molecules which is about $10^3$ 
cm$^{-3}$. As a consequence, the inter-clump gas already at the 1 pc scale 
appears to be too diffuse to excite the CO molecules.

Alternatively, the non-detection of CO emission might be ascribed to the 
peculiar physical conditions that reign in the HVCs of Complex~C. First, the 
sub-solar metallicity of $0.1-0.3$ dex measured by \citet{tripp03} in these 
HVCs might be decisive. As stated in Sect.~\ref{direct}, \citet{richter03b} 
have invoked the argument that the lower dust-to-gas ratio associated with 
low metallicity media leads to the suppression of the most efficient H$_2$ 
formation process onto dust grains in order to explain the non-detection of 
diffuse H$_2$ in Complex~C. However, the unknown dust properties in HVCs, e.g. 
the grain size distribution and the grain surface, make the implications of low
metallicities on the H$_2$ grain formation rate very uncertain. It has, in 
addition, been empirically shown that the lack of heavy elements also appears 
to directly affect the relationship between the tracer molecule CO and H$_2$ in 
the way that large quantities of H$_2$ molecular gas are not necessarily traced 
by large CO emission. In low metallicity media, an underabundance of dust may 
result in more intense radiation fields, and therefore while H$_2$ survives 
because of self-shielding, this does not stop photodissociation of CO
\citep[e.g.][]{maloney88}.

\citet{leroy07} recently derived a new estimation of the CO-to-H$_2$ conversion 
factor over the entire Small Magellanic Cloud (SMC), characterized by a low 
metallicity similar to that of the HVCs in Complex~C. They found $X({\rm CO}) = 
13\times 10^{21}$ cm$^{-2}$~(K~km~s$^{-1}$)$^{-1}$ in agreement with the value 
previously obtained by \citet{israel97} in the SMC, namely about 60 times 
higher than the conversion factor $X({\rm CO}) = 1.9-2.3\times 10^{20}$ 
cm$^{-2}$~(K~km~s$^{-1}$)$^{-1}$ determined for molecular clouds in the 
Galactic plane with approximately solar metallicities 
\citep{bloemen89,strong96}. This implies that for the same H$_2$ column density 
the corresponding integrated intensity of CO is 60 times {\it lower} in low 
metallicity than in solar metallicity molecular clouds. As a consequence, the 
detection of molecular gas in low metallicity HVCs suffers from the low CO 
intensity, and may result in a drastic lack of sensitivity for detecting CO 
emission despite a possibly large amount of H$_2$ molecular gas in HVCs. This 
can perfectly reconcile the presence of cold dust emission as claimed by MD05 
and the CO emission non-detection. MD05 estimated the total gas column density 
of the HVCs to be more than 5 times higher than observed in \ion{H}{i}. This 
cold gas ($T_{\rm HVC} = 10.7$ K) component if entirely associated with 
molecular gas implies a high H$_2$ column density of at least 
$N({\rm H}_2)\approx 3.5\times 10^{20}$ cm$^{-2}$. Using the CO-to-H$_2$ 
conversion factor for SMC metallicity molecular clouds, the corresponding CO 
intensity still is very weak, $I({\rm CO})\approx 0.027$ K~km~s$^{-1}$, namely 
one order of magnitude below our CO detection limit.

Secondly, the very low temperature of $T_{\rm HVC} = 10.7^{+0.9}_{-0.8}$~K
estimated in the HVCs of Complex~C by MD05 from the infrared dust emission 
might have severe consequences. Indeed, solving a Clausius-Clapeyron type 
equation for the critical vapor pressure of CO ice sublimation 
$\ln P = a/T + b$ using the CO triple point ($T_3 = 68.14$ K, $P_3 = 
0.1535\times 10^6$ dyn) and the CO critical point ($T_c = 132.91$ K, $P_c = 
34.987\times 10^6$ dyn)\footnote{Encyclop\'edie des gaz, Gas Encyclopaedia, 
L'Air Liquide, 1976, Elsevier, Amsterdam}, we find $P = 1.04\times 10^{10}
\exp(-759/T)$ dyn. Below 20~K the CO molecule can be in a solid state even 
under interstellar medium pressure conditions. This implies that CO may 
condense onto dust grains and hence be depleted in gas-phase observations. 
\citet{leger83} has examined this CO sublimation effect in dark molecular 
clouds, and current measurements in starless/pre-stellar cores show evidence 
for CO depletion by a factor of 10 \citep[e.g.][]{bacmann02,tafalla02}. 
Similarly, such CO depletion levels can easily be expected in HVCs if the 
temperatures are at least as low as 20 K in these clouds.

%
%________________________________________________________________

\section{Conclusions}\label{conclusions}

The recent detection of cold dust emission in the HVCs of Complex~C by MD05 
brought a new surge of hope for the presence of large amounts of molecular gas 
in HVCs and its possible detection via CO emission. We thus undertook a survey 
for molecular gas toward the HVCs observed by MD05 with the IRAM 30~m telescope
at an angular resolution of 22 arcsec. We obtained two mosaics of 25 
$^{12}$CO(1--0) spectra covering a $1'\times 1'$ area each and centered on two 
HVC \ion{H}{i} cores with column densities $N$(\ion{H}{i}) $\sim 7\times 
10^{19}$ cm$^{-2}$. No firm CO emission line at the HVC LSR velocity was 
detected. The best baseline rms value of 11 mK obtained leads to a 5~$\sigma$ 
CO intensity detection limit of $I({\rm CO}) < 0.16$ K~km~s$^{-1}$. The 
improved sensitivity and angular resolution by a factor of two and three, 
respectively, compared with the CO observations made by \citet{wakker97b} in 
HVCs, show that this still is not sufficient for detecting the CO emission in 
HVCs.

The non-detection of both the CO emission and the diffuse H$_2$
\citep{richter01b,wakker06} in the HVCs of Complex~C do not provide at a first
glance any evidence for large amounts of molecular gas in these clouds, and
hence do not support the findings by MD05. Although the reliability of their 
analysis may be questioned and requires confirmation by additional 
measurements, our negative CO detection does not, however, allow their results
to be refuted, because the physical conditions at play are far too complex and
still not well constrained. We can invoke different configurations which enable 
reconciliation of the CO emission non-detection with the presence of molecular 
gas and hence of cold dust emission:

\begin{enumerate}

\item[(i)] Our CO data may still suffer strong dilution effects: \\ The 
typical clouds resolved in the $22''$ beam width are of 0.5 pc if the HVCs in 
Complex~C are at a distance of about 5 kpc. They enter the size range observed 
in molecular clouds. However, taking at face value the \citet{heithausen04} 
results obtained in cirrus clouds, the molecular clumps with H$_2$ column 
densities higher than our detection limit, $N({\rm H}_2) = 3\times 10^{19}$ 
cm$^{-2}$, should still have about 20 times smaller sizes compared to our 
resolution. This suggests that most of the molecular gas may be confined in 
very small and dense clumps in HVCs which are then highly diluted in our 
observations.

\item[(ii)] The CO non-detection may be the result of a too diffuse inter-clump 
medium, filling a typical molecular cloud, to excite the CO molecules: \\ The 
global ensemble of molecular clumps resolved in our beam are of the order of 
1~pc. Since these clumps are undetected, they have an average surface density 
lower than our detection limit. This corresponds to a volume density lower than 
20 cm$^{-3}$ at 5~$\sigma$, namely lower than the minimal density required for 
the CO excitation.

\item[(iii)] The sub-solar metallicity of $0.1-0.3$ dex measured in the HVCs of
Complex~C may play an important role: \\ Partly because the lower dust-to-gas 
ratio may lead to the suppression of the most efficient H$_2$ formation process 
onto dust grains. And more specifically, because as \citet{leroy07} showed, the 
CO-to-H$_2$ conversion factor measured at the SMC metallicity is 60 times 
higher than at the solar metallicity. This implies that for a given H$_2$ 
column density, the corresponding integrated CO intensity will be 60 times 
lower in low metallicity molecular clouds. As a consequence, despite the large 
molecular gas content estimated by MD05 from the cold dust emission detected in 
the HVCs of Complex~C, the expected CO intensity is $I({\rm CO})\approx 0.027$ 
K~km~s$^{-1}$, namely one order of magnitude below our CO detection limit.

\item[(iv)] CO may be partly depleted in gas-phase observations:\\ MD05
determined a very low temperature of $\sim 10.7$~K for the dust in the HVCs of 
Complex~C. If correct, this implies the presence of molecular gas cold enough 
to allow CO condensation onto dust grains under ISM pressure conditions.

\end{enumerate}

These results illustrate the difficulty of determining the molecular gas 
content in the HVCs. The Atacama Large Millimeter Array, ALMA, should achieve 
at least a 10 times better sensitivity and angular resolution than the current 
millimeter instruments, and will hence hopefully satisfy the requirements for 
detecting CO emission in physical media with characteristics similar to those 
of HVCs. The determination of the physical conditions and origins of HVCs is 
crucial to better understand their implications in the formation and evolution 
of galaxies that may be in multiple forms, for example as gas fuel for star 
formation, as potential tracers of small dark matter halos, and as low 
metallicity gas for chemical evolution.

%
%________________________________________________________________

\begin{acknowledgements}

The authors wish to thank the IRAM 30~m staff for their valuable support and 
advice. M.D.-Z. is, in particular, grateful to Sergio Mart\'in for allowing her 
to benefit of his large millimeter observation experience. M.D.-Z. also extends 
special thanks to Brad K. Gibson for helpful discussion on high-velocity clouds 
during his stay at Geneva Observatory.

\end{acknowledgements}

%
%________________________________________________________________

%
%________________________________________________________________

\end{document}